\documentclass[a4paper,fleqn,10pt]{article}
\usepackage{amsmath}
\usepackage{graphicx}
\usepackage[merge,numbers,compress]{natbib}
\usepackage{booktabs}
\usepackage{xcolor}
\usepackage{todonotes}
\usepackage{xspace}
\usepackage{dcolumn}
\usepackage{placeins}
\usepackage{makecell}
\usepackage[colorlinks=true,citecolor=blue!50!black,linkcolor=black]{hyperref}
\usepackage{caption}
\usepackage
[subrefformat=parens,position=top,skip=-15pt,margin=15pt,justification=justified,singlelinecheck=false]
{subcaption}
\usepackage{authblk}

\usepackage{mciteplus}
\mciteErrorOnUnknownfalse 
\usepackage{geometry}
\usepackage{sectsty}
\usepackage{siunitx}
\let\spreprint\empty
\newcommand{\preprint}[1]{\def\spreprint{\protect#1}}
\let\sinstitute\empty

\makeatletter
\renewcommand{\maketitle}{\begingroup
  \null\thispagestyle{empty}%
    \ifx\spreprint\empty
      \vskip 5ex
    \else
      \flushright\large\spreprint\vskip 2ex
    \fi
    \vskip 5ex
    \flushleft
      {\sffamily\bfseries\huge\@title}\vskip 2ex
      \@author\vskip 2ex
      \ifx\sinstitute\empty
      \else
        {\small\sinstitute}
      \fi
    \vskip 5ex
  \endgroup
}
\makeatother

\renewenvironment{abstract}{\begin{center}
  {\large\sffamily\bfseries Abstract: }
  \begin{minipage}[t]{0.75\textwidth}
}{\end{minipage}\end{center}\vskip 10ex}

\allsectionsfont{\normalfont\sffamily\bfseries}
\subsectionfont{\normalfont\sffamily\bfseries}
\subsubsectionfont{\normalfont\sffamily\bfseries}

\makeatletter
\DeclareRobustCommand*{\bfseries}{%
  \not@math@alphabet\bfseries\mathbf
  \fontseries\bfdefault\selectfont
  \boldmath
}
\makeatother

\newcommand{\la}{\lambda_\alpha}

\def\beq{\begin{equation}}  
\def\eeq{\end{equation}}
\def\({\left(}
\def\){\right)}
\def\[{\left[}
\def\]{\right]}

\newcommand{\alphas}{\ensuremath{\alpha_\text{s}}\xspace}

\newcommand{\pt}{\ensuremath{p_\text{T}}\xspace}

\newcommand{\sherpa}{S\protect\scalebox{0.8}{HERPA}\xspace}
\newcommand{\pythia}{P\protect\scalebox{0.8}{YTHIA}\xspace}

\newcommand{\herwig}{H\protect\scalebox{0.8}{ERWIG}\xspace}

\newcommand{\comix}{C\protect\scalebox{0.8}{OMIX}\xspace}

\newcommand{\Caesar}{C\protect\scalebox{0.8}{AESAR}\xspace}
\newcommand{\fastjet}{F\protect\scalebox{0.8}{AST}J\protect\scalebox{0.8}{ET}\xspace}
\newcommand{\rivet}{R\protect\scalebox{0.8}{IVET}\xspace}
\newcommand{\recola}{R\protect\scalebox{0.8}{ECOLA}\xspace}
\newcommand{\collier}{C\protect\scalebox{0.8}{OLLIER}\xspace}
\newcommand{\OpenLoops}{O\protect\scalebox{0.8}{PEN}L\protect\scalebox{0.8}{OOPS}\xspace}

\newcommand{\mgamcnlo}{M\protect\scalebox{0.8}{ADGRAPH}{5\_aMC@NLO}\xspace}

\newcommand{\softdrop}{\textsf{SoftDrop}\xspace}

\newcommand{\muR}{\ensuremath{\mu_{\text{R}}}}
\newcommand{\muF}{\ensuremath{\mu_{\text{F}}}}
\newcommand{\muQ}{\ensuremath{\mu_{\text{Q}}}}

\newcommand{\zcut}{\ensuremath{z_{\text{cut}}}}
\newcommand{\alphaS}{\alpha_\text{s}\xspace}

\newcommand{\NLO}{\text{NLO}\xspace}    
\newcommand{\NLL}{\text{NLL}\xspace}
\newcommand{\NLLp}{\ensuremath{\text{NLL}^\prime}\xspace}
\newcommand{\NLOpNLL}{\ensuremath{\NLO+\NLL}\xspace}
    
\newcommand{\NLOpNLLp}{\ensuremath{\NLOpNLL^\prime}\xspace}
\newcommand{\NLOpNLLpNP}{\ensuremath{\NLOpNLL^\prime+\text{NP}}\xspace}

\newcommand{\BSZ}{BSZ\xspace}

\newcommand{\DIRE}{\text{\textsc{DIRE}}\xspace}

\newcommand{\MCatNLO}{\text{\textsc{MC@NLO}}\xspace}
\newcommand{\SHMCatNLO}{\text{\textsc{SH-MC@NLO}}\xspace}

\def\draftdate{\relax}
\def\mda{\relax}
\def\mua{\relax}
\def\mla{\relax}
\def\draft{
\def\thtystars{******************************}
\def\sixtystars{\thtystars\thtystars}
\typeout{}
\typeout{\sixtystars**}
\typeout{* Draft mode!
         For final version remove \protect\draft\space in source file *}
\typeout{\sixtystars**}
\typeout{}
\def\draftdate{\today}
\def\mua{\marginpar[\boldmath\hfil$\uparrow$]%
                   {\boldmath$\uparrow$\hfil}\color{black}%
                    \typeout{marginpar: $\uparrow$}\ignorespaces}
\def\mda{\color{red}\marginpar[\boldmath\hfil$\downarrow$]%
                   {\boldmath$\downarrow$\hfil}%
                    \typeout{marginpar: $\downarrow$}\ignorespaces}
\def\mla{\marginpar[\boldmath\hfil$\rightarrow$]%
                   {\boldmath$\leftarrow $\hfil}%
                    \typeout{marginpar: $\leftrightarrow$}\ignorespaces}
\def\Mua{\marginpar[\boldmath\hfil$\Uparrow$]%
                   {\boldmath$\Uparrow$\hfil}\color{black}%
                    \typeout{marginpar: $\uparrow$}\ignorespaces}
\def\Mda{\color{red}\marginpar[\boldmath\hfil$\Downarrow$]%
                   {\boldmath$\Downarrow$\hfil}%
                    \typeout{marginpar: $\downarrow$}\ignorespaces}
\def\Mla{\marginpar[\boldmath\hfil\textcolor{red}{$\Rightarrow$}]%
                   {\boldmath\textcolor{red}{$\Leftarrow $}\hfil}%
                    \typeout{marginpar: $\leftrightarrow$}\ignorespaces}
\overfullrule 5pt
\oddsidemargin 15mm
\marginparwidth 29mm
}

\numberwithin{equation}{section}

\usepackage{color}
\definecolor{darkblue}{rgb}{0,0,0.5}
\definecolor{darkred}{rgb}{0.5,0,0}
\definecolor{darkgreen}{rgb}{0,0.5,0}

\DeclareMathOperator{\Tr}{Tr}

\pdfoutput=1

\hypersetup{
  pdfauthor={Steffen Schumann et al.},
  pdftitle={Phenomenology of jet angularities at the LHC}
}

\preprint{MCNET-21-36\\IPPP/21/57}

\usepackage[symbol]{footmisc}

\author[1,2]{Daniel~Reichelt\footnote[1]{Email: \texttt{daniel.reichelt@durham.ac.uk}}}
\author[3]{Simone~Caletti\footnote[2]{Email: \texttt{simone.caletti@ge.infn.it}}}
\author[3]{Oleh~Fedkevych\footnote[3]{Email: \texttt{oleh.fedkevych@ge.infn.it}}}
\author[3]{Simone~Marzani\footnote[4]{Email: \texttt{simone.marzani@ge.infn.it}}}
\author[2]{Steffen Schumann\footnote[5]{Email: \texttt{steffen.schumann@phys.uni-goettingen.de}}}
\author[4]{Gregory~Soyez\footnote[6]{Email: \texttt{gregory.soyez@ipht.fr}}}

\affil[1]{\parbox[t]{\textwidth}{Institute for Particle Physics Phenomenology, Department of Physics, Durham University,\newline
  South Road, Durham DH1 3LE, United Kingdom}}
\affil[2]{\parbox[t]{\textwidth}{Institut f{\"u}r Theoretische Physik, Georg-August-Universit{\"a}t
  G\"ottingen,\newline
  Friedrich-Hund-Platz 1, 37077 G\"ottingen, Germany}}
\affil[3]{\parbox[t]{\textwidth}{Dipartimento di Fisica, Universit\`a di Genova and INFN, Sezione di Genova,\newline Via Dodecaneso 33, 16146, Genoa, Italy}}
\affil[4]{\parbox[t]{\textwidth}{Institut de Physique Th\'eorique, Paris Saclay University, CNRS, CEA,\newline Orme des Merisiers, B\^at 774, F-91191 Gif-sur-Yvette, France}}

\title{Phenomenology of jet angularities at the LHC}
\begin{document}
\maketitle

\begin{abstract}
We compute resummed and matched predictions for jet angularities in
  hadronic dijet and $Z$+jet events with and without grooming the candidate jets
  using the \softdrop technique. 
  Our theoretical predictions also account for non-perturbative
  corrections from the underlying event and hadronisation through
  parton-to-hadron level transfer matrices extracted from dedicated
  Monte Carlo simulations with \sherpa.
  Thanks to this approach we can account for non-perturbative
  migration effects in both the angularities and the jet transverse
  momentum.
  We compare our predictions against recent measurements from the CMS
  experiment. This allows us to test the description of quark- and
  gluon-jet enriched phase-space regions separately.
  We supplement our study with \sherpa results based on the matching
  of NLO QCD matrix elements with the parton shower.
  Both theoretical predictions offer a good description of the data,
  within the experimental and theoretical uncertainties.
  The latter are however sizeable, motivating higher-accuracy
  calculations.
\end{abstract}


\clearpage
\vspace{10pt}
\noindent\rule{\textwidth}{1pt}
\tableofcontents
\noindent\rule{\textwidth}{1pt}
\vspace{10pt}

\section{Introduction}

The LHC data-taking campaigns during run I and II recorded huge amounts of
high-quality data which can be explored to perform high-precision tests of the
Standard Model. Recent studies have demonstrated the particular potential of jet
cross-section measurements. For example, jet data can be used in determinations
of the strong coupling constant, see
\textit{e.g.}~\cite{Britzger:2017maj,CMS:2013vbb, ATLAS:2017qir,
  ATLAS:2015yaa, CMS:2014mna}, and form an important input for constraining
parton distribution functions (PDFs)~\cite{ATLAS:2021qnl, ATLAS:2013pbc,
  CMS:2014qtp, CMS:2016lna, Ball:2021leu,Bailey:2020ooq,AbdulKhalek:2020jut, Harland-Lang:2017ytb,
  Pumplin:2009nk, Watt:2013oha}.
Information on the physical processes underlying jet formation can be gained by
studying so-called jet substructure observables (for recent reviews see
\textit{e.g.}~\cite{Larkoski:2017jix,Kogler:2018hem,Marzani:2019hun}).
In this context, a popular jet substructure technique used to remove radiation in
phase-space regions dominated by non-perturbative physics is  \softdrop
grooming~\cite{Larkoski:2014wba}, including the modified Mass Drop
Tagger (mMDT) as a special case~\cite{Butterworth:2008iy,Dasgupta:2013ihk}.
Physics observables measured on \softdrop jets have been shown
to be less affected by non-perturbative corrections and to be amenable
to precision calculations in
QCD~\cite{Marzani:2017mva,Marzani:2017kqd,Frye:2016aiz,Kardos:2020gty,Kardos:2020ppl,Larkoski:2020wgx}.
This has triggered the attention of the jet substructure community,
both theorists and
experimentalists~\cite{ATLAS:2017zda,CMS:2018vzn,CMS:2021iwu}, over the
past few years.

One of the central goals of jet substructure analyses is to determine
differences between jets originating from quarks versus gluons. This
information can be exploited, for instance, to enhance sensitivity in
searches for new physics. Consequently, the consistent modelling of
quark and gluon jets has been an active area of research over the past
years, see for example~\cite{Larkoski:2014pca,Andersen:2016qtm,
  Gras:2017jty, Mo:2017gzp, Reichelt:2017hts, Reichelt:2017wwt,
  Larkoski:2019nwj,Amoroso:2020lgh, Caletti:2021ysv}. An often-studied
benchmark, widely used in many of the above references, is the
so-called generalised jet angularities~\cite{Berger:2003iw,
  Almeida:2008yp}. A sound theoretical understanding of these
angularities is thus critical to the task of quark--gluon
discrimination. These observables, and closely related ones, have
hence been studied in a variety of theoretical
frameworks~\cite{Berger:2003iw, Berger:2002ig, Banfi:2004yd,
  Ellis:2010rwa, Larkoski:2013eya, Larkoski:2014uqa, Hornig:2016ahz,
  Kang:2018qra, Kang:2018vgn, Aschenauer:2019uex, Caletti:2021oor,
  Zhu:2021xjn}. Recent measurements of jet angularities at the LHC
have for example been presented in~\cite{ATLAS:2019kwg, ALICE:2021njq,
  CMS:2018ypj,CMS:2021iwu}.

In this paper we focus on the wealth of data presented by the CMS
experiment in~\cite{CMS:2021iwu}, which measured angularities, with
and without \softdrop grooming, on jets produced either in association
with a $Z$ boson or in dijet events.  In an earlier article,
\cite{Caletti:2021oor}, we already presented \NLOpNLLp accurate
predictions for the $Z$+jet case, derived using the resummation plugin
to the \sherpa event generator
framework~\cite{Gerwick:2014gya,Sherpa:2019gpd}.  Here, we aim to
extend this study to the case of dijet production. In this context, we
introduce an improved prescription to apply non-perturbative
corrections, extracted from Monte Carlo (MC) simulations, to our
resummed analytic predictions.
Since $Z$+jet and dijet events receive different contributions from
quark- and gluon-initiated jets, a combined study of (groomed) jet
shapes such as the angularities in both channels is expected to shed light
on our fundamental understanding of jet formation from hard quarks and gluons. 
To that aim, besides providing results for the angularity differential
distributions, we also consider the angularities mean values in quark-
and gluon-dominated phase-space regions of $Z$+jet and dijet final
states.

The article is organised as follows: in Sec.~\ref{sec:definitions} we introduce
the considered angularity observables and define the fiducial phase space used in
our study of jets in $Z$+jet and dijet production. In Sec.~\ref{sec:method} we introduce
the theoretical methods used for making predictions at \NLOpNLLp accuracy, based on
the implementation of the \Caesar formalism in \sherpa. Here, we also present our new
approach to account for non-perturbative corrections based on transfer matrices
extracted from MC simulations, see Sec.~\ref{sec:res_np}. We present our
final \NLOpNLLpNP results in Sec.~\ref{sec:results}, alongside
with MC predictions from \sherpa, and compare those against data from the
CMS experiment. We compile our conclusions and give an outlook in Sec.~\ref{sec:conclusions}.

\section{Phase space and observable definition}\label{sec:definitions}

To be able to directly compare to the CMS measurement presented
in~\cite{CMS:2021iwu}, based on proton--proton collisions at a
centre-of-mass energy of $\sqrt{s} = \SI{13}{TeV}$, the exact
definition of angularities and the event-selection criteria have to be
carefully matched to the experimental setup. For completeness, this
section summarises the final definitions and cuts and we refer the
reader to Ref.~\cite{CMS:2021iwu} for additional details and motivations.

The events considered are conceptually triggered either by an
approximately on-shell $Z$ boson, decaying into a pair of muons, accompanied by
a hard jet ($Z$+jet case), or by two high-$\pt$ jets (dijet case). The precise phase space for the two cases is
defined by the selection criteria summarised in table~\ref{tab:cms_cuts}. Substructure
observables are then calculated for the jet with the highest $\pt$ in the
$Z$+jet case. For dijet events, the two jets with the largest transverse momenta are
considered and the measurement is performed separately for the more central
(smaller rapidity $|y|$) and more forward (larger $|y|$) of those jets. For the
$Z$+jet process $p_{T,\text{jet}}$ bins in the interval $[50, 1500]$~GeV were
considered, while in the dijet case this range was extended to $[50, 4000]$~GeV.

The substructure observables we are interested in belong to the family of jet
angularities. These probe both the angular and the transverse momentum distribution
of particles within a given jet. They are defined from the momenta of
jet constituents as follows:
\begin{equation}\label{eq:ang-def}
\la^\kappa= \sum_{i \in \text{jet}}\left(\frac{p_{T,i}}{\sum_{j \in \text{jet}} p_{T,j}}\right)^\kappa\left(\frac{\Delta_i}{R} \right)^\alpha\,.
\end{equation}
Here
\begin{equation}\label{eq:dist-def}
\Delta_i=\sqrt{(y_i-y_\text{jet})^2+(\phi_i-\phi_\text{jet})^2}\,, 
\end{equation}
is the azimuth-rapidity distance of particle $i$ from the jet axis,
which is determined using the Winner-Take-All (WTA) recombination
scheme~\cite{Larkoski:2014uqa} for $\alpha\leq 1$, and using the
default $E$-scheme axis otherwise. $R$ denotes the radius parameter
used for the anti-$k_t$~\cite{Cacciari:2008gp} jet clustering.  We
restrict our analysis to the infrared-and-collinear-safe case
$\kappa = 1$ with $\alpha = 0.5,1,2$, while Ref.~\cite{CMS:2021iwu} in
addition considered the infrared-unsafe combinations
$\kappa=0,\,\alpha=0$ and $\kappa=2,\,\alpha=0$. For the different
choices of $\alpha$ we follow the common naming
convention~\cite{Larkoski:2014pca,Badger:2016bpw,CMS:2021iwu}, and
refer to $\lambda^1_{0.5}$ as Les Houches Angularity (LHA),
$\lambda^1_1$ as jet width, and $\lambda^1_2$ as jet thrust.

The CMS collaboration also measured the same jet angularities based on
the jet constituents obtained after \softdrop grooming, as implemented
in the \fastjet~\cite{Cacciari:2011ma} ~\verb|contrib| package. The
\softdrop parameters are chosen as $\beta=0$ and
$z_\text{cut}=0.1$. All results are provided either using all the
particles in the jet (both charged and neutral), or using only the
charged jet constituents. 
Note that the $p_{T,\text{jet}}$ bin is always determined from the
full jet, without any grooming or restriction on the charge of
particles.

\begin{table} 
\begin{center}
	\begin{tabular}{ | c | c |}
    \hline
        Process 		& Selection cuts 		\\ \hline
		
		$Z$+jet 	    & \makecell[l]{Require at least two muons  
		                           with  $p_{T,\mu} > 26~\mathrm{GeV},\;|\eta_\mu|<2.4,$ \\
		                           oppositely charged with invariant mass $m_{\mu^+\mu^-} \in [70, 110]$ GeV and $p_{T,\mu^+\mu^-}>30$ GeV. \\
		                           The leading (highest $\pt$)
                                           $R=0.4$ (or $R=0.8$) anti-$k_t$ jet has
                                           to satisfy $|y_{\rm jet}| < 1.7$, \\
		                           $\left|\phi_{\mu^+\mu^-} - \phi_{\rm jet}\right| > 2$ and \\ 
		                           $\left|p_{T,\rm jet} - p_{T, \mu^+\mu^-}\right|  / \left|p_{T,\rm jet} + p_{T, \mu^+\mu^-}\right| < 0.3$\,. }\\ \hline
		 Dijet			& \makecell[l]{Require at least two $R=0.4$
                   (or $R=0.8$) anti-$k_t$ jets with $\pt > 30~\text{GeV}$.\\ The two leading (highest
                   $\pt$) jets have to satisfy $|y_{\rm jet}| < 1.7$, $p_{T,\rm jet} > 30$ GeV, \\
		                              $\left|\phi_{j_1} - \phi_{j_2}\right| > 2$ and \\
		                              $\left|p_{T,j_1} - p_{T,j_2}\right| / \left|p_{T,j_1} + p_{T,j_2}\right| < 0.3$\,. }\\ \hline	
	\end{tabular}
	\captionof{table}{Event-selection cuts used by the CMS collaboration in~\cite{CMS:2021iwu}.
          The muons used to reconstruct the $Z$ boson are considered as ``bare'', \emph{i.e.}\ without QED emissions,
          and are excluded from the input to the jet clustering.}
	\label{tab:cms_cuts}
\end{center}
\end{table}

\section{Theoretical framework for jet-angularity predictions}\label{sec:method}

We here consider two sets of theoretical predictions for jet angularities in $Z$+jet and dijet production
to confront with the CMS data. The first are simulations with the \sherpa multi-purpose event generator.
The jet-angularity distributions presented in~\cite{CMS:2021iwu} were compared
against several event-generator predictions. The highest accuracy was reached in two simulations
based on LO merging~\cite{Alwall:2007fs,Hoeche:2009rj} of tree level matrix
elements with additional jets dressed with parton showers, one using
\mgamcnlo~\cite{Alwall:2014hca} in conjunction with
\pythia~\cite{Sjostrand:2014zea}, the other one using
\sherpa~\cite{Gleisberg:2008ta, Sherpa:2019gpd}.
In addition, results obtained from the parton showers of \pythia,
\herwig~\cite{Bellm:2015jjp,Bellm:2019zci} and \sherpa, based on Born matrix
elements, were presented. We here provide
\sherpa predictions at NLO accuracy for the $Z$+jet and $jj$ processes, which
were not covered in~\cite{CMS:2021iwu}.

The second set of predictions is obtained from \NLOpNLLp semi-analytic
calculations, based on the implementation of the \Caesar resummation
formalism~\cite{Banfi:2004yd} in the \sherpa
framework. In~\cite{CMS:2021iwu} the CMS results for $Z$+jet
production were already compared against our \NLOpNLLp predictions
from~\cite{Caletti:2021oor}.  We here extend this comparison to the
case of dijet production. In addition, we introduce a more physical
way to apply non-perturbative corrections to our calculations, that we
refer to as \emph{transfer-matrix approach} which we also apply 
to both the $Z$+jet and dijet cases.

In the following, we describe our setup for the MC simulations with
\sherpa, then introduce the necessary additions to our resummation
framework to carry out the \NLOpNLLp calculation for the dijet case,
and finally introduce the transfer-matrix approach.

\subsection{Monte Carlo simulations with \sherpa}\label{sec:sherpasetup}

We compile hadron-level predictions for jet angularities with the
\sherpa~\cite{Gleisberg:2008ta, Sherpa:2019gpd} event generator, version 2.2.11,
using the NNPDF-3.0 NNLO PDF set \cite{NNPDF:2014otw}. We consider inclusive dijet and
$\mu^+\mu^-$+jet production based on the \sherpa implementation of the MC@NLO formalism~\cite{Hoeche:2012yf},
dubbed \SHMCatNLO in what follows, matching the NLO QCD matrix elements for two-jet
and $\mu^+\mu^-$+jet production with the \sherpa Catani--Seymour dipole shower~\cite{Schumann:2007mg}.
The involved QCD one-loop amplitudes we obtain from \OpenLoops~\cite{Buccioni:2019sur},
using the \collier library~\cite{Denner:2016kdg} for the evaluation of tensor and scalar integrals.
The central values for the perturbative scales entering the calculation are set to  
\begin{eqnarray}
  \mu_{\text{F}}=\mu_{\text{R}}=\mu_\text{Q}=\sqrt{m^2_{\mu^+\mu^-}+p^2_{T,\mu^+\mu^-}}\,\quad \quad&:&\quad Z+\text{jet process}\,,\label{eq:scalesZj}\\
  \mu_{\text{F}}=\mu_{\text{R}}=H_T/2\,,\quad \mu_{\text{Q}}=p_{T,\text{jet}}/2\quad&:&\quad \text{dijet process}\,.\label{eq:scalesjj}
\end{eqnarray}
To estimate the perturbative uncertainties of our predictions, we
perform on-the-fly~\cite{Bothmann:2016nao} $7$-point
variations~\cite{Cacciari:2003fi} of the factorisation
($\mu_{\text{F}}$) and renormalisation ($\mu_{\text{R}}$) scales in
the matrix elements and the parton shower, while we keep the
parton-shower starting scale ($\mu_{\text{Q}}$) fixed as it is related
to the dipole kinematics of the parton shower.
As error estimate we then consider the envelope of the results for
\begin{equation}\label{eq:7pointvar}
  \{(\tfrac{1}{2}\muR, \tfrac{1}{2}\muF)$, $(\tfrac{1}{2}\muR,\muF),(\muR,\tfrac{1}{2}\muF)$, $(\muR,\muF)$, $(\muR,2\muF),(2\muR,\muF)$, $(2\muR,2\muF)\}\,.
\end{equation}
The \sherpa Underlying Event (UE) simulation is based on its
implementation of the Sj\"ostrand--Zijl multiple-parton interaction
model~\cite{Sjostrand:1987su}.
To account for the parton-to-hadron transition we use the \sherpa
cluster fragmentation~\cite{Winter:2003tt}.

Later, when we will use \sherpa to account for non-perturbative
corrections to the analytic calculation, we include an estimate of the
uncertainty related to the hadronisation model.
This uses the interface to \pythia 6.4~\cite{Sjostrand:2006za},
provided with \sherpa, with the Lund string fragmentation
model~\cite{Andersson:1983ia, Sjostrand:1984ic} as an alternative. In
both models the default set of tuning parameters is used,
see~\cite{Sherpa:2019gpd} for details.  However, we additionally vary
the default for the rescale-exponent parameter
$\alpha^\text{MPI}_\text{min}$ used to model the dependence of the
cutoff regulator $p_{T,\text{min}}$ for the $2\to 2$ cross-section
integration in the UE model on the hadronic centre-of-mass energy,
\emph{i.e.}\
\begin{equation}
  p_{T,\text{min}}(E_{\text{cms}})  = p_{T,\text{min}}(E_\text{ref})\left(\frac{E_\text{cms}}{E_\text{ref}}\right)^{\alpha^\text{MPI}_\text{min}}\,,
\end{equation}
where $E_\text{ref}=1.8\,\text{TeV}$ and $p_{T,\text{min}}(E_\text{ref})=2.895\,\text{GeV}$.
As new central value we here consider
\begin{equation}
  \alpha^\text{MPI}_\text{min,central} = 0.16\,,
\end{equation}  
corresponding to a somewhat enhanced UE activity with respect to the
default in public \sherpa-2.2.11, where
$\alpha^\text{MPI}_\text{min,def} = 0.244$.
This is motivated by a preliminary study comparing \sherpa predictions
for the jet angularities in particular for the low-$p_{T,\text{jet}}$
region against the data from~\cite{CMS:2021iwu}.  As systematic up-
and down-variations of the MPI activity we consider
$\alpha^\text{MPI}_\text{min,up}=0.08$ and
$\alpha^\text{MPI}_\text{min,down}=0.24$.

\subsection{\NLOpNLLp resummation for dijet final states}\label{sec:res_setup}

Throughout this paper, we target a description of the angularity
distribution at the \NLOpNLLp accuracy, whereby the exact distribution
at order $\alphaS^2$ (NLO)\footnote{Note, in this counting the powers of $\alphaS$
  from the underlying Born process, \emph{i.e.}\ $\alphaS^2$ for
  dijet and $\alphaS^1$ for $Z$+jet, are not included.} is matched to the all-order
resummation, which is relevant for small values of the observable at hand, 
carried out at the \NLLp accuracy.

The NLL resummation is performed using the \sherpa implementation of the
\Caesar formalism~\cite{Banfi:2004yd}. This plugin to \sherpa was
first presented in~\cite{Gerwick:2014gya} and further developed in
subsequent work, \emph{e.g.}\ to introduce new matching schemes~\cite{Baberuxki:2019ifp},
to include the \softdrop phase-space constraints~\cite{Baron:2020xoi}, or to support some non-global
configurations in Ref.~\cite{Caletti:2021oor}.
The master formula for the all-order resummation can be written as the
following sum over partonic channels $\delta$:
\begin{align}\label{eq:CAESAR_master}
  \Sigma_\text{res}(v)
  & = \sum_\delta \Sigma^\delta_\text{res}(v)\,,\\
  \Sigma^\delta_\text{res}(v)
  & = \int \mathop{d\mathcal{B}^\delta}
    \frac{\mathop{d\sigma^\delta}}{\mathop{d\mathcal{B}}}
    \exp\left[-\sum_{l\in\mathcal{B}^\delta}R^{\mathcal{B}^\delta}_l(L)\right]
    \mathcal{S}^{\mathcal{B}^\delta}(L)\,
    \mathcal{P}^{\mathcal{B}^\delta}(L)\,
    \mathcal{F^{\mathcal{B}^\delta}}(L)\,
    \Theta_\text{hard}^{\mathcal{B}^\delta}\,.
\end{align}
In this expression, $\mathop{d\sigma^\delta}/\mathop{d\mathcal{B}}$ is
the Born-level cross section, obtained through the \comix
matrix element generator~\cite{Gleisberg:2008fv}; the collinear
radiators $R_l$ are directly obtained from the \Caesar
paper~\cite{Banfi:2004yd} or extended to include the effect of
\softdrop~(see \emph{e.g.}\ Refs.~\cite{Larkoski:2014wba,Baron:2020xoi});
$\mathcal{S}$ is the soft function including global and non-global
colour evolution; $\mathcal{P}$ includes corrections associated with
PDF evolution for initial-state radiation (which are absent in our
case, \emph{i.e.}\ $\mathcal{P}=1$); $\mathcal{F}$ is the multiple-emission function, which for our
application is simply given
by $e^{-\gamma_ER^\prime}/\Gamma(1+R^\prime)$ with $R^\prime=\partial_LR$
since angularities are additive observables. Finally, $\Theta_\text{hard}$
implements the cuts on the Born-level phase space, \emph{cf.}\ table~\ref{tab:cms_cuts}.
The phase-space integration as well as the separation over different
partonic channels is greatly facilitated by
the integration of the plugin into the \sherpa framework.
With this approach, one reaches NLL accuracy including an estimate of
the resummation scale uncertainty (introducing a rescaling parameter
$x_L$), as well as a treatment of the endpoint corrections and
associated uncertainties.
Modulo an enhanced treatment of non-global logarithms to account for
the more complicated colour structure of dijet events that we discuss
below, our usage of the resummation plugin to \sherpa is the same as in
our study of angularities in $Z$+jet events and we therefore refer the
reader to Ref.~\cite{Caletti:2021oor} for details.
In particular, through the iterative flavour-$k_t$ clustering
procedure~\cite{Banfi:2006hf} (referred to as BSZ algorithm in what follows),
our automated resummation plugin allows us to keep track of the jet flavour 
(see again~\cite{Caletti:2021oor} for a detailed description of the algorithm),
yielding the so-called \NLLp resummation accuracy.

Let us now discuss our treatment of soft logarithms and, in
particular, of non-global logarithms.
In practice, the soft function $\mathcal{S}$ can be factorised in a
global contribution $\mathcal{S}_\text{global}$ and a
non-global~\cite{Dasgupta:2001sh,Dasgupta:2002bw} part
$\mathcal{S}_\text{non-global}$. 
Although we perform the resummation of non-global logarithms in the
large-$N_C$ limit, we can account, to all orders, for the full set of
colour correlations induced by the first emission.  In order to
achieve this we note that, for a given Born-level partonic channel
$\mathcal{B}_\delta$, the overall contribution from soft logarithms
can be written as
\begin{equation}\label{eq:Sglobal}
  \mathcal{S}^{\mathcal{B}_\delta}(t(L))
  =\Tr\left[H
    e^{-t\left(\mathbf{\Gamma}^{\mathcal{B}_\delta}\right)^\dagger}ce^{-t\mathbf{\Gamma}^{\mathcal{B}_\delta}}\right]/\Tr\left[cH\right]\,,
\end{equation}
with
\begin{equation}
  t\mathbf{\Gamma}^{\mathcal{B}_\delta}=\sum_{i>j\in\mathcal{B}_\delta}
  \mathbf{T}_i\cdot\mathbf{T}_j (I_{ij}t + f_{ij}(t))\,,
  \qquad\text{ and }\qquad
  t(L) = \int^{\mu_Q}_{\mu_Qe^{-L}}
  \frac{dk_{t}}{k_{t}}\frac{\alphaS\left(k_{t}\right)}{\pi} = \frac{-\ln(1-2\alphas\beta_0L)}{2\pi\beta_0}\,,
\end{equation}
where $\alphaS=\alphaS(\muQ)$. Note that these contributions arise
only at NLL accuracy and so the actual scale choice for $\alphaS$ is
subleading. In practice, we evaluate $\alphaS$ at $\mu_\text{R}$.
Here, $H$ and $c$ are the hard function and colour metric for this
channel. The $I_{ij}t$ coefficients in
$t\mathbf{\Gamma}^{\mathcal{B}_\delta}$ correspond to global contributions,
stemming from soft-wide-angle radiation. These depend only on the jet
radius $R$ and are computed in the small-$R$ limit.
They are taken from~\cite{Dasgupta:2012hg} and we note that, compared
to the $Z$+jet case, the dijet case also has to include dipoles involving
the recoiling jet.

The non-global contributions to $\mathcal{S}$ are encoded in the
functions $f_{ij}(t)$ in $t\mathbf{\Gamma}^{\mathcal{B}_\delta}$.
They are computed independently for each dipole using a Monte Carlo
approach as described in~\cite{Dasgupta:2001sh}.\footnote{This
  procedure requires a cutoff $\theta_\text{min}$ on collinear
  radiation. In practice, we compute $f(t)$ for different cuts and
  extrapolate numerically to $\theta_\text{min}\to 0$.}
The main difference compared to our previous work on $Z$+jet events is
the fact that one also has to include the $f_{ij}(t)$ corresponding to
dipole configurations involving the recoiling jet.
On top of the standard dependence on $t(L)$ and on the jet radius,
these dipoles involving the recoiling jet also depend on the rapidity
separation between the measured jet and the recoiling one (see
\emph{e.g.}~\cite{Dasgupta:2012hg}).
This approach allows us to compute the non-global soft function for
ungroomed angularities.
For \softdrop groomed distributions, the non-global factor
remains the same for $v\ge z_\text{cut}$ and saturates at that value,
\emph{i.e.}\ $\mathcal{S}_\text{non-global}^\text{(groomed)}(v) =
\mathcal{S}_\text{non-global}^\text{(ungroomed)}(\text{max}(v,z_\text{cut}))$\,.
Finally, we note that while for $Z$+jet processes the matrix structure of
Eq.~\eqref{eq:Sglobal} disappears, the situation for dijet events is
more complex and the full matrix structure needs to be taken into account.

The above-described resummed observable calculation is matched to exact distributions at
NLO. These are obtained using \sherpa with \recola~\cite{Actis:2016mpe,Biedermann:2017yoi}
and \OpenLoops~\cite{Buccioni:2019sur} as one-loop generators in conjunction with
the \collier library~\cite{Denner:2016kdg}, and using Catani--Seymour dipole
subtraction~\cite{Catani:1996vz,Gleisberg:2007md}. All tree-level 
contributions are obtained from the built-in \comix~\cite{Gleisberg:2008fv} generator.
For a generic angularity $\lambda$, the matching is done at the level
of cumulative distributions $\Sigma^\delta(\lambda)$ defined for each
flavour channel $\delta$.
For this, we expand the (fixed-order) jet cross section
$\sigma_\text{fo}(\pt)$ as well as the fixed-order and resummed
cumulative distributions, $\Sigma_\text{fo}(\lambda)$ and
$\Sigma_\text{res}(\lambda)$ in series of $\alphaS$:
\begin{align}
  \sigma^\delta_\text{fo}(\pt) & = \sigma^{\delta(0)} + \sigma^{\delta(1)}_\text{fo} + \sigma^{\delta(2)}_\text{fo} + \dots\,,\\
  \Sigma^\delta_\text{fo}(\lambda) & = \sigma^{\delta(0)} + \Sigma^{\delta(1)}_\text{fo} + \Sigma^{\delta(2)}_\text{fo} + \dots\,,\\
  \Sigma^\delta_\text{res}(\lambda) & = \sigma^{\delta(0)} + \Sigma^{\delta(1)}_\text{res} + \Sigma^{\delta(2)}_\text{res} + \dots\,,
\end{align}
where the $\pt$ dependence is left implicit, the number in
parentheses indicates the respective order in $\alphaS$ (without
including the Born-level factors of $\alphaS$ and, possibly,
$\alpha_\text{EW}$), and $\sigma^{\delta(0)}$ is the Born-level jet
cross section in the specified flavour channel.
Our final matched expression for the cumulative distribution reads:
\begin{equation}\label{eq:matching}
  \Sigma^\delta_\text{matched}(\lambda)=\Sigma^\delta_\text{res}(\lambda)
  \left(
    1+\frac{\Sigma^{\delta(1)}_\text{fo}-\Sigma^{\delta(1)}_\text{res}}{\sigma^{\delta(0)}}
    -\frac{\bar\Sigma^{\delta(2)}_\text{fo}+\Sigma^{\delta(2)}_\text{res}}{\sigma^{\delta(0)}}
    -\frac{\Sigma^{\delta(1)}_\text{res}}{\sigma^{\delta(0)}}
     \frac{\Sigma^{\delta(1)}_\text{fo}-\Sigma^{\delta(1)}_\text{res}}{\sigma^{\delta(0)}}
  \right)\,,
\end{equation}
where
$\bar\Sigma^{(2)}_\text{fo}=\sigma_\text{fo}^{(2)}-\Sigma^{(2)}_\text{fo}=\int_\lambda^1
d\sigma^{(2)}$ can be calculated without the need for the two-loop
virtual corrections to the jet cross section.

To estimate the perturbative uncertainties of our predictions, we
again consider $7$-point variations of the factorisation
and renormalisation scales in the matrix elements and the resummation,
\emph{cf.}\ Eq.~\eqref{eq:7pointvar}. The central values in the $Z$+jet case are
taken from \cite{Caletti:2021oor}, \emph{i.e.}\ the transverse momentum of the $Z$ boson,
multiplied by the jet radius in the case of $\mu_\text{Q}$. Similarly, in the
dijet case we use $\mu_\text{R}=\mu_\text{F}=\mu_\text{Q}/R=H_T/2$,
closely matching the scale choice in the \SHMCatNLO simulations see
Eq.~\eqref{eq:scalesjj}. 
When implementing the matching between our fixed-order and resummed
results, we treat the endpoint of the resummed expression and of its expansion by setting
\begin{equation}\label{eq:endpoint+xL}
  L = \frac{1}{p}\ln\left[1+\left(\frac{x_L}{v}\right)^p-\left(\frac{x_L}{v_\text{max}}\right)^p\right]\,,
\end{equation}
with $v_\text{max}$ the kinematic endpoint of the fixed-order
distribution.
Varying $x_L$ between $\tfrac{1}{2}$ and $2$ then allows us to
quantify the resummation uncertainty (we keep $p=1$ fixed). As total
perturbative uncertainty estimate we consider the envelope of all
$(\mu_R,\mu_F)$ and $x_L$ variations. 

\subsection{Non-perturbative corrections in the transfer-matrix approach}\label{sec:res_np}

So far we have explained our analytic description of jet angularities
in perturbative QCD.
Clearly, these observables probe QCD dynamics in the infrared regime
and are therefore also sensitive to non-perturbative (NP)
corrections.
At a hadron collider such as the LHC, this includes the Underlying
Event --- \emph{i.e.}\ remnant--remnant interactions --- on top
of the parton-to-hadron transition.
Because jet angularities are IRC-safe observables, NP effects can be
considered as power corrections and are therefore expected to become
less important as one increases the transverse momentum of the jet
they are measured on.
However, at moderate $p_{T,\text{jet}}$ their effect can be sizeable
and it is crucial to include them in any realistic phenomenological
prediction.
Additionally, the size of the jet-radius parameter impacts the
susceptibility to NP corrections.

In previous
works~\cite{Marzani:2017mva,Marzani:2017kqd,Marzani:2019evv,Caletti:2021oor},
we have adopted a rather simple procedure to account for NP
effects.
We have computed predictions for a given jet-angularity distribution
in the considered fiducial phase space, \emph{i.e.}\ applying the
phase-space constraints on jets and restricting to a given $p_{T,\text{jet}}$
bin, both at parton and hadron level, using a general-purpose MC
generator.
The ratio between these two distributions was then used to correct our
resummed and matched predictions for that particular observable and
phase-space region.
While this model captures the dominant NP effects of redistributing
the partonic jet-constituent momenta onto hadrons, it is still rather
crude, as it does not fully account for alterations of the underlying
parton-level event kinematics due to NP effects. In particular, both
hadronisation and UE affect the transverse-momentum distribution of
jets, potentially leading to migration between the considered
$p_{T,\text{jet}}$-bins.
By neglecting this important feature of NP corrections, this approach
turns out to be rather sensitive to cutoff-effects in the employed
parton shower. In particular, NP corrections can become pathological
for observable values corresponding to scales below the shower
cutoff. This generator dependence can be partially overcome by
considering several MC programs and using the spread of the extracted
NP ratios as an uncertainty on the NP corrections, as was done in
Ref.~\cite{Caletti:2021oor}, and consequently used in the comparison
to CMS data in~\cite{CMS:2021iwu}.

\subsubsection{The transfer-matrix approach}

In this new study, we attempt to overcome the limitations of the method
for extracting NP corrections we previously used.
To this end we develop and implement a more realistic and detailed
model to include NP corrections, in which we account for the change of
parton-level event kinematics due to hadronisation and the UE.
In particular, considering double-differential measurements in
jet-transverse momentum and the angularity observables, we derive
non-perturbative transfer matrices which account for the alteration of
both $p_{T,\text{jet}}$ and the angularity.
Through the migration in transverse momentum we significantly reduce
the sensitivity to phase-space restrictions and non-perturbative
parameters in the underlying MC simulations, in particular to the
parton-shower cutoff parameter.
We here present our approach in full generality, applicable to an
arbitrary set of observables, measured in a multi-differential way. We
thereby aim to keep track both of the migration in the underlying
event-kinematical variables, used to define the fiducial phase space,
that get partially integrated over, as well as changes in the actual
observable of interest, \emph{e.g.}\ a specific angularity variable.

Let us consider a scattering process which results in a partonic
configuration $\mathcal{P}$. Through NP effects the set of parton
momenta is then mapped onto a hadron-level configuration
$\mathcal{H}\left(\mathcal{P} \right)$.  The map $\mathcal{H}$, which
does not need to be fully specified at this point, accounts for
hadronisation and UE corrections.
It could be derived from field-theoretical considerations (see for
example Refs.~\cite{Hoang:2019ceu,Pathak:2020iue} for recent work on
\softdrop observables) or it can be extracted from any given parton-shower
simulation interfaced to a model of NP phenomena.
For a given configuration $\mathcal{P}$ or
$\mathcal{H}\left(\mathcal{P} \right)$, we then measure a set of $m$
observables, $\vec{V}\left(\mathcal{P}\right)$ or
$\vec{V}\left(\mathcal{H}\left(\mathcal{P}
  \right)\right)$.\footnote{Here, we have chosen the same set of
  observables $\vec{V}$ on the parton- and hadron-level
  configurations for simplicity. It is of course trivial to extend this to
  differing sets of observables for parton and hadron level, for example using
  additional auxiliary observables to parameterise the parton-level phase
  space, or adding selection criteria like, for instance, particle charge, at hadron
  level.}
We define the transfer operator as the conditional probability to
measure a hadron-level set of observables $\vec{v}_h$, evaluated on
$\mathcal{H}\left(\mathcal{P} \right)$, given that the parton-level
observables were $\vec{v}_p$:
\begin{equation} \label{transfer-operator}
  \mathcal{T}(\vec{v}_h|\vec{v}_p) = \frac{\int \mathop{d\mathcal{P}}
    \frac{\mathop{d\sigma}}{\mathop{d \mathcal{P}}}
    \delta^{(m)}\left(\vec{v}_p-\vec{V}\left(\mathcal{P}\right)\right) \delta^{(n)} \left(\vec{v}_h-\vec{V}\left(\mathcal{H}\left(\mathcal{P}\right)\right)\right)}
  {\int \mathop{d\mathcal{P}}
    \frac{\mathop{d\sigma}}{\mathop{d \mathcal{P}}}
    \delta^{(m)}\left(\vec{v}_p-\vec{V}\left(\mathcal{P}\right)\right)}\,.
\end{equation}
This way, the multi-differential distribution for the set of hadron-level observables $\vec{v}_h$ can be written as
\begin{equation}
  \frac{\mathop{d^m \sigma^\text{HL}} }{\mathop{dv_{h,1} \dots dv_{h,m}}} = \int \mathop{d^m \vec{v}_p}\,  \mathcal{T}(\vec{v}_h|\vec{v}_p) \, \frac{\mathop{d^m\sigma^\text{PL} }}{\mathop{dv_{p,1} \dots dv_{p,m}}}\,.
\end{equation}
When performing numerical studies, we often work with binned distributions, \emph{i.e.}\ we consider binned cross sections
obtained by integrating the multi-differential distribution over hypercubes in the observables' space. If we consider, for
instance, the parton-level case, the cross section in any given hyper-bin $p$ is written as 
\begin{align}
   \Delta \sigma^\text{PL}_{p} &= \int \mathop{d\mathcal{P}}
  \frac{\mathop{d\sigma}}{\mathop{d \mathcal{P}}}
  \Theta_{p}\left(\mathcal{P}\right)\,,
\end{align} 
where 
\begin{align} \label{thetap}
  \Theta_{p}\left(\mathcal{P}\right)=
    \prod_{i=1}^m \theta(V_i(\mathcal{P})-v^\text{min}_{p,i})\theta(v^\text{max}_{p,i}-V_i(\mathcal{P}))\,.
\end{align} 
If we now consider a binned distribution at hadron level, the transfer
operator from parton-level bin $p$ to a given hadron-level bin $h$
becomes a matrix of the form
\begin{equation}
  \mathcal{T}_{hp} = \frac{\int \mathop{d\mathcal{P}}
    \frac{\mathop{d\sigma}}{\mathop{d \mathcal{P}}}
    \Theta_p\left(\mathcal{P}\right)\Theta_h\left(\mathcal{H}\left(\mathcal{P}\right)\right)
  }
  {\int \mathop{d\mathcal{P}}
    \frac{\mathop{d\sigma}}{\mathop{d \mathcal{P}}}
    \Theta_p\left(\mathcal{P}\right)
  }\,,
\end{equation}
with
\begin{align} \label{thetah}
  \Theta_{h}\left(\mathcal{\mathcal{H}(\mathcal{P})}\right)=
   \prod_{i=1}^m \theta \left( V_i\left(\mathcal{\mathcal{H}(\mathcal{P})}\right)-v^\text{min}_{h,i}\right) \theta \left(v^\text{max}_{h,i}-V_i\left(\mathcal{\mathcal{H}(\mathcal{P})}\right)\right)\,.
\end{align}
Consequently, the final hadron-level distribution in the hyper-bin $h$
is obtained by the weighted sum of all parton-level contributions
\begin{equation}\label{eq:Tfinal}
  \mathop{\Delta\sigma_h^\text{HL}} = \sum_{p} \mathcal{T}_{hp} \mathop{\Delta\sigma_p^\text{PL}}\,.
\end{equation}
We note that the $\Theta$-functions introduced in Eqs.~(\ref{thetap})
and~(\ref{thetah}) can also be used to represent event-selection cuts,
such as the ones reported in Tab.~\ref{tab:cms_cuts}, that define the
fiducial region of a measurement, at parton- and hadron level,
respectively.
We further note that the parton- and hadron-level bins do not
necessarily have to be the same. For example, one would typically
define underflow and overflow bins at parton level so as to include
their contribution to the final hadron-level predictions.

The elements of the transfer matrices appearing in
Eq.~\eqref{eq:Tfinal} can easily be extracted from a multi-purpose
generator in a single run, given that events are accessible at
different stages of the simulation process.
Note however that, while the parton shower and hadronisation are
treated in a factorised form in all multi-purpose event
generators~\cite{Buckley:2011ms}, this is not necessarily the case
for the UE.
In particular \pythia~\cite{Sjostrand:2006za,Sjostrand:2014zea} makes
use of an interleaved evolution of the initial-state shower and the
secondary interactions~\cite{Sjostrand:2004ef,Corke:2010yf}.
Accordingly, in a full event simulation within such model there is no
notion of an intermediate parton-level final state that is directly
comparable to a resummed calculation.  However, in \sherpa the parton
showers off the hard process and the simulation of multiple-parton
interactions are fully separated, \emph{i.e.}\ the UE is simulated
only after the shower evolution of the hard interaction is
completed. The secondary scatterings then get showered and ultimately the
partonic final state consisting of the showered hard process and
multiple-parton interactions gets hadronised. In what follows we
therefore determine the transfer matrices using the \sherpa generator.

Of course, the matrix $\mathcal{T}_{hp}$ depends on the model or the
specific generator used to derive them. However, we argue that a
milder dependence than with the naive
parton-to-hadron level ratios we previously adopted can be expected.
As the matrices are based on the conditional probabilities for transitions
between bins, \emph{e.g.}\ in $p_{T,\text{jet}}$, they only depend on
the parton shower insofar as it determines what exact phase space
those probabilities are averaged over.

It is interesting to note that, the task of correcting PL predictions
for NP effects closely resembles the situation of accounting for
detector effects in an experimental analysis. For this purpose
proposals have recently been made (see for example
Refs.~\cite{Arratia:2021otl,Howard:2021pos,Bellagente:2020piv,Andreassen:2019cjw,Bellagente:2019uyp})
to use machine-learning techniques to accomplish such a mapping for
either binned or unbinned distributions.  One can envisage that
similar methods can be applied for the problem of non-perturbative
corrections, possibly even in an invertible way, relating the observed
hadronic final state to an underlying parton-level
prediction.\footnote{In practice, treating non-perturbative corrections in an
  invertible way may be complicated by the fact that the mapping
  between partons and hadrons is not one-to-one.
  There exist however approaches allowing for maps which are
  many-to-many, such as the one using conditional invertible networks
  (see e.g.\ Ref.~\cite{Bellagente:2020piv}).}

\subsubsection{Transfer matrices for jet-angularity observables}

For our study of jet angularities we follow the above strategy to determine
$\mathcal{T}_{hp}$ from MC simulations with
\sherpa, and use them to correct our binned perturbative \NLOpNLLp
predictions. To this end we employ the \rivet analysis
tool~\cite{Buckley:2010ar,Bierlich:2019rhm}, which allows us to
directly access the {\sc{HepMC}}~\cite{Dobbs:2001ck,Buckley:2019xhk}
event record and extract the intermediate truth-level information on
the event evolution after parton showering, prior to UE and
hadronisation. This enables us to consistently derive the desired
migration matrices for all angularity observables (with and without
grooming, both jet radii, and based on all/charged hadrons) for a
particular UE parameter set and hadronisation model in a single
generator run for $Z$+jet and dijet production, respectively. To
assess the systematic uncertainties related to the UE activity and
hadronisation, we derive alternative transfer matrices for the up and down
variation of the $\alpha^\text{MPI}_\text{min}$ parameter and using the string
fragmentation model, \emph{cf.} Sec.~\ref{sec:sherpasetup}. In total, we
consider the envelope of
\begin{equation}
  (\text{model},\alpha^\text{MPI}_\text{min}) = \{(\text{cluster},0.16),(\text{string},0.16),(\text{cluster},0.08),(\text{cluster},0.24)\}~,
\end{equation}
using cluster hadronisation with $\alpha^\text{MPI}_\text{min}=0.16$ as
our default.

Ideally, the uncertainties on the parton-to-hadron transfer matrix
should include a contribution associated with the choice of the
parton-shower cutoff scale, especially if one then wants to use the
transfer matrix with analytic parton-level results which do not have
an explicit cutoff.
Varying the parton-shower cutoff would however require a full
re-tuning of the non-perturbative parameters of the Monte Carlo, a
task which is clearly beyond the scope of the present study.
In practice, our use of two hadronisation models, cluster and string
fragmentation, should at least partially include the effect of varying
the parton-shower stopping scale, at a much lower cost.

In order to illustrate our transfer-matrix method, we discuss some
concrete examples. First, we consider the change in transverse
momentum for the leading jet in $Z$+jet production. We apply the
transfer matrix extracted from the \SHMCatNLO simulation for the
$Z$+jet event selection for ungroomed jets, based on all hadrons, to
our parton level (PL) \NLOpNLLp resummed calculation.
Thus, we consider $\vec{v}_p$ and $\vec{v}_h$ in
Eq.~(\ref{transfer-operator}) as two-component arrays, with elements
given by the jet transverse momentum and the jet thrust, at
parton level,
$\vec{v}_p=\{p^\text{PL}_{T,\text{jet}},\lambda^{1,\text{PL}}_2\}$, and
at hadron level,
$\vec{v}_h=\{p^\text{HL}_{T,\text{jet}},\lambda^{1,\text{HL}}_2\}$.

We first focus on the jet-$p_T$ distribution.
Integrating out the angularity at hadron level, our transfer-matrix approach
yields the following result:
\begin{align}
  \frac{\mathop{d \sigma^\text{HL}} }{\mathop{dp^\text{HL}_{T,\text{jet}}}} &= \int
  \mathop{d\lambda^{1,\text{HL}}_2} \frac{\mathop{d^2\sigma^\text{HL}} }{\mathop{dp^{\text{HL}}_{T,\text{jet}}d\lambda^{1,\text{HL}}_2}}\nonumber\\
  &= \int
  \mathop{d\lambda^{1,\text{HL}}_2} \int
  \mathop{dp_{T,\text{jet}}^\text{PL}}\mathop{d\lambda_2^\text{1,PL}}  
  \mathcal{T}\left(\{p^\text{HL}_{T,\text{jet}},\lambda^{1,\text{HL}}_2\}|\{p_{T,\text{jet}}^\text{PL},\lambda_2^\text{1,PL}\}\right)
  \, \frac{\mathop{d^2\sigma^\text{PL}
    }}{\mathop{dp_{T,\text{jet}}^\text{PL}d\lambda_2^\text{1,PL}}}\nonumber\\
  &= \int
  \mathop{dp_{T,\text{jet}}^\text{PL}}\mathop{d\lambda_2^\text{1,PL}}  
  \mathcal{T}\left(\{p^\text{HL}_{T,\text{jet}}\}|\{p_{T,\text{jet}}^\text{PL},\lambda_2^\text{1,PL}\}\right)
  \, \frac{\mathop{d^2\sigma^\text{PL}
    }}{\mathop{dp_{T,\text{jet}}^\text{PL}d\lambda_2^\text{1,PL}}}\,,\label{lambda-pt}
\end{align}
where we consider binned distributions, and multiply
our result for $\lambda^1_2$ at \NLOpNLLp accuracy by the matrix form
of
$\mathcal{T}(\{p^\text{HL}_{T,\text{jet}}\}|\{p^\text{PL}_{T,\text{jet}},\lambda^{1,\text{PL}}_2\})$.
We note that Eq.~(\ref{lambda-pt}) still non-trivially depends on the
assumed \emph{double}-differential distribution of
$\sigma^{\text{PL}}$, even though in itself it is only differential in
one observable. 
This happens because the above expression uses different parton-level distributions
for the determination of the transfer matrix $\mathcal{T}$, here \SHMCatNLO simulations,
and for the parton-level cross section $\sigma^{\text{PL}}$, given by the \NLOpNLLp prediction.

\begin{figure}
  \centering
  \includegraphics[width=0.9\textwidth]{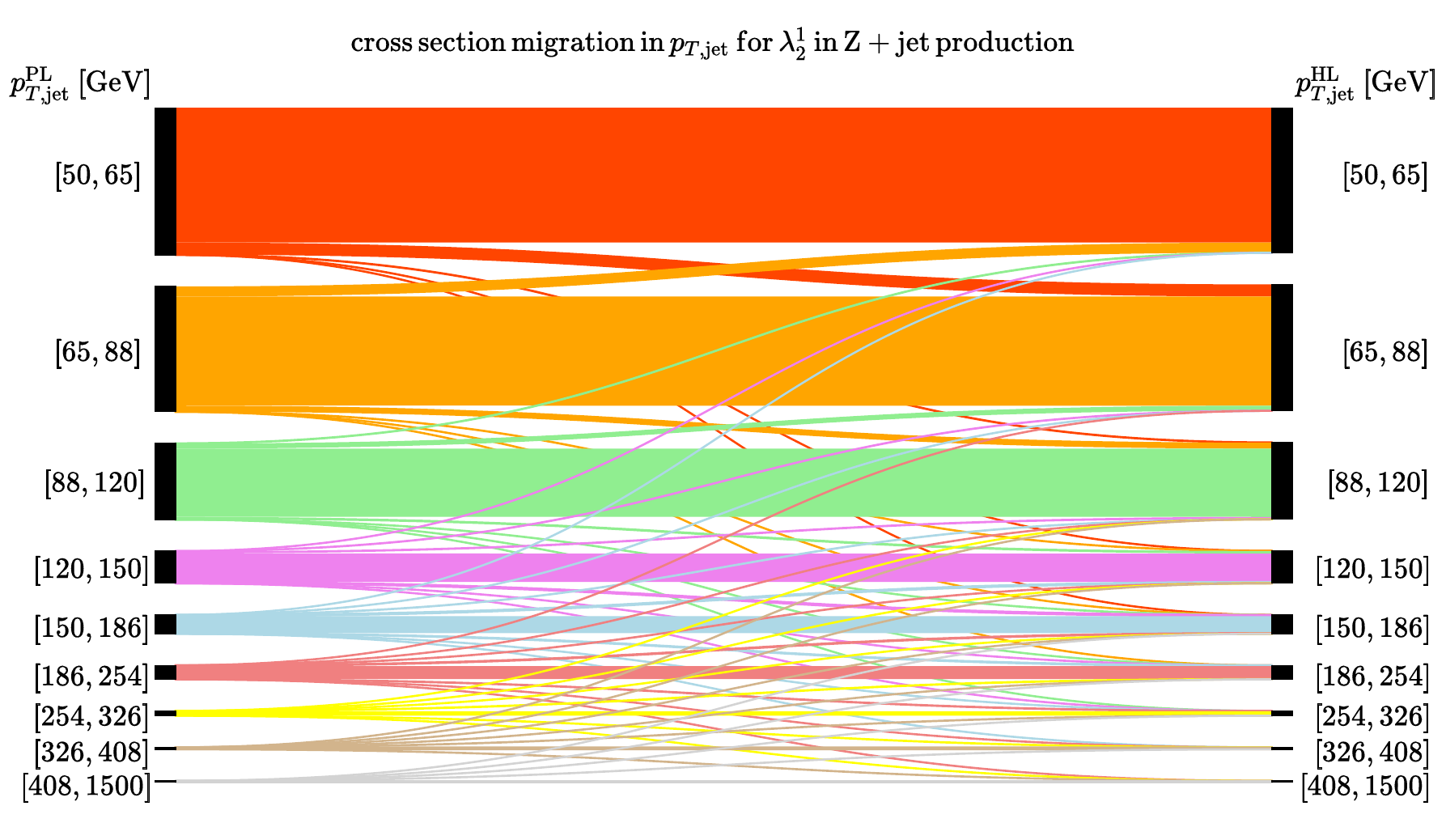}
  \caption{Illustration of cross-section migration between parton level and hadron level with respect
    to the jet transverse momentum for the double-differential \NLOpNLLp resummed distribution of
    $\lambda^1_2$ in $Z$+jet production. The used jet-radius parameter is $R=0.4$, the employed
    transfer matrix corresponds to the case of \emph{all} hadrons contributing at hadron level.} \label{fig:pt-migr}
\end{figure}

It should be noted that there is an additional contribution from events where the
parton level does not pass some of the event-selection cuts, like the lowest
transverse momentum bin edge but also any other cut in principle, whereas the
corresponding hadron-level event actually does pass the cuts. We have found that this
contribution is basically negligible starting from the $p_{T,\text{jet}}\in
[88-120]$~GeV bin. For the Z+jet case, we added an underflow bin, dependent on a
technically necessary cut $p_{T,\text{jet}} > 30$~GeV on matrix-element level,
and confirmed that this contribution is indeed dominated by lower
$p_{T,\text{jet}}$ regions. In the following, we will consider distributions
starting from $p_{T,\text{jet}} > 88$~GeV and neglect this additional contribution.

The result is illustrated in Fig.~\ref{fig:pt-migr}, where we show the
set of transverse-momentum bins measured by CMS and show the effect of
bin migration from parton- to hadron level on the respective \NLOpNLLp
cross section for anti-$k_t$ jets with $R=0.4$. For example, the
hadron-level cross section for
$p_{T,\text{jet}}\in [65,88]\,\text{GeV}$ receives non-negligible
contributions from parton-level events with smaller transverse
momentum, \emph{e.g.}\ through an additive contribution from the UE,
as well as higher slices, \emph{e.g.}\ through the widening of jets
due to hadronisation.  It is evident that these effects, which were
completely ignored in our previous study, can be rather sizeable,
especially at low and moderate $p_{T,\text{jet}}$, and are therefore
important to consider in the derivation of NP corrections. Note that we do not
illustrate the flow in or out of the underflow bin at $p_{T,\text{jet}} <
50$~GeV for simplicity. The net flow over this border appears to be negligible.
In principle, a similar study could be performed for all the other
variables that are used to characterise the event kinematics and to
define the fiducial region, for example the jet rapidity. However, we
expect that their impact is much less pronounced and, therefore, in
order to keep the dimensionality of the transfer matrices as low as
possible, we have ignored them.

\begin{figure}
  \centering
  \includegraphics[width=.49\textwidth]{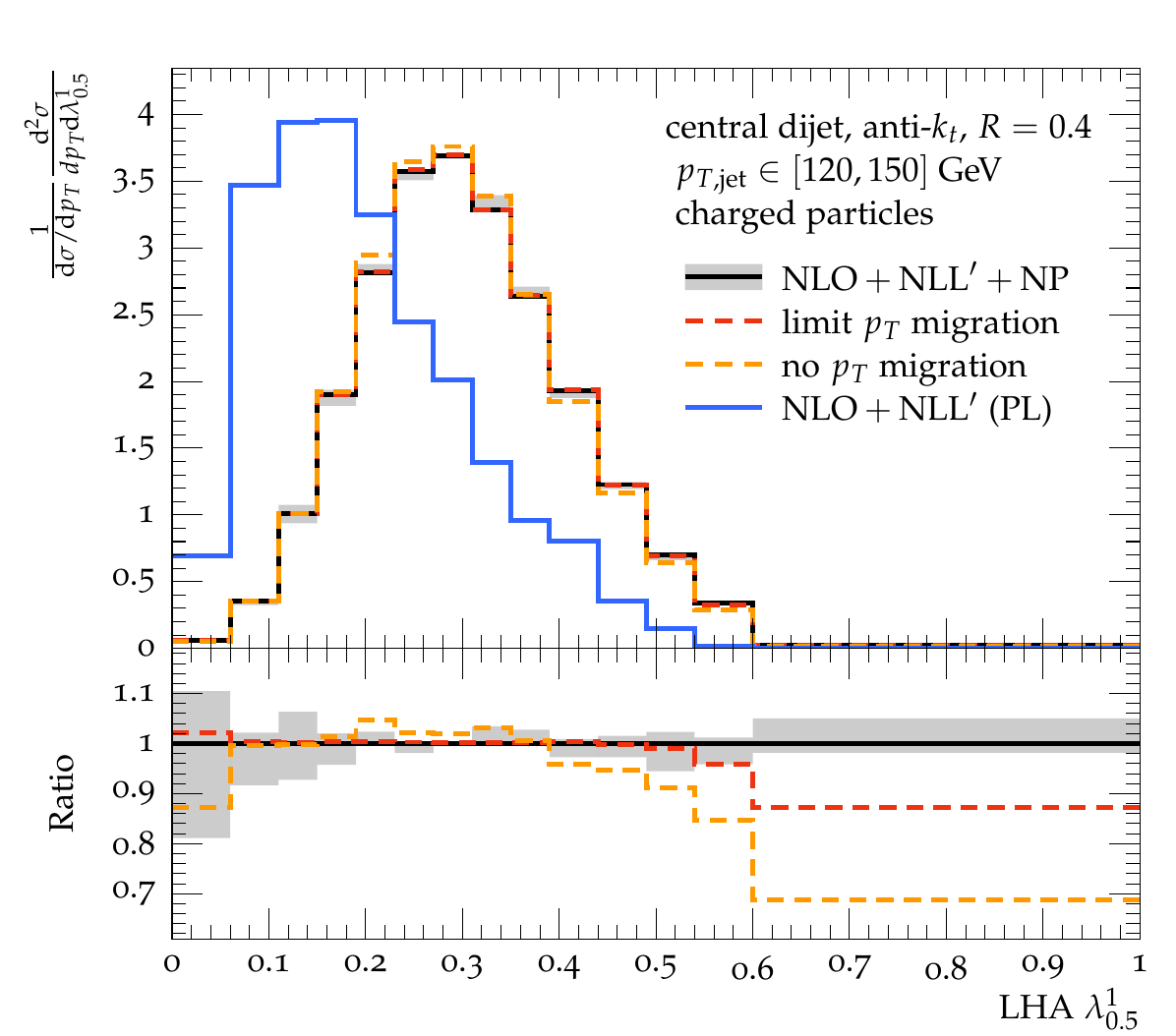}
  \includegraphics[width=.49\textwidth]{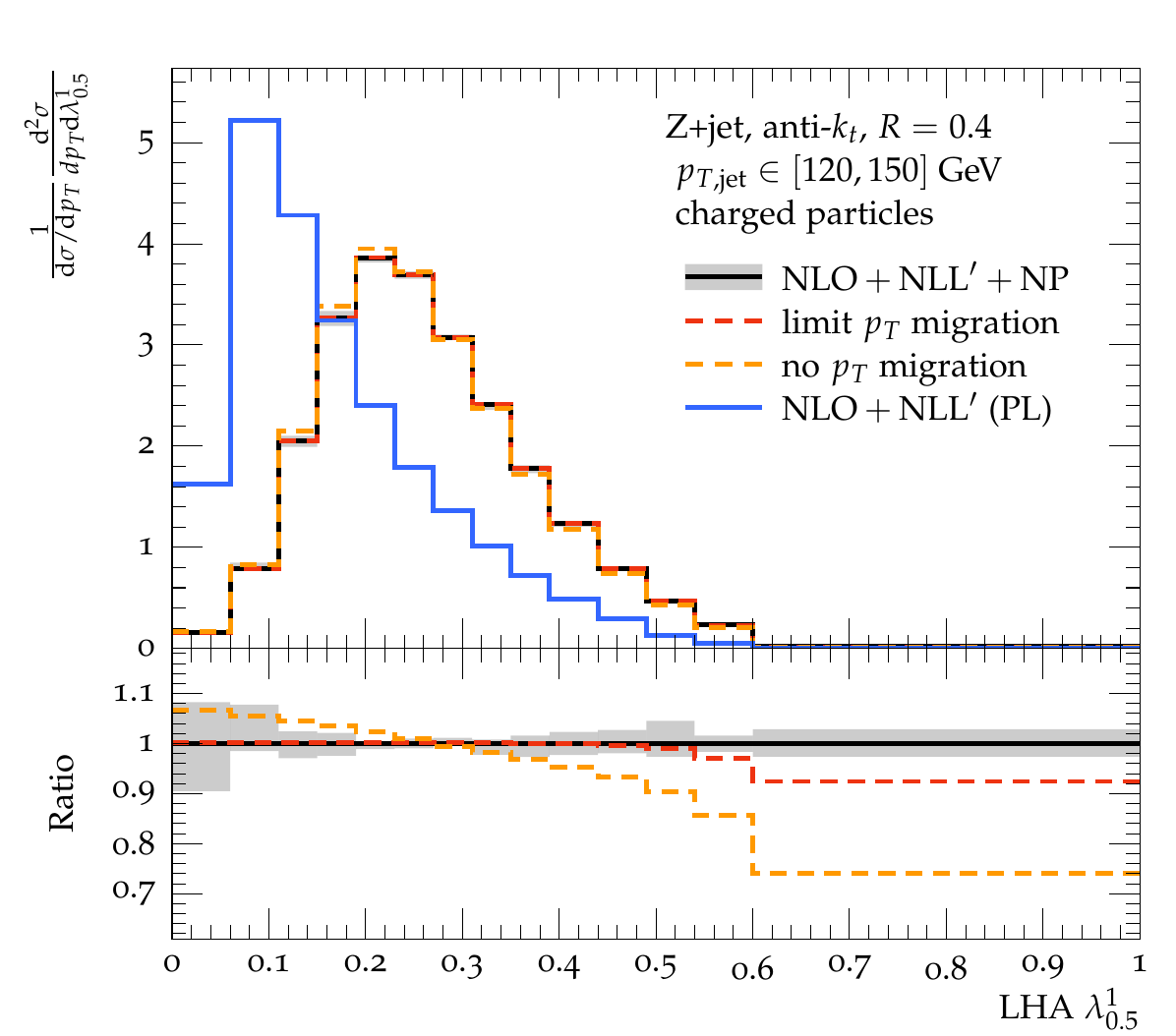}
  \caption{Effect of NP corrections on the $\lambda^1_{0.5}$ observable with
    $p_{T,\text{jet}}\in [120,150]\,\text{GeV}$ for the central dijet (left) and
    $Z$+jet (right) selection. Shown are the purely
    perturbative result (blue), the distribution after transferring events
    between different observable bins within that $p_{T,\text{jet}}$ bin (orange), after
    including migration from the neighbouring $p_{T,\text{jet}}$ bins (red) and our final
    prediction at hadron level (black). The uncertainty band corresponds to
    variations in the UE model.}\label{fig:pt-obs-migr1}
\end{figure}

\begin{figure}
  \centering
  \includegraphics[width=.49\textwidth]{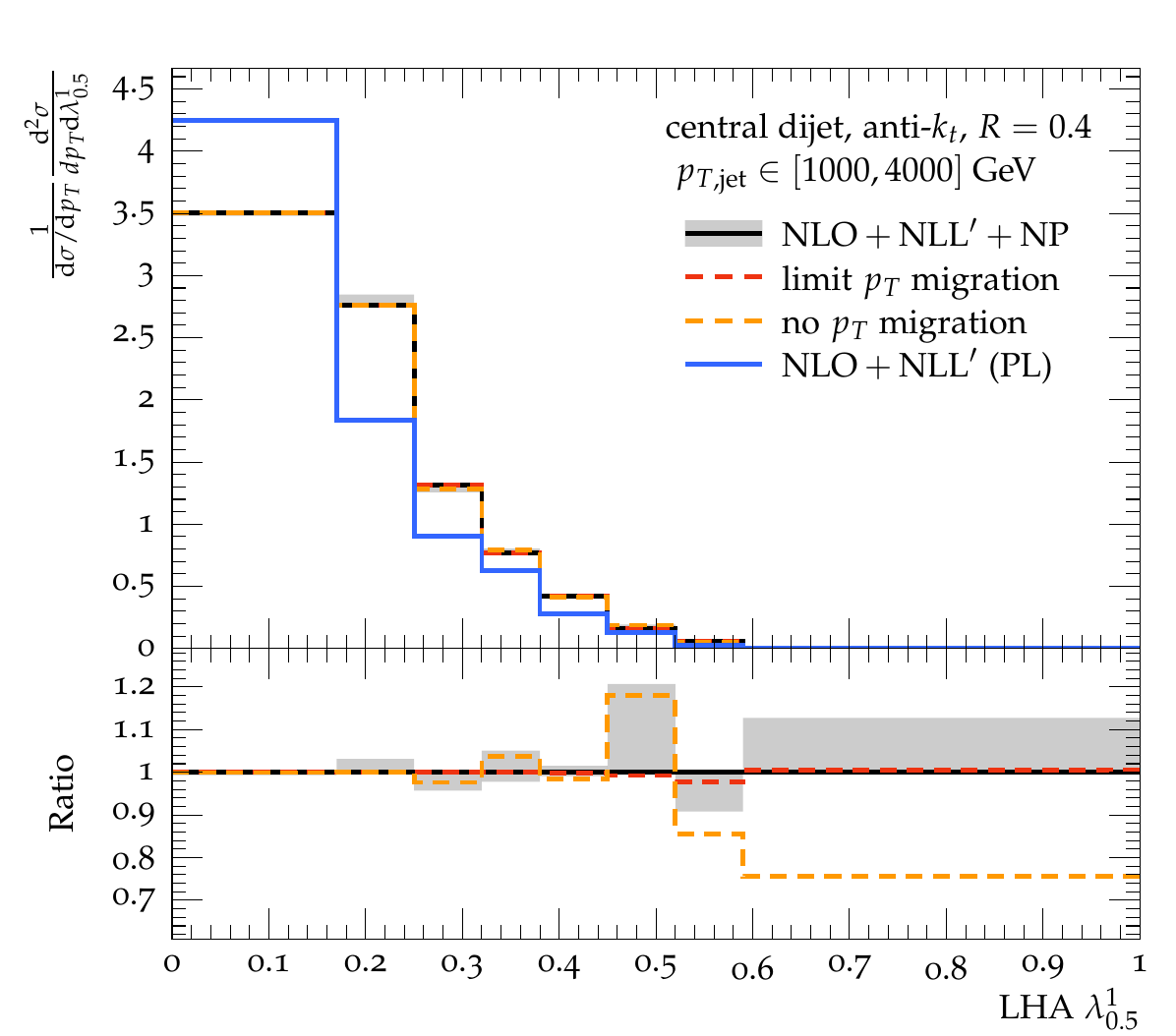}
  \includegraphics[width=.49\textwidth]{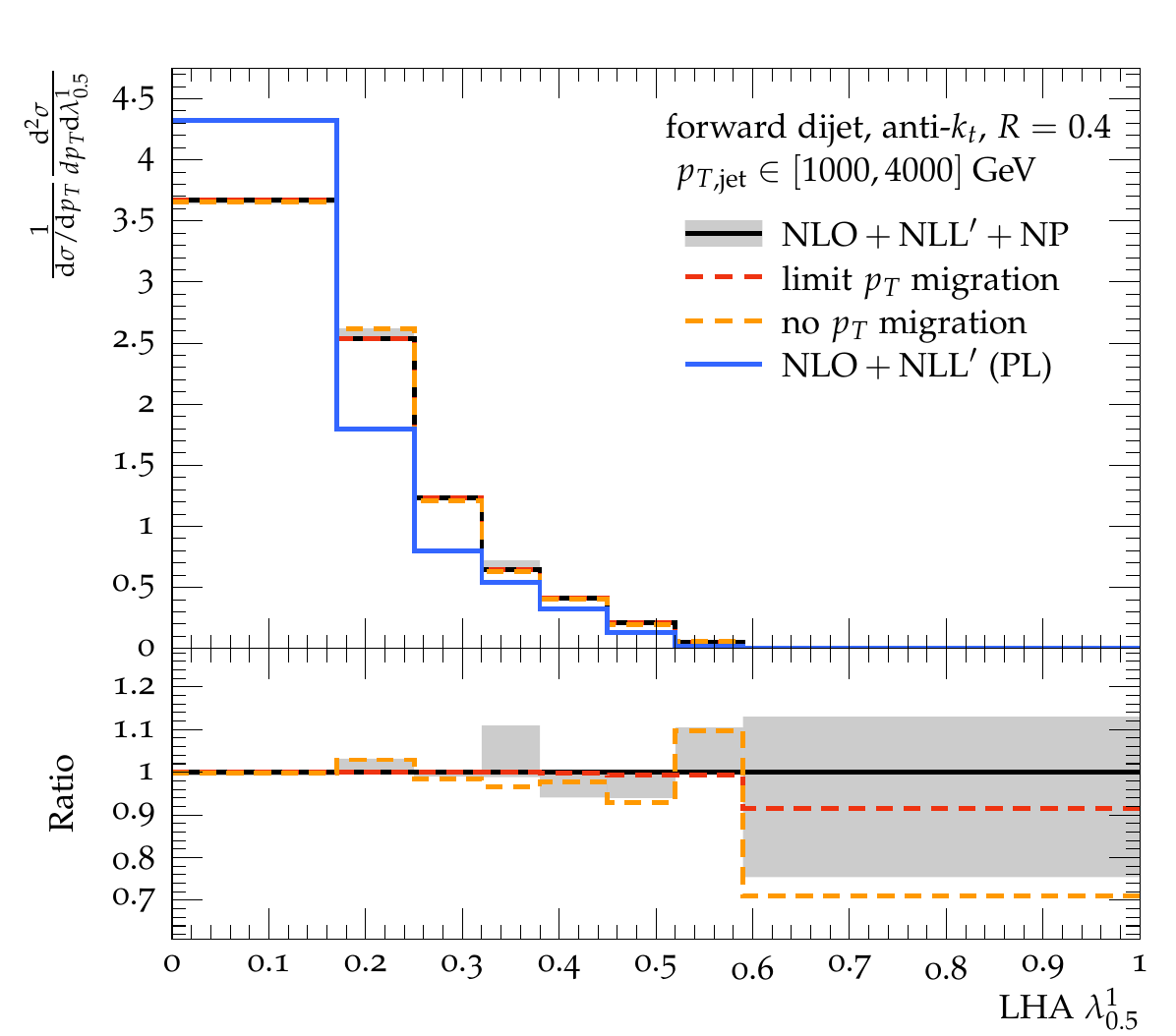}
  \caption{Same as Fig.~\ref{fig:pt-obs-migr1} but for jets with
    $p_{T,\text{jet}}\in [1000,4000]\,\text{GeV}$ in the central (left) and forward (right)
    dijet selection.}\label{fig:pt-obs-migr2}
\end{figure}

As second step in our assessment of NP corrections, we study their
effect on angularity distributions.
While at parton level the angularities are measured
on all the partons in the jet, at hadron level they can either be
measured on all the particles or using charged particles only.
To illustrate the effect of NP corrections, we consider in
Figs.~\ref{fig:pt-obs-migr1} and~\ref{fig:pt-obs-migr2} the LHA,
$\lambda^1_{0.5}$, measured on ungroomed jets, using different
approximations for the $p_{T,\text{jet}}$ migration from parton to
hadron level: including only the contribution of the transfer matrix
for the same $p_{T,\text{jet}}$ bin (dashed orange curve), including
as well migration from the two neighbouring bins (dashed red curve),
and including the full transfer matrix (solid black curve).
Each time, the same approximation is used for both the numerator and
the denominator of the normalised distribution.
For reference, the parton-level result is shown by the solid blue line.
Fig.~\ref{fig:pt-obs-migr1} concentrates on a medium-$p_{T,\text{jet}}$ bin,
 $p_{T,\text{jet}}\in[120,150]\,\text{GeV}$, for both the
$Z$+jet and the central dijet selection and Fig.~\ref{fig:pt-obs-migr2}
shows the corresponding results for the highest $p_{T,\text{jet}}$
interval, $p_{T,\text{jet}}\in[1000,4000]\,\text{GeV}$.

%
In Fig. \ref{fig:pt-obs-migr1}, we see that including only the same
$p_{T,\text{jet}}$ bin already accounts for the bulk of the
non-perturbative correction, and that NP effects cause sizeable
migrations between the different angularity bins.
However, we see a non-negligible contribution of events from the
neighbouring $p_{T,\text{jet}}$ bins.
The effect of  $p_{T,\text{jet}}$ migration beyond the two
closest bins differs from the full correction only at large
values of the angularity.
For the final prediction, using the full transfer matrix, 
we also derive an uncertainty estimate, shown as a grey band,
corresponding to the variation of the UE
$\alpha^\text{MPI}_\text{min}$ parameter as described above. We
observe that, despite the large overall effect of the NP corrections,
they are fairly stable under the considered variations.  The observed
uncertainty bands do not fully cover up the case where we discard the
effect of $p_{T,\text{jet}}$ migration from the neighbouring bins, in
particular for large $\lambda^1_{0.5}$ values. However, as one might
expect, NP effects decrease when considering the higher
$p_{T,\text{jet}}$ bin in Fig. \ref{fig:pt-obs-migr2}. Despite being
also much smaller at high $p_{T,\text{jet}}$, the migration between
different observable bins is still by no means negligible. The effect
of migration of events from lower $p_{T,\text{jet}}$ regions does
however appear to be very small in this case, and is mostly covered by
the UE variations.

\begin{figure}
  \centering
  \includegraphics[width=.48\textwidth,page=1]{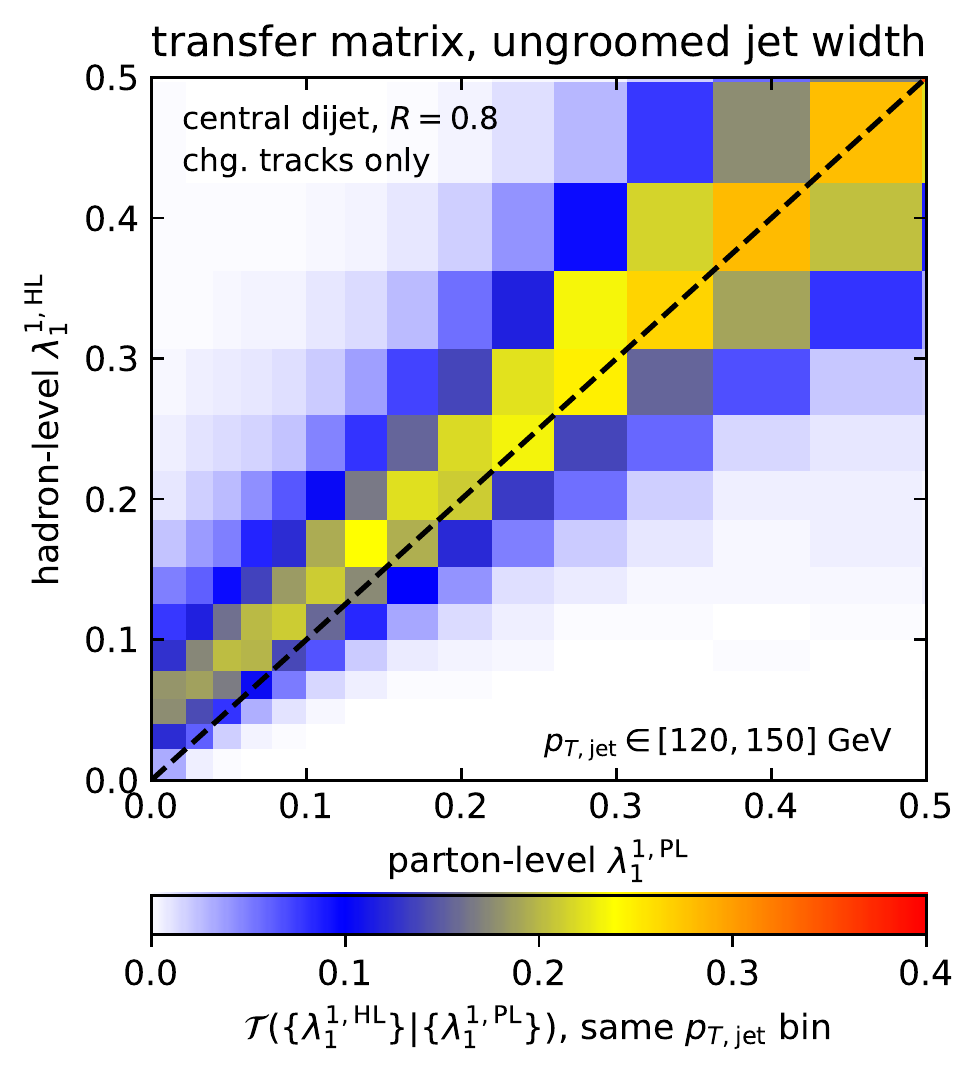}
  \includegraphics[width=.48\textwidth,page=2]{figures/transfer-matrix.pdf}
  \caption{Transfer matrix
    $\mathcal{T}(\{\lambda_1^{1,\text{HL}}\}|\{\lambda_1^{1,\text{PL}}\})$
    for the jet width, $\lambda^1_1$, for central dijet events with
    $R=0.8$ and $p_{T,\text{jet}}\in[120,150]$~GeV.
    The angularities at hadron level are measured on charged
    particles only.
    For simplicity of the presentation, we here assumed no $p_{T,\text{jet}}$ migration.
    The left (right) plot corresponds to ungroomed (groomed)
    jet width.
  } \label{fig:transfer-matrix-lambda}
\end{figure}

To better illustrate the effect of hadronisation in our
transfer-matrix approach, we directly plot the transfer matrix itself
in Fig.~\ref{fig:transfer-matrix-lambda}.
We select jets with $R=0.8$ and $p_{T,\text{jet}}\in[120,150]$~GeV in
central dijet events and measure the jet width, $\lambda_1^1$, either
on ungroomed jets (left plot) or on groomed jets (right plot). At
hadron level, the jet width is measured on charged particles only.
The transfer matrix for ungroomed jets shows clear signs of bin
migration, especially at low-to-mid angularities where the
parton-level values predominantly get pushed to larger values of the jet width at
hadron level.
If we instead apply grooming, the transfer matrix appears to be much
more diagonal, albeit with longer tails away from the
diagonal.\footnote{At larger $p_{T,\text{jet}}$, not shown here, the
  transfer matrix for groomed jets shows clear signs of migration from
  relatively large partonic values of the angularity to very small
  hadron-level values. These are likely related to subjets passing the
  \softdrop condition at parton level being pushed below the \softdrop
  cut after hadronisation, an effect which is typically included in
  analytic treatments of non-perturbative corrections of \softdrop
  observables~\cite{Dasgupta:2013ihk,Marzani:2017kqd,Hoang:2019ceu}.}

\begin{figure}
  \centering
  \includegraphics[width=.49\textwidth]{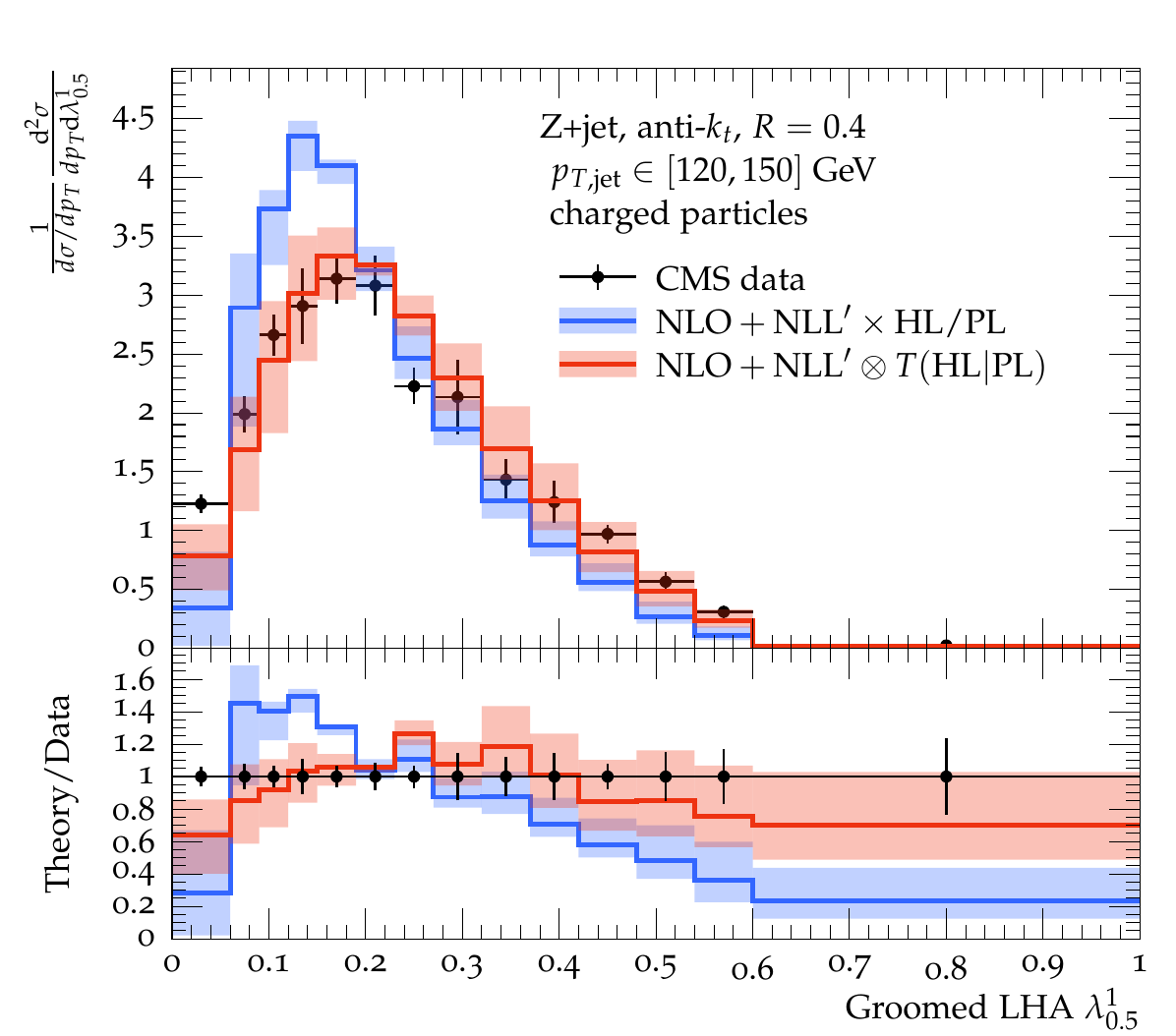}
  \includegraphics[width=.49\textwidth]{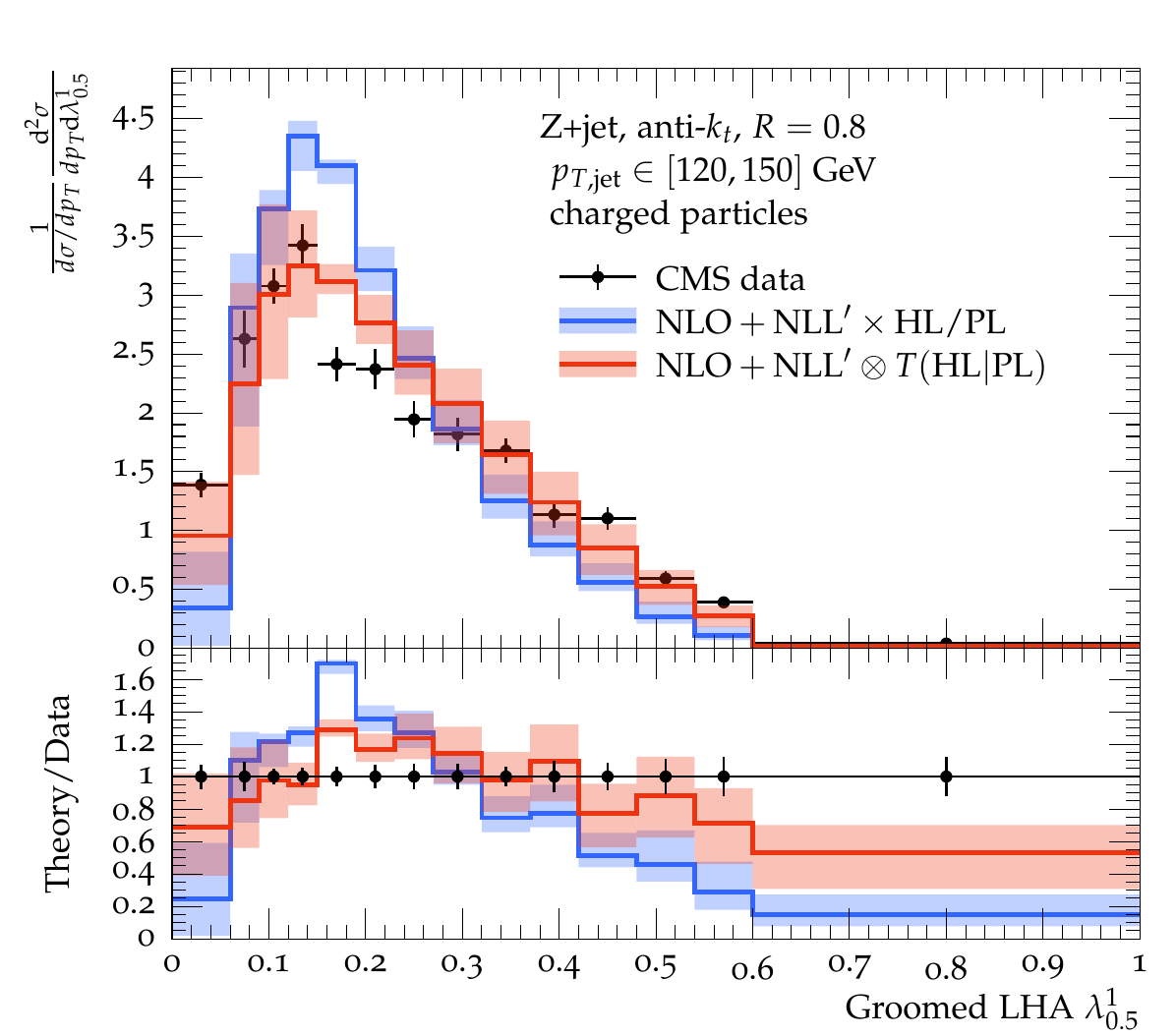}
  \caption{Hadron-level predictions for the groomed Les Houches
    Angularity $\lambda^1_{0.5}$ in $Z$+jet production, measured on
    the charged hadrons in $R=0.4$ (left) and $R=0.8$ (right) jets
    with $p_{T,\text{jet}}\in [120,150]\,\text{GeV}$. Results are
    obtained based on the \NLOpNLLp perturbative predictions including
    NP correction using the HL/PL ratio ($\times\text{HL}/\text{PL}$)
    and the new transfer-matrix approach
    ($\otimes T(\text{HL}|\text{PL})$).} \label{fig:NP-comp}
\end{figure}

Finally, it is interesting to compare our new method with the simpler
approach we have used in earlier studies, which consisted in
accounting for NP contributions via a bin-by-bin HL/PL ratio.  In
Fig.~\ref{fig:NP-comp} we perform such comparison for an observable
known to be rather sensitive to NP contributions, namely the groomed
LHA $\lambda_{0.5}^1$ measured on charged tracks on the hardest jet in
$Z$+jet events~\cite{Caletti:2021oor}. We again consider the moderate
$p_{T,\text{jet}}$ region, here both for $R=0.4$ (left), and $R=0.8$
(right). The old approach, used in~\cite{Caletti:2021oor}, is shown in
blue,\footnote{Note that, for a meaningful comparison between the two
  methods, both the approach based on the HL/PL ratio and the one
  based on transfer matrices are derived using the \sherpa generator
  with variations of the UE parameter $\alpha^\text{MPI}_\text{min}$.}
while the results obtained with the new transfer-matrix approach are
shown in red.  For both predictions we estimate theoretical
uncertainties, illustrated by the bands, corresponding to the envelope
of the $7$-point scale variations, the alternative $x_L$-parameter
settings, and variations of the $\alpha^\text{MPI}_\text{min}$
parameter of the UE model. We see that the difference in the nominal
predictions is rather substantial, and that the new treatment of NP
corrections yields a significantly better agreement with the CMS data,
shown in black, although visible differences remain for the $R=0.8$
case. One notices that the uncertainty estimates for the results based
on the transfer-matrix approach are somewhat larger than for the old
ratio method. A source of this increase is the larger range of
kinematics probed when allowing for migration from lower and higher
$p_{T,\text{jet}}$ slices.

We conclude this section by noting that a similar procedure to account
for NP effects was employed by the ALICE collaboration in
Ref.~\cite{ALICE:2021njq}, where a measurement of various angularity
observables in inclusive jet production in $pp$ collisions at
$\sqrt{s}=\SI{5.02}{TeV}$ was reported. These data, based on
\softdrop-groomed charged-tracks jets, was compared to analytical
predictions at \NLLp accuracy obtained from
SCET~\cite{Kang:2018qra,Kang:2018vgn}.  Two techniques for correcting
the analytic predictions for NP effects and the selection of
charged-particle jets were considered. The first being based on
``folding'' the \NLLp result with a response matrix extracted from MC
simulations that maps parton-level jets with
$(p^{\text{PL}}_T,\lambda^{\text{PL}})$ to hadron-level jets with
$(p^{\text{HL}}_T,\lambda^{\text{HL}})$, thereby accounting for
hadronisation corrections only. To incorporate UE effects an
additional bin-wise correction has been applied. As an alternative a
NP shape-function approach~\cite{Kang:2018vgn} to simultaneously
correct for hadronisation and the UE has been employed. We refer
to~\cite{ALICE:2021njq} for additional details.

\FloatBarrier

\section{Results for jet angularities in dijet and $Z$+jet production}\label{sec:results}

In this section, we present the results obtained from the calculation
detailed in section~\ref{sec:method}, \emph{i.e}\ \NLOpNLLp accurate
predictions accounting for NP corrections through the transfer-matrix
approach. We start in section~\ref{sec:results-perturbative} with a
few considerations at the purely perturbative level. In
section~\ref{sec:results-hadron}, we then discuss our results with NP
corrections included, and present full hadron-level predictions at NLO
QCD accuracy from \sherpa.

\subsection{Selected parton-level results}\label{sec:results-perturbative}

\begin{figure}
  \centering
  %
  \begin{subfigure}[t]{0.32\textwidth}
    \includegraphics[width=\textwidth]{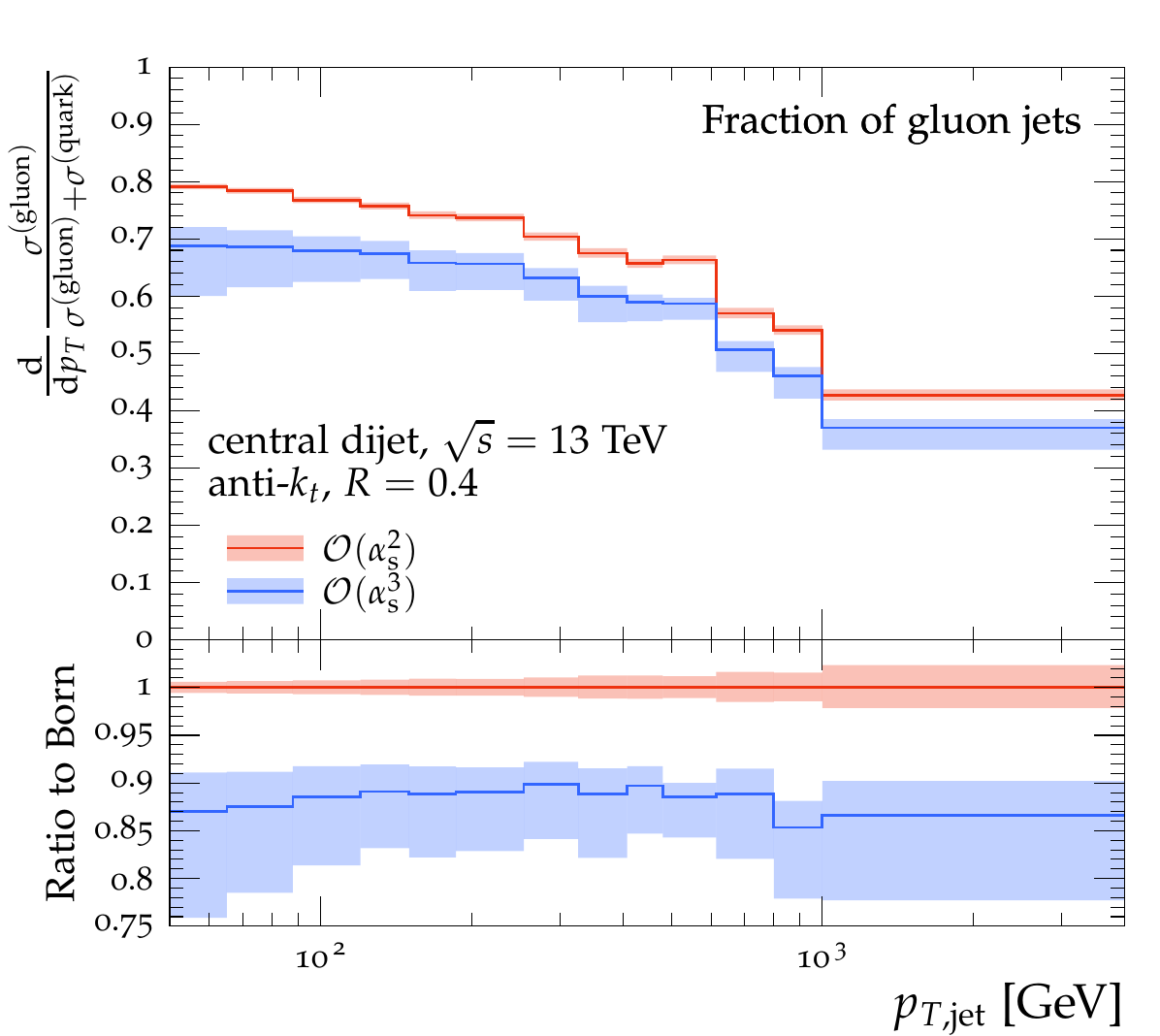}
    \caption{central dijets}\label{fig:gluonfractions-centraljj}
  \end{subfigure}
  \hfill
  \begin{subfigure}[t]{0.32\textwidth}
    \includegraphics[width=\textwidth]{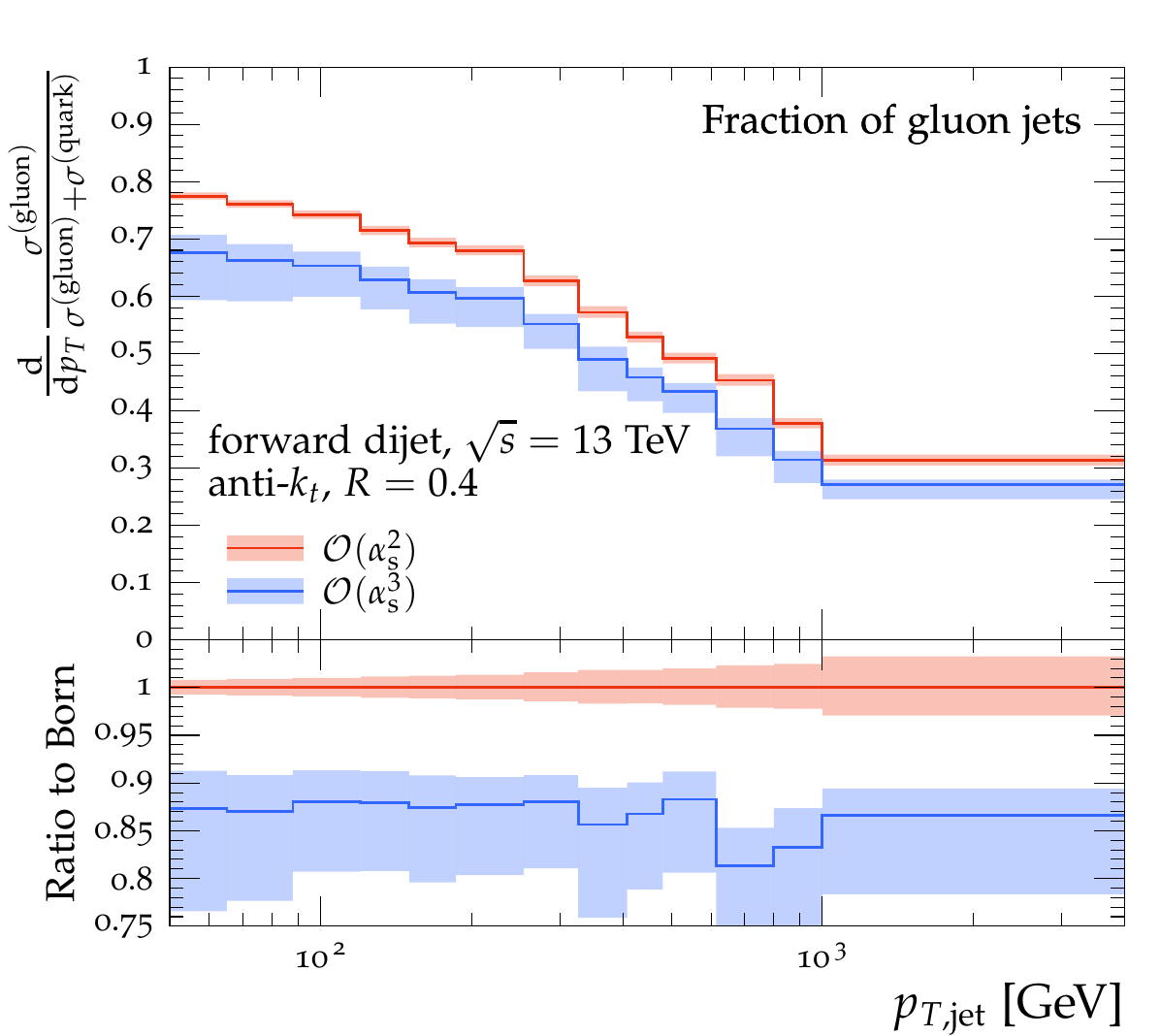}
    \caption{forward dijets}\label{fig:gluonfractions-forwardjj}
  \end{subfigure}
  \hfill
  \begin{subfigure}[t]{0.32\textwidth}
    \includegraphics[width=\textwidth]{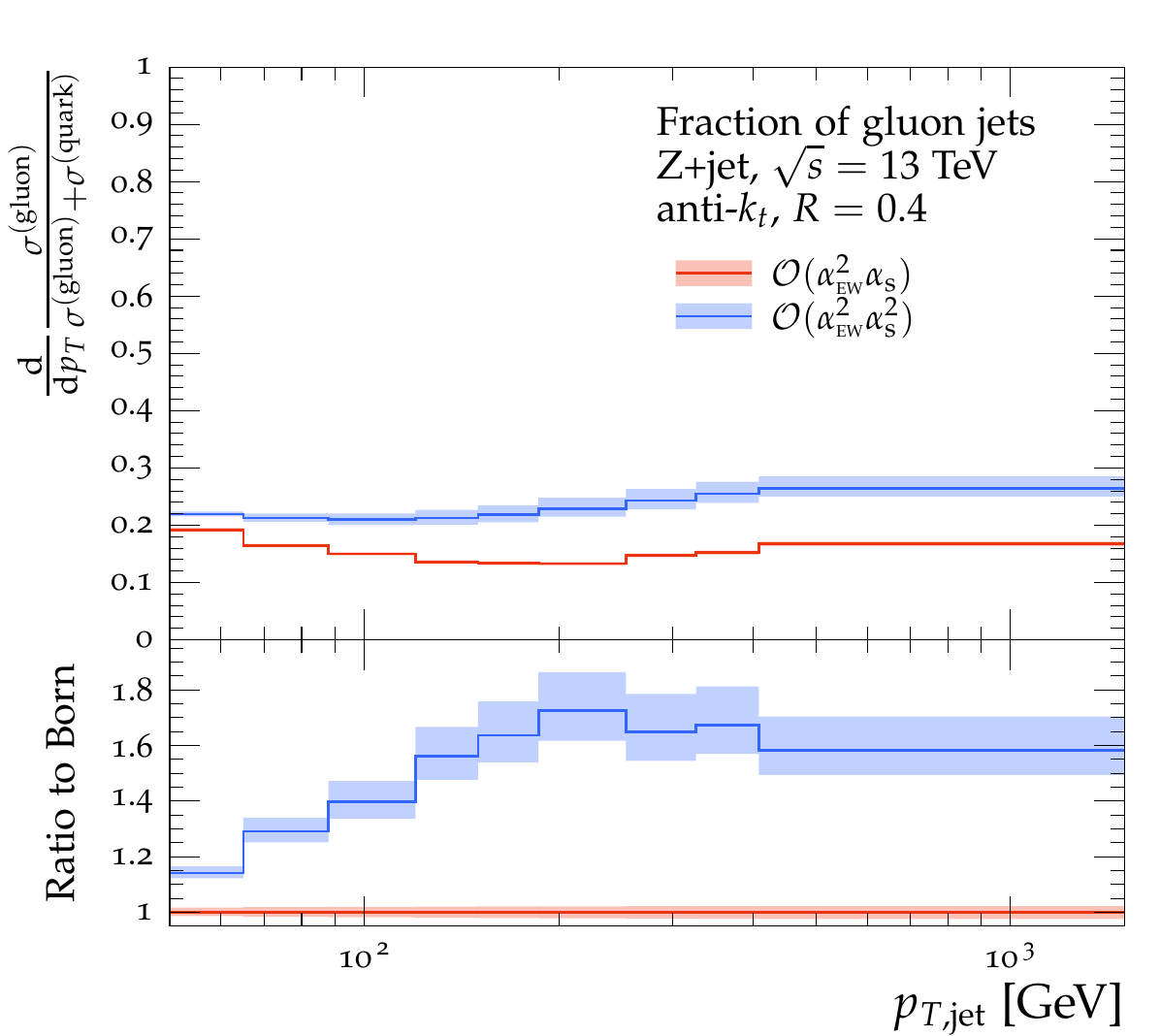}
    \caption{$Z$+jet}\label{fig:gluonfractions-Zj}
  \end{subfigure}\\
  \vspace*{0.3cm}
  \caption{Fraction of gluon jets as a function of the jet transverse
    momentum $p_{T,\text{jet}}$ in dijet (left and middle) and $Z$+jet (right)
    processes. For the dijet production we present results for the
    leading central and forward jet, respectively.
    We show predictions at Born (red bands) and at one-loop level (blue bands).
    Flavour tagging for the considered anti-$k_t$ jets relies on the algorithm
    described in~\cite{Caletti:2021oor}, that employs the BSZ flavour-$k_t$ algorithm,
    and in turn defines the partonic channels $\delta$ in our implementation
    of NLL \Caesar resummation in \sherpa.
  }\label{fig:gluonfractions}
\end{figure}

As discussed in section~\ref{sec:res_setup}, the implementation of the
\Caesar resummation in the \sherpa plugin~\cite{Gerwick:2014gya}
relies on a separation into different flavour channels $\delta$. Since
we obtain fully exclusive parton-level events from \comix, we can
easily apply the (BSZ) flavour-clustering algorithm from
Ref.~\cite{Banfi:2006hf} so as to identify the flavour assignment of a
given phase-space point. We use this algorithm iteratively, following
the procedure introduced in~\cite{Caletti:2021oor} for the leading jet
in the $Z$+jet case. This can trivially be extended to the
central/forward jet in dijet production. It allows us to talk about
the flavour of a particular anti-$k_t$ jet of radius $R$ in an IR-safe
way.  Note that in practice we run the \emph{bland} variant of the
algorithm whereby each jet gets identified as either quark- or
gluon-like. While flavour identification is a necessity in the context
of our matching scheme (at least at LO in the angularity
distribution), these results also provide a well defined way to
analyse the flavour decomposition of jet cross sections. Such
truth-level flavour assignment could for example be used as
well-defined input, \emph{e.g.}\ for machine-learning based methods,
or as benchmark for flavour-tagging
algorithms~\cite{Lonnblad:1990bi,ATLAS:2017dfg,Komiske:2016rsd,Romero:2021qlf,Cheng:2017rdo,Kasieczka:2018lwf,Lee:2019ssx}.

We start our discussion by considering the fractions of gluon and quark anti-$k_t$ jets
in the $Z$+jet and dijet final states. These are compiled in Fig.~\ref{fig:gluonfractions}
for the central- and forward jet in dijet events, as well as in $Z$+jet production,
as a function of the jet transverse momentum, $p_{T,\text{jet}}$.
On each plot, two curves are shown together with their respective
scale uncertainties from the $7$-point scale variation: gluon
fractions at Born level (red bands), \emph{i.e.}\
$\mathcal{O}(\alphaS^2)$ for dijets and
$\mathcal{O}(\alpha_\text{EW}^2\alpha_s)$ for $Z$+jet\footnote{Note
  that, as described in Sec.~\ref{sec:definitions}, we always include
  the decay of the $Z$ boson into muons in the calculation.}, and for
the NLO QCD matrix element (blue bands), at $\mathcal{O}(\alphaS^3)$
and $\mathcal{O}(\alpha_\text{EW}^2\alphaS^2)$, respectively. Note
that the (NLO) one-loop result corresponds to the LO accuracy for
jet-angularity distributions.
One sees that for low-$\pt$ dijet events, both the central and
forward jets have a large fraction of gluon jets ($\sim 70\%$), and that high-$\pt$
dijet events and $Z$+jet events at any $p_{T,\text{jet}}$, with gluon fractions of
about $20{-}30\%$, are dominated by quarks. 

A striking feature of Fig.~\ref{fig:gluonfractions} is the size of the
NLO QCD corrections. For dijet production, we see an (absolute)
decrease of about 15\% of the gluon fraction, regardless of the
transverse momentum and rapidity of the jet, marginally larger than
the estimated one-loop perturbative uncertainty.
For $Z$+jet events, NLO corrections cause a $2{-}10$\% (absolute)
increase with a clearly-visible dependence on the jet $p_T$.
This is in line with earlier observations that the NLO corrections to
the gluon fraction in $Z$+jet events are both larger than in dijet
events and show a stronger jet-$p_T$ dependence (see
\emph{e.g.}~\cite{Rubin:2010xp,Frye:2016aiz}).
It is also worth commenting on the relative size of the theoretical uncertainty,
which appears to increase when going from Born-level to NLO QCD.  The reason for this
counterintuitive behaviour can be traced back to the fact that the considered
gluon fractions are determined by cross-section ratios.
At Born-level the dependence on $\alphaS$, and hence on the renormalisation scale,
exactly cancels between numerator and denominator, leaving only a rather weak
dependence on the factorisation scale. 

Our results can be directly compared to the corresponding gluon-jet
fractions reported by CMS in~\citep{CMS:2021iwu} (see Fig.~2 therein and the
corresponding discussion), although that study was done using a
different definition of gluon jets based on \mgamcnlo\!\!+\pythia
simulations as well as a series of other generators.
Qualitatively, the gluon fractions presented in~\cite{CMS:2021iwu} show the
same pattern as the ones obtained here.
However, with the BSZ-based approach we find gluon fractions that are
generally smaller than the ones reported by CMS. This is especially the case in
Fig.~\ref{fig:gluonfractions-Zj} for the $Z$+jet process where CMS
found gluon fractions reaching almost 40\% at large $p_{T,\text{jet}}$.
This is in agreement with their comment that other generators
predict up to 25\% smaller gluon fractions in the $Z$+jet sample,
which is also where the largest NLO corrections are observed.

\begin{figure}
  \centering
  \includegraphics[width=.49\textwidth]{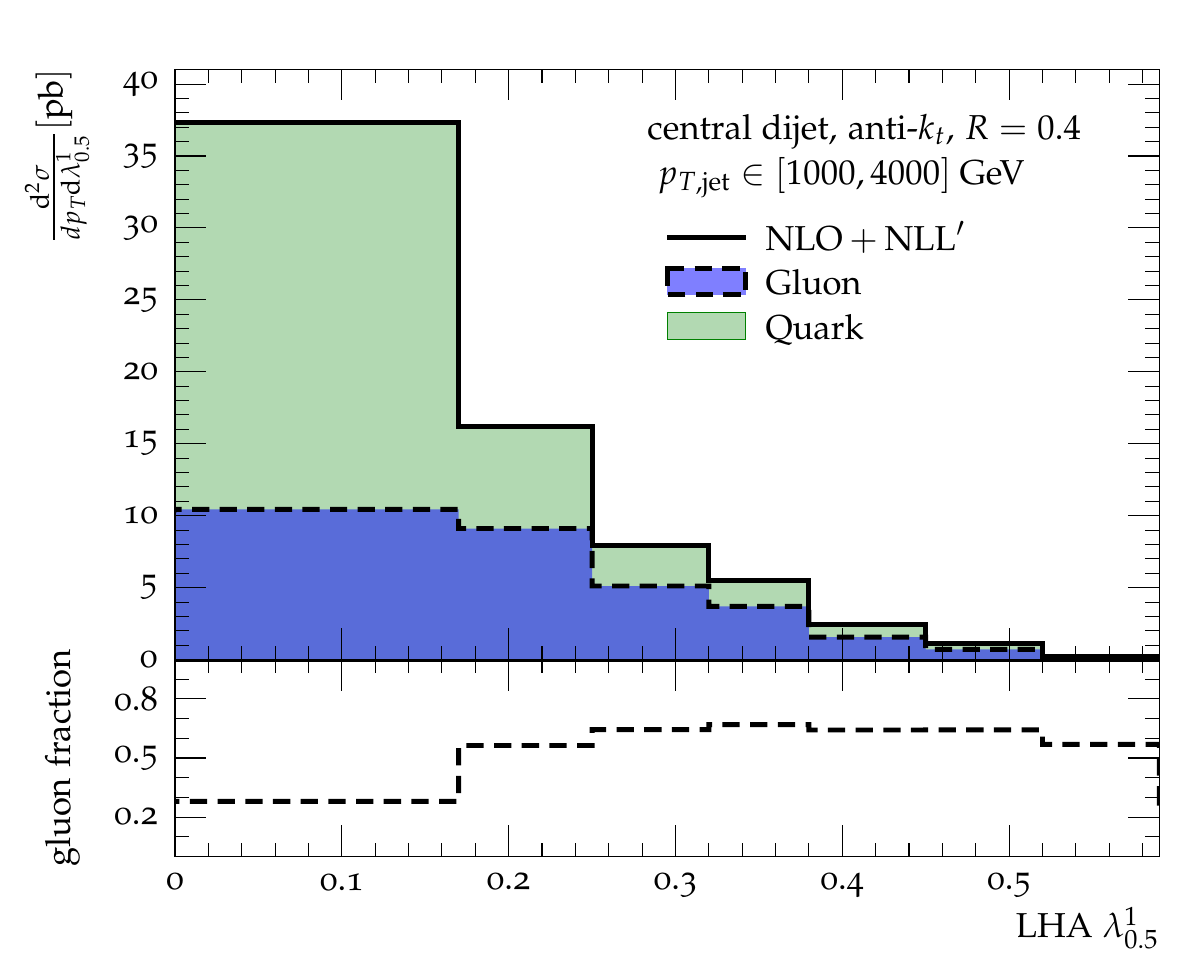}
  \includegraphics[width=.49\textwidth]{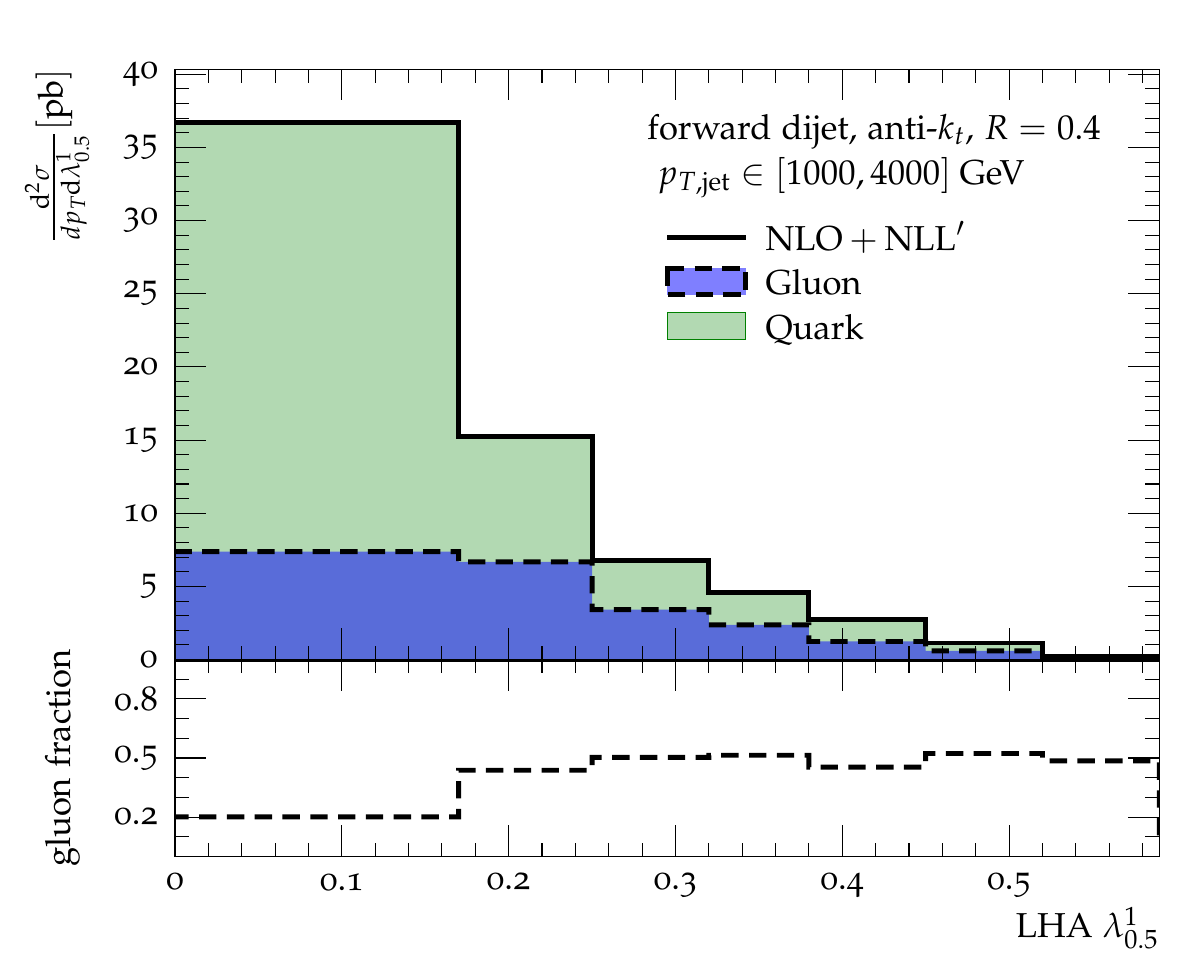}
  \caption{\NLOpNLLp accurate predictions for the quark- and gluon-jet contribution
    to the Les Houches Angularity $\lambda^1_{0.5}$ for the central (left) and forward
    (right) dijet event selection, with jets in the range
    $p_{T,\text{jet}}\in[1000,4000]~\text{GeV}$. The lower panels show the gluon
    fractions in the respective angularity bins.
}\label{fig:flav_sep_dijet}
\end{figure}

\begin{figure}
  \centering
  \includegraphics[width=.49\textwidth]{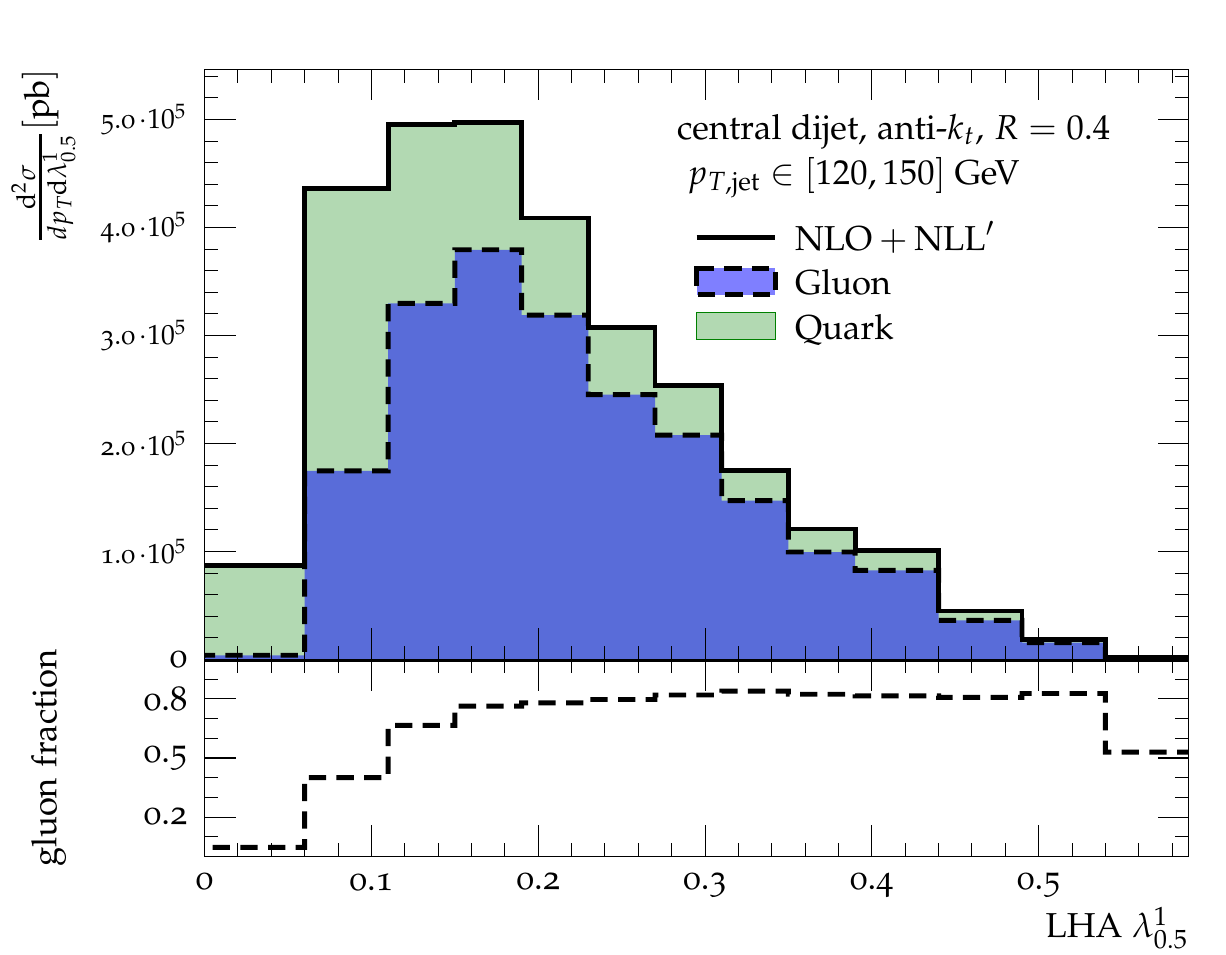}
  \includegraphics[width=.49\textwidth]{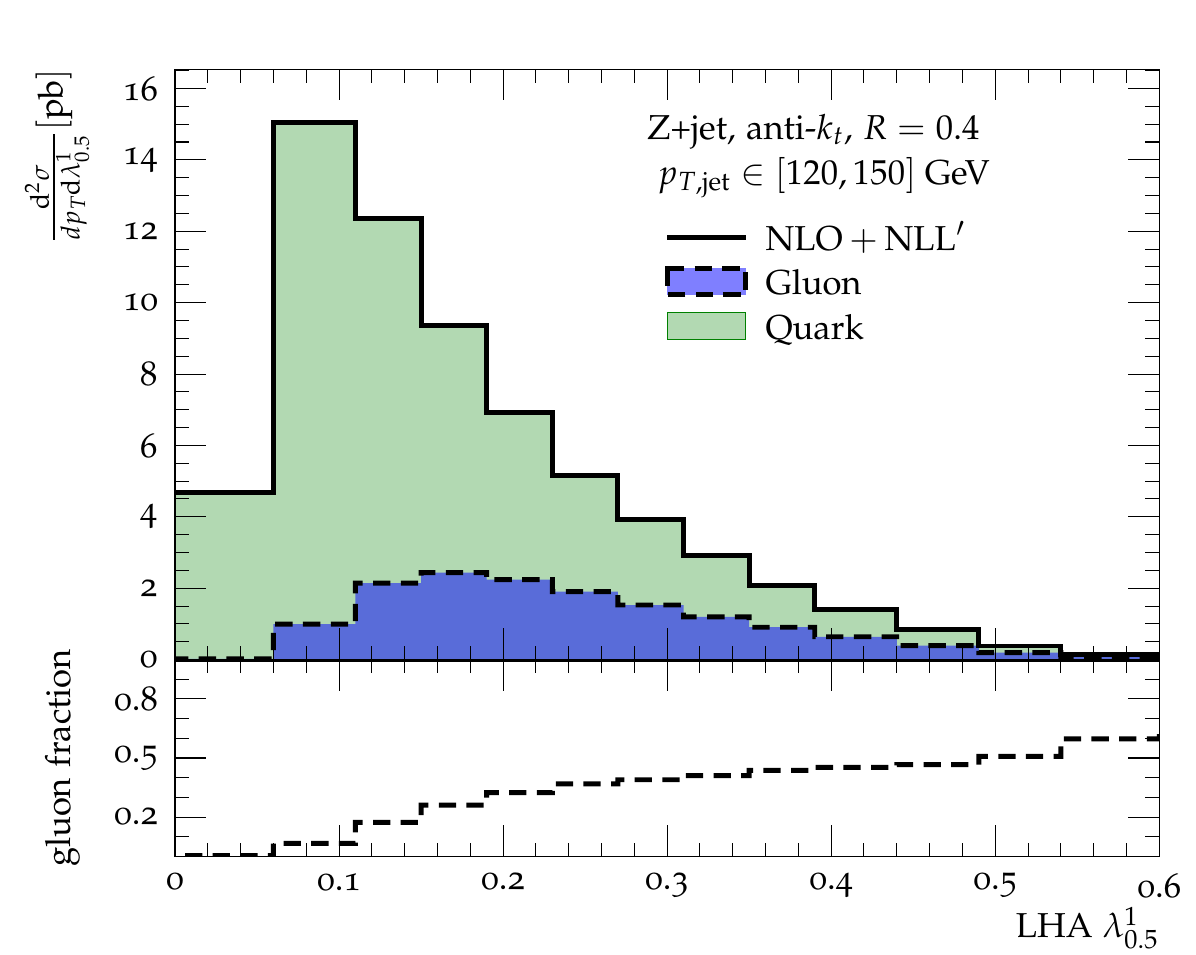}
  \caption{\NLOpNLLp accurate predictions for the quark- and gluon-jet
    contribution to the Les Houches Angularity $\lambda^1_{0.5}$ for the central
    dijet (left) and the $Z$+jet (right) event selections, with jets
    in the range $p_{T,\text{jet}}\in[120,150]~\text{GeV}$. The lower panels show the gluon
    fractions in the respective angularity bins.
  }\label{fig:flav_sep_cen_zjet}
\end{figure}

Next, Figs.~\ref{fig:flav_sep_dijet} and~\ref{fig:flav_sep_cen_zjet}
show examples of our matched \NLOpNLLp results for the LHA
$\lambda^1_{0.5}$ distribution, for different $p_{T,\text{jet}}$ and event
selections. 
In each case, we have separated the total cross section into
contributions from quark and gluon jets. The total result is given by the
sum of both components, indicated by the solid (black) line. The
fraction of gluon jets in each bin is shown in the lower panels.
The selected $p_{T,\text{jet}}$ ranges are chosen such that they
coincide with the gluon- and quark-enriched samples studied by CMS
in~\cite{CMS:2021iwu} (see also table~\ref{tab:final} in the following
section).
The results confirm the findings from Fig.~\ref{fig:gluonfractions}: we
indeed see that the high-$\pt$ dijets and the $Z$+jet sample
for $p_{T,\text{jet}}\in[120,150]~\text{GeV}$ are dominated by quark jets, and the
low-$\pt$ dijet events are instead dominated by gluon jets.
It is also clearly visible in these figures that the gluon
distributions contribute at larger values of the angularity variable
than the quark distributions, indicative of their potential as
a quark--gluon discriminator.

\subsection{Hadron-level results}\label{sec:results-hadron}

\begin{figure}[t!]
  \centering
  \includegraphics[width=0.32\textwidth]{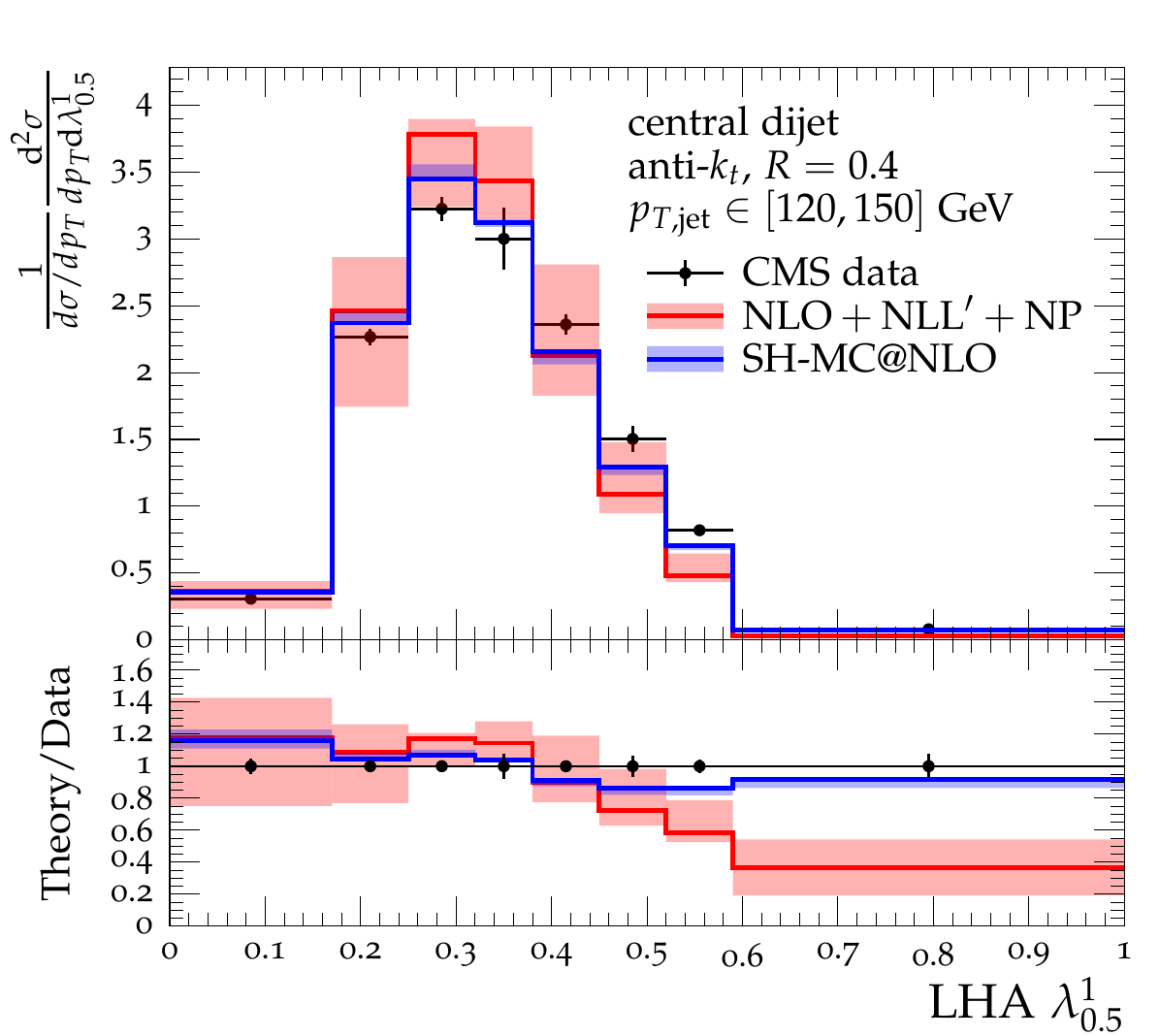}
  \includegraphics[width=0.32\textwidth]{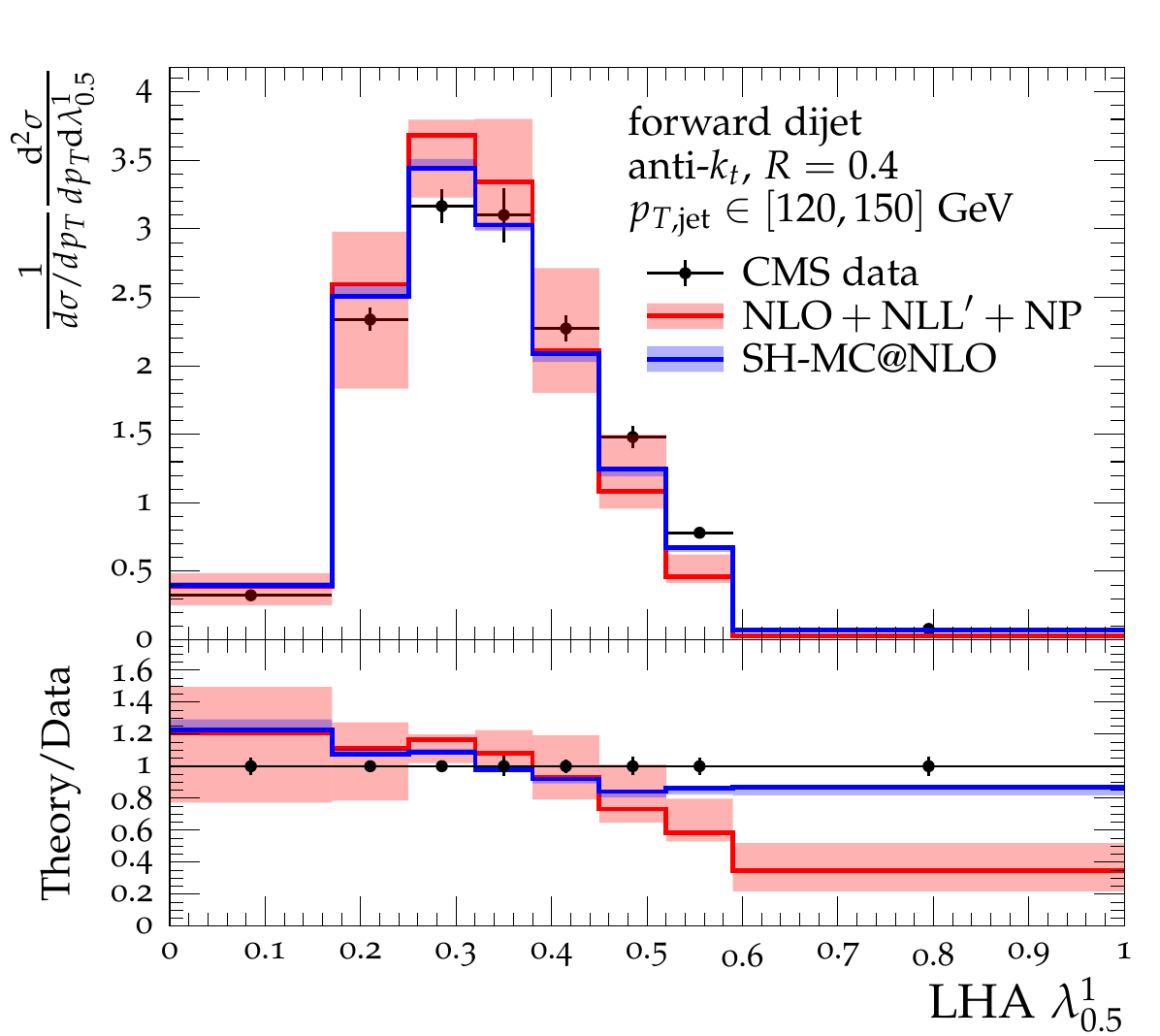}
  \includegraphics[width=0.32\textwidth]{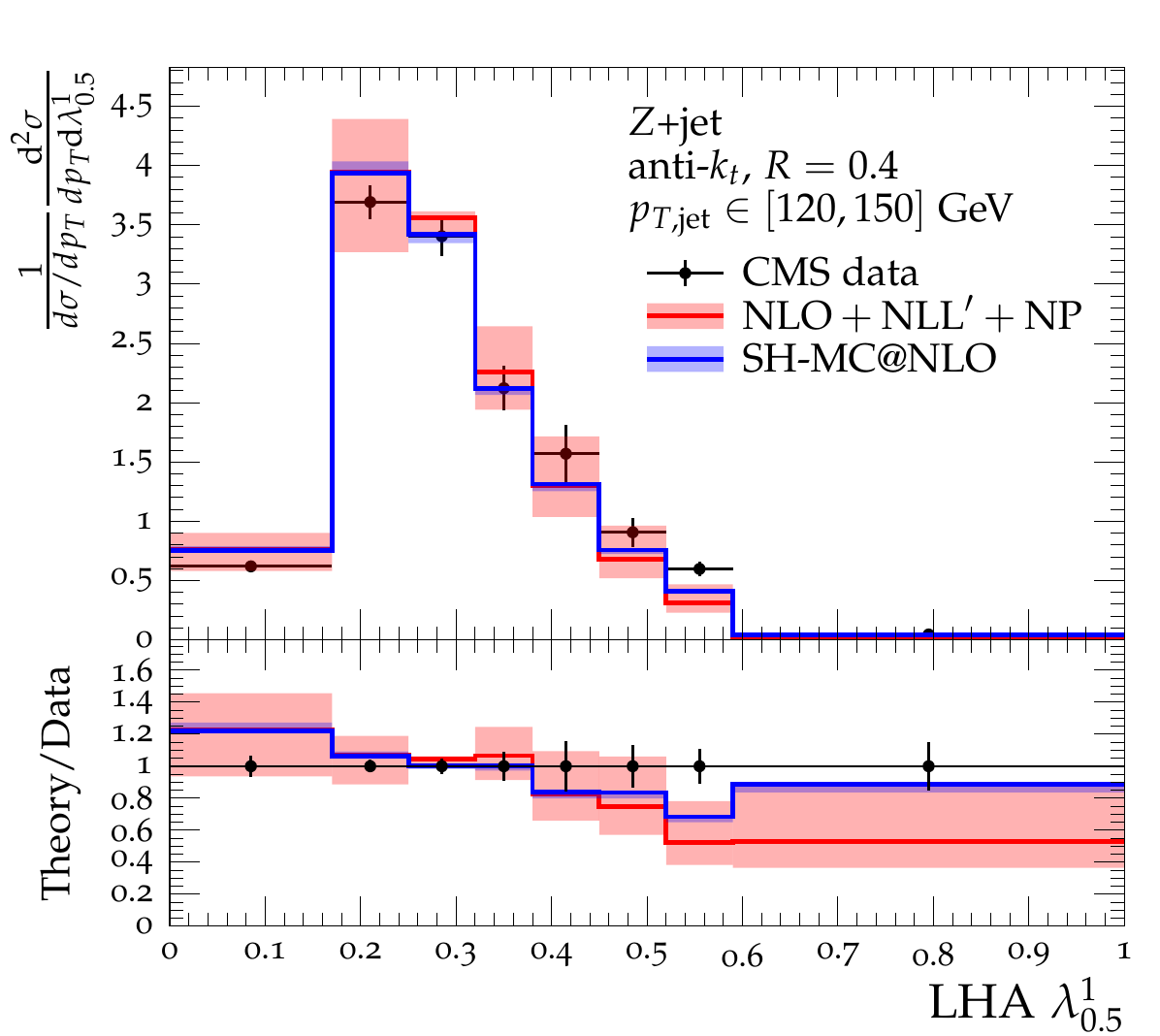}
  \includegraphics[width=0.32\textwidth]{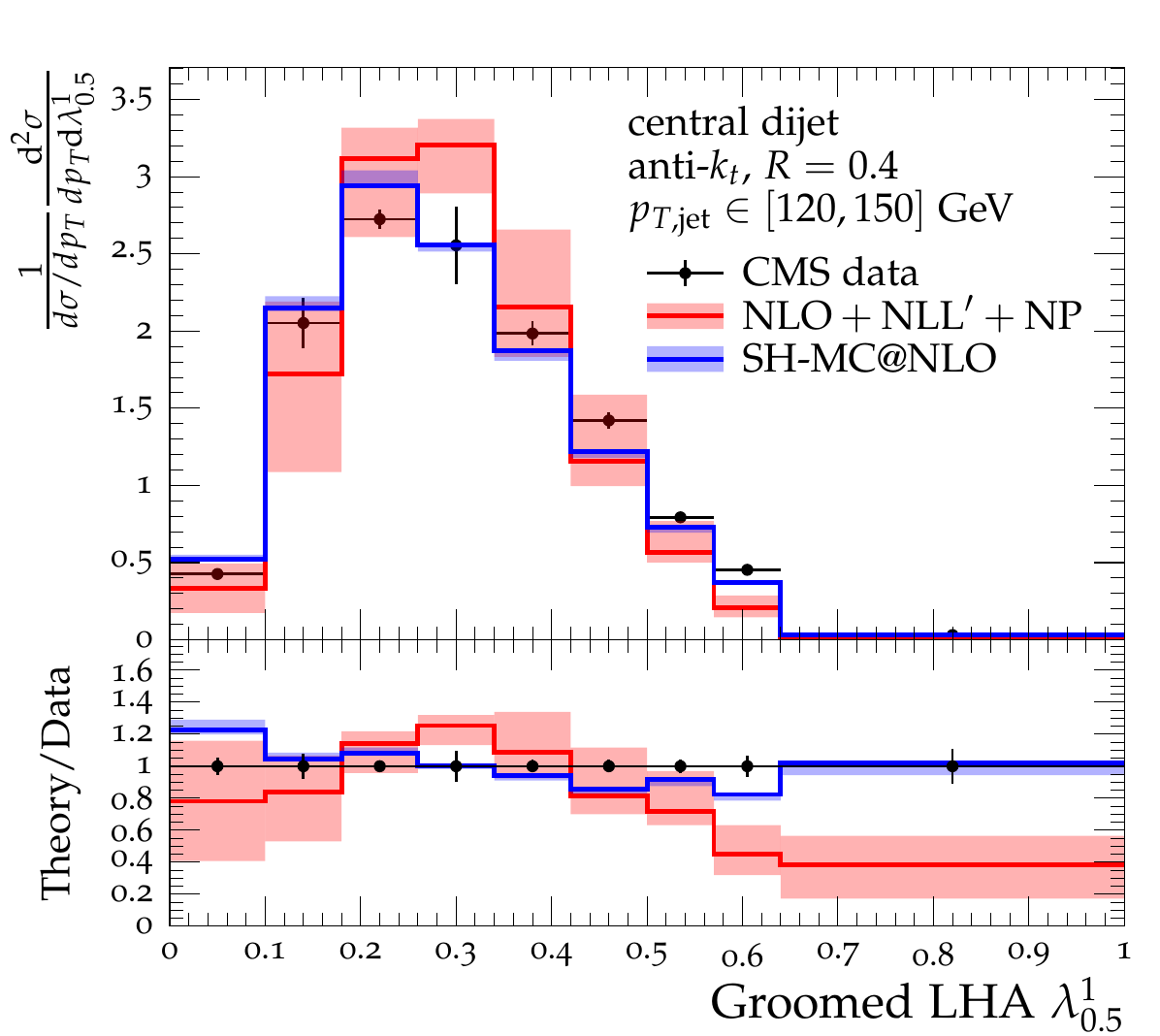}
  \includegraphics[width=0.32\textwidth]{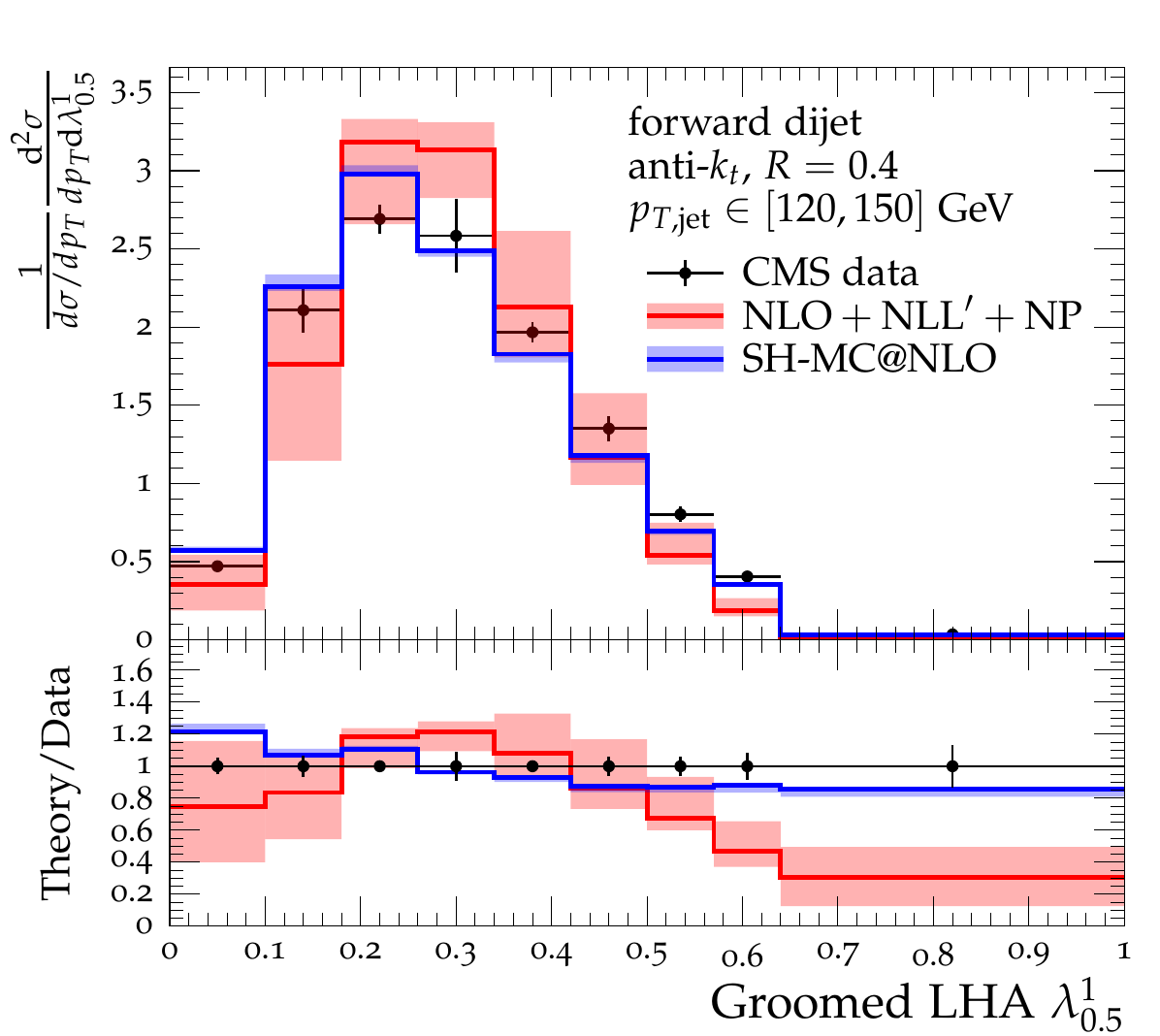}
  \includegraphics[width=0.32\textwidth]{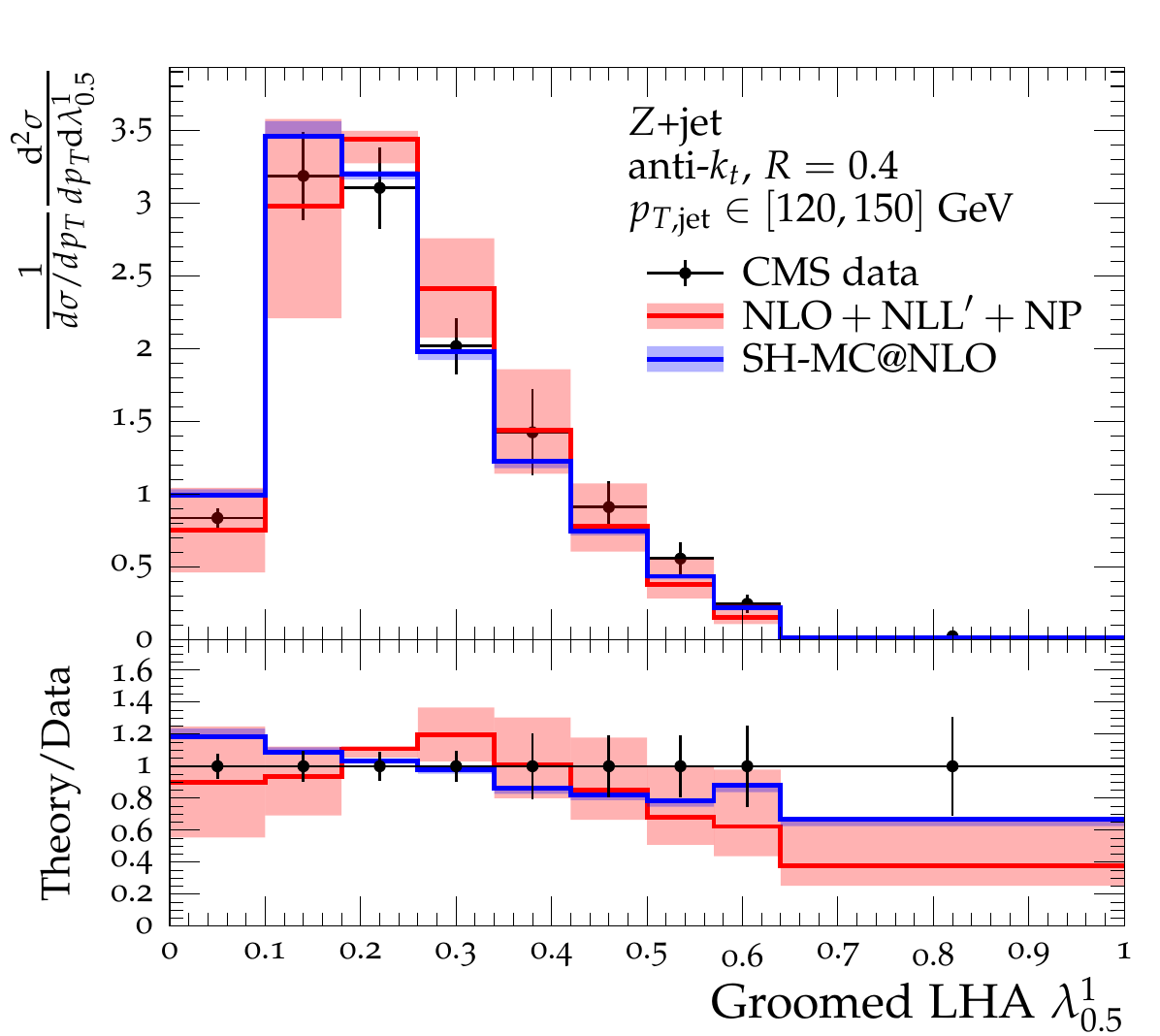}
  \caption{\NLOpNLLpNP and hadron-level \SHMCatNLO predictions for the
    differential cross section in the Les Houches Angularity $\lambda^1_{0.5}$
    for ungroomed (top row) and groomed (bottom row) $R=0.4$ anti-$k_t$ jets
    with $p_{T,\text{jet}}\in [120,150]~\text{GeV}$, compared to data from
    CMS~\cite{CMS:2021iwu}. The left and
    middle panel correspond to the central and forward jet in dijet events, respectively,
    the right one to the leading jet in $Z$+jet production. The NP corrections
    to the perturbative \NLOpNLLp prediction have been obtained with the transfer-matrix approach.
  }\label{fig:dist_LHA}
\end{figure}

In this section we present our final \NLOpNLLpNP predictions, with NP
corrections implemented through the transfer-matrix approach, for jet
angularities in both dijet and $Z$+jet production and compare them to
the results obtained by the CMS collaboration. To avoid the
proliferation of plots, we concentrate mostly on the LHA
$\lambda^1_{0.5}$ measured on (groomed or ungroomed) $R=0.4$
anti-$k_t$ jets at moderate transverse momentum, namely
$p_{T,\text{jet}}\in [120,150]~\text{GeV}$.
As before, we consider separately the central and forward jet
in dijet events, as well as the leading jet in $Z$+jet events, we thereby
restrict to the data for \emph{all} hadrons. Corresponding
results for the jet width $\lambda^1_{1}$ and jet thrust $\lambda^1_{2}$ we
compile in App.~\ref{app:other-ang}. The full set of predictions, \emph{i.e.}\
for all the $p_{T,\text{jet}}$ slices considered by CMS in~\cite{CMS:2021iwu},
for jet radius $R=0.8$, as well as based on charged tracks, and their
comparison to the data from CMS are also publicly
available~\cite{resultslink}\footnote{Our results are also attached to this
paper as supplementary material.}.
To estimate the theoretical uncertainty of our \NLOpNLLpNP predictions
we consider $7$-point variations of the factorisation and renormalisation
scales, as well as variations of the $x_L$ parameter, see section~\ref{sec:res_setup}
for details. Besides the NP corrections corresponding to the default parameter
set in \sherpa, we also consider up- and down variations of the UE activity and
use Lund string fragmentation as alternative hadronisation model,
\emph{cf.}\ section~\ref{sec:sherpasetup}. This way we derive four alternative
hadron-level distributions for \emph{each} combination of scale- and $x_L$-variations. The
final uncertainty is then obtained by taking the envelope of all those variations.

Alongside the \NLOpNLLpNP results we present hadron-level simulations with
\sherpa at NLO QCD accuracy, as described in section~\ref{sec:sherpasetup}. The perturbative
uncertainty of the \SHMCatNLO predictions are taken as the envelope of $7$-point
variations of $\muF$ and $\muR$ in the matrix elements and the parton shower. 

In Fig.~\ref{fig:dist_LHA} we present \NLOpNLLpNP and \SHMCatNLO predictions for
normalised differential cross sections in $\lambda^1_{0.5}$ and compare them with
the CMS experimental data. The top row of plots thereby corresponds to ungroomed
jets, while the bottom row ones are obtained with \softdrop ($\beta=0,\,z_\text{cut}=0.1$)
applied to the jets prior to the observable evaluation. The left-hand plots correspond
to LHA measurements on the central jet, the middle ones
on the forward jet in dijet events, respectively, and the right-hand ones on
the leading jet in $Z$+jet production.

Overall, our resummed and matched predictions when corrected of NP effects (shown in red)
provide a good description of the hadron-level data, with the exception of the
last (and in some cases the second to last) bin at large values of the angularity.
The corresponding region of phase space is outside the jurisdiction of the
all-orders calculation and one might have hoped that it would be well-described by the NLO contribution.
However, the last bin contains the kinematic endpoint of the fixed-order calculation,
accordingly, this part of the distribution is very sensitive to the effect of multiple
emissions. Indeed the \SHMCatNLO predictions (in blue) are able to populate this region
of phase space through additional parton-shower real radiation, resulting in a better
description of the data.
In the future, it would be interesting to see if higher-order,
\emph{e.g.}\ NNLO, corrections will yield an improved description in this
large-angularity region.
For the groomed distribution, we also see that the MC simulation provides
a better description of the peak region than the analytic calculation.
One should however bear in mind that all the $\lambda_{0.5}^1$ bins,
except the lowest one, are in a region with
$\lambda_{0.5}^1\ge \zcut$, \emph{i.e.}\ not directly affected by
\softdrop at NLL accuracy. In this region, non-trivial subleading
effects can have a sizeable impact (see also
Refs.~\cite{Marzani:2019evv,Benkendorfer:2021unv} for discussions and
potential improvements).

A crucial difference between the resummed and the \SHMCatNLO results
is the size of the theoretical uncertainty, which is much larger for
the \NLOpNLLpNP calculation. We have investigated this feature by
decomposing the total uncertainty into its various perturbative and
non-perturbative contributions. This clearly led us to the conclusion
that this effect is dominated by the resummation-scale variation,
\emph{i.e.}\ the variation of $x_L$ in Eq.~\eqref{eq:endpoint+xL}. A
systematic reduction of this uncertainty would require to improve the
accuracy of the resummation, \emph{i.e.}\ to include NNLL
contributions.  We will comment on the feasibility of this calculation
in the conclusions.  
We furthermore note that in the here shown MC simulations no
corresponding variation is performed (see
section~\ref{sec:sherpasetup}).
It would be interesting to study systematic variations
when using a different shower model, \emph{e.g.}\ the \DIRE cascade~\cite{Hoche:2015sya}
available in the \sherpa framework.

\begin{figure}
  \centering
  \includegraphics[width=0.32\textwidth]{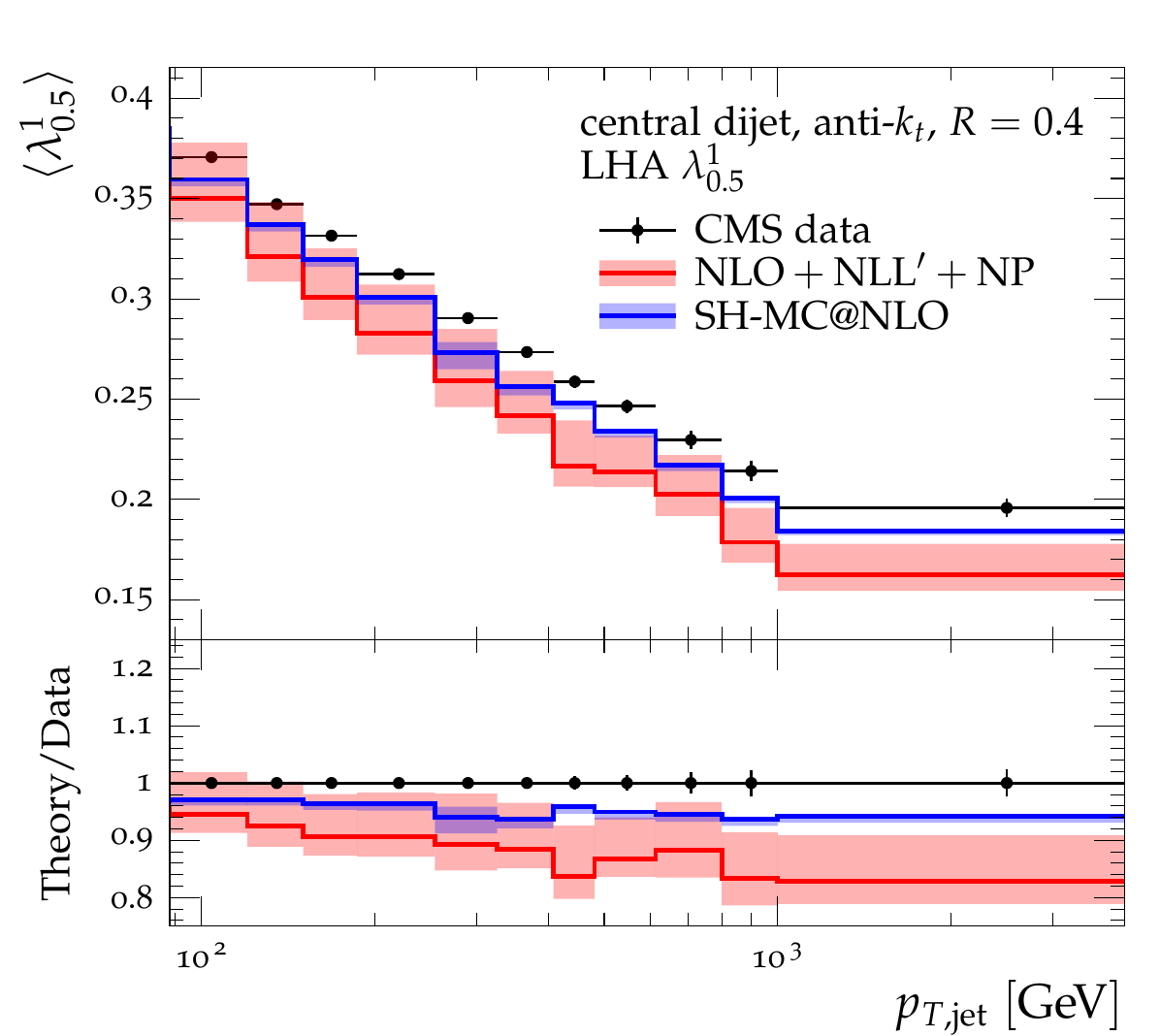}
  \includegraphics[width=0.32\textwidth]{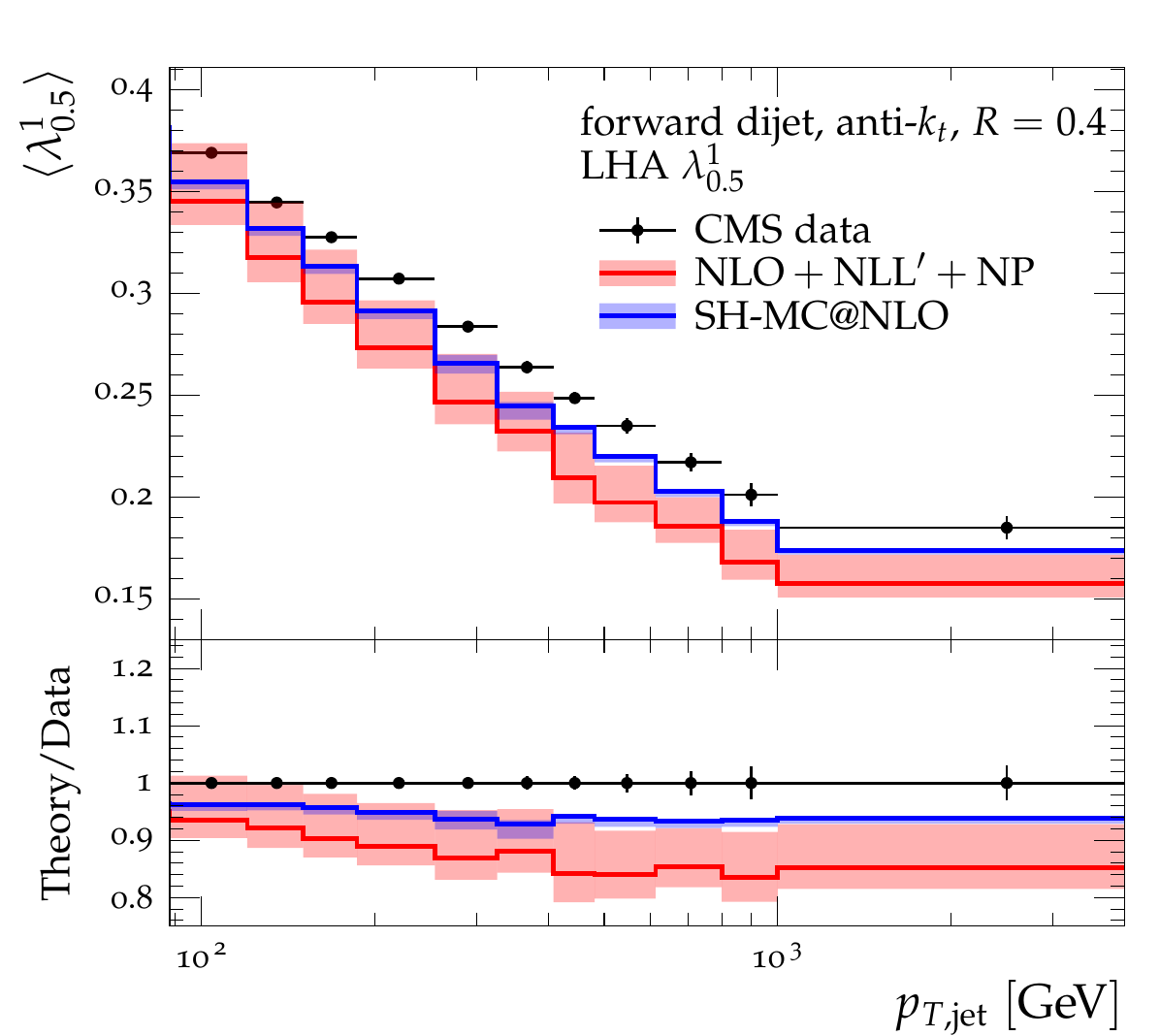}
  \includegraphics[width=0.32\textwidth]{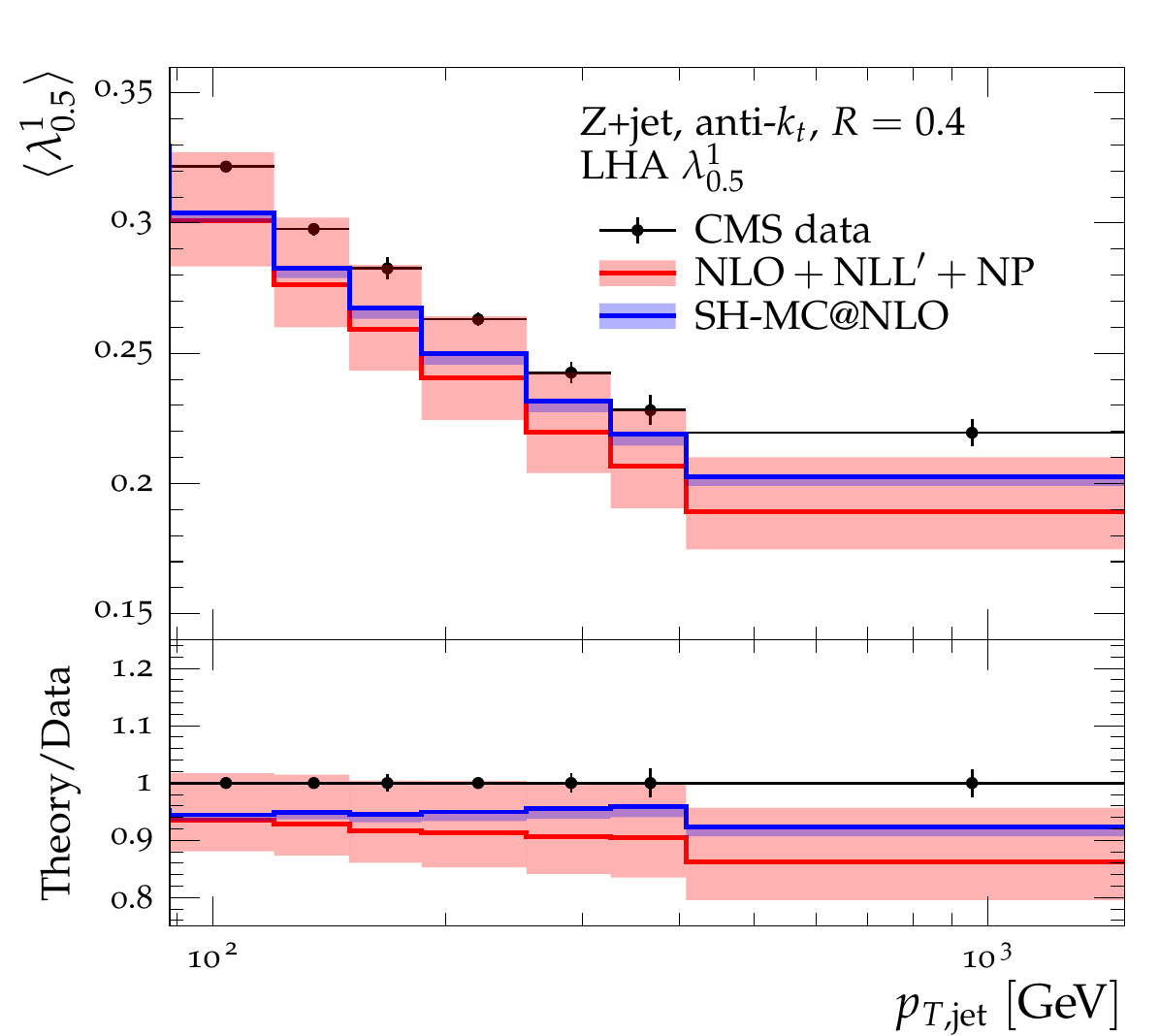}
  \includegraphics[width=0.32\textwidth]{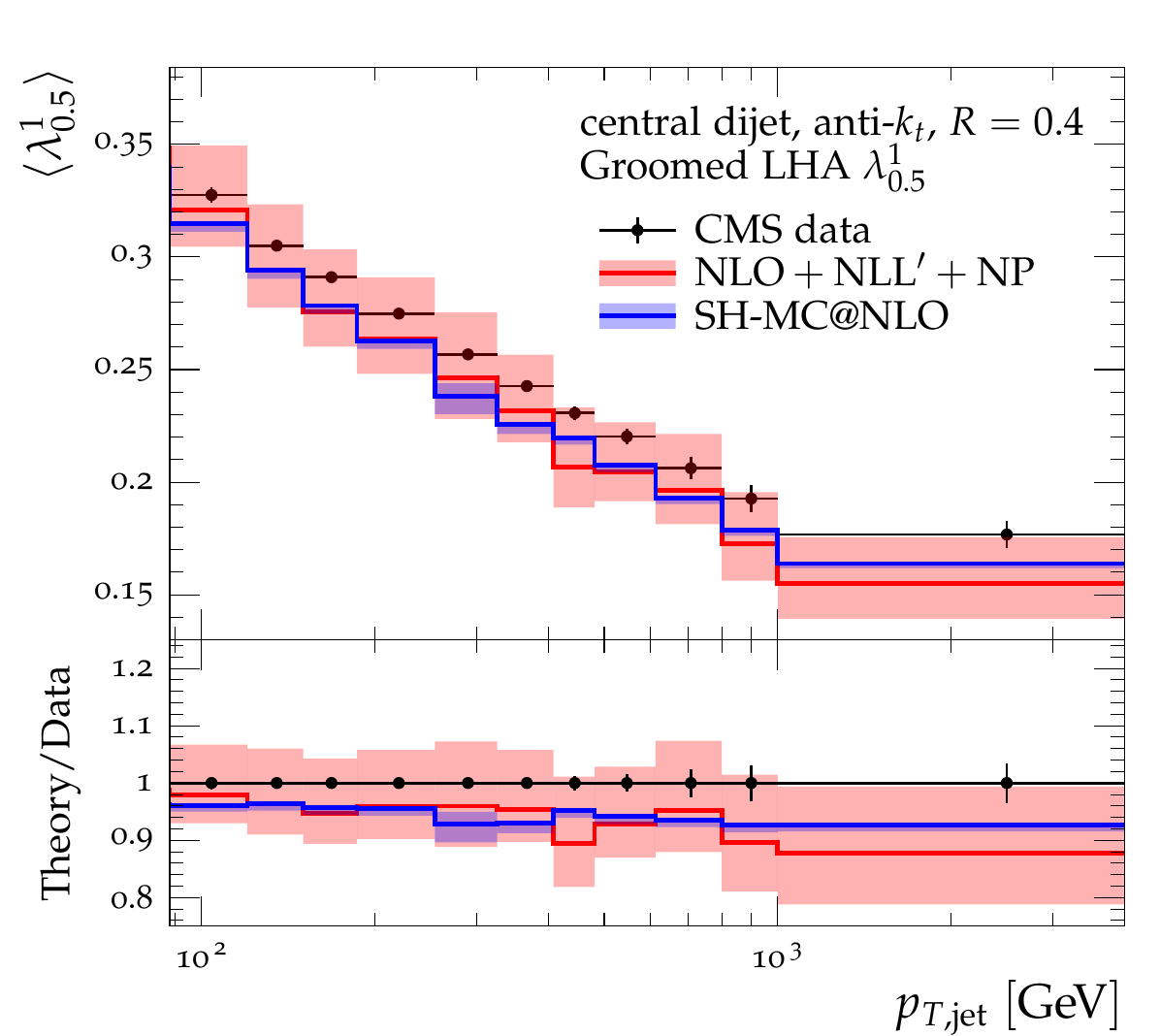}
  \includegraphics[width=0.32\textwidth]{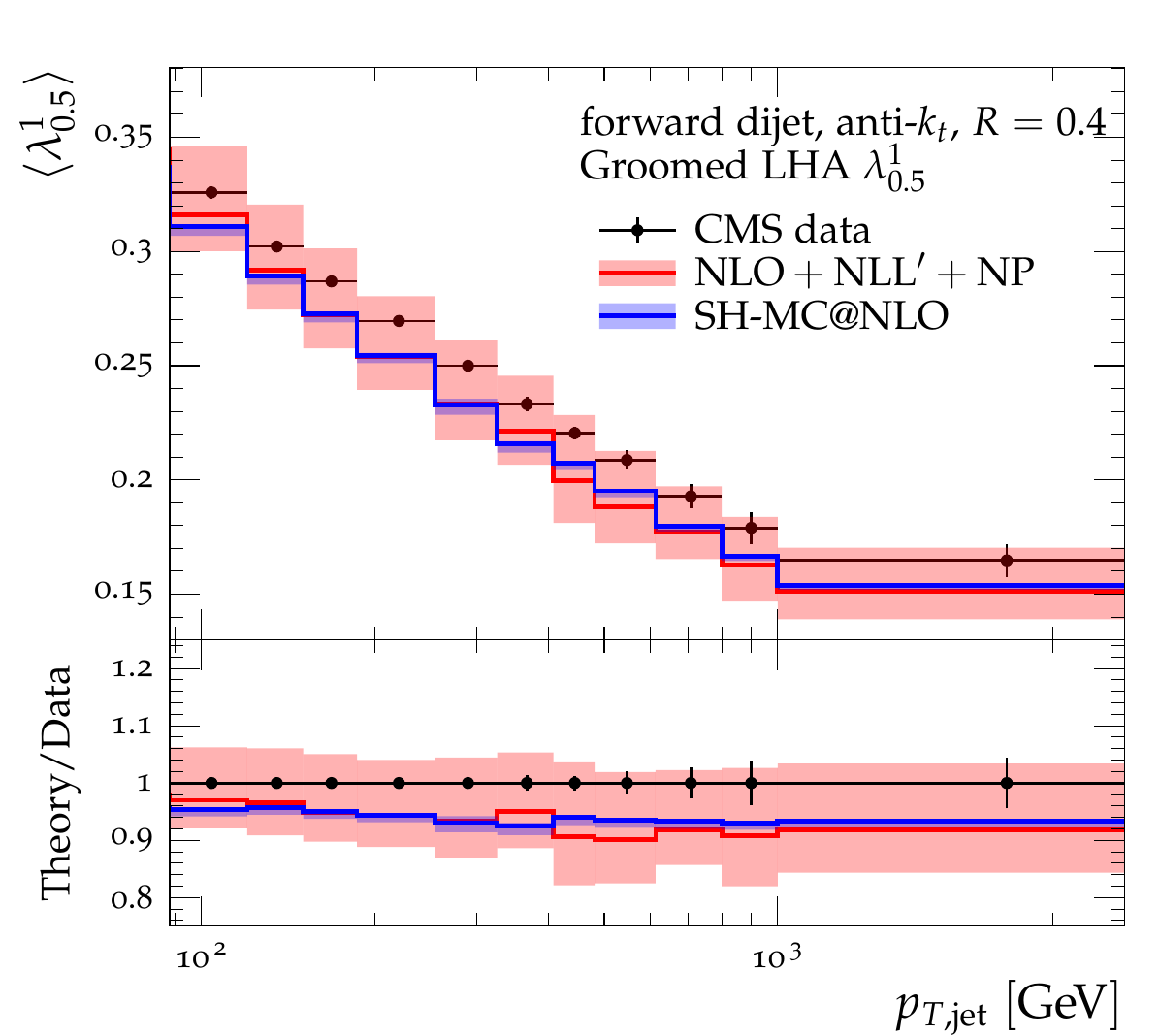}
  \includegraphics[width=0.32\textwidth]{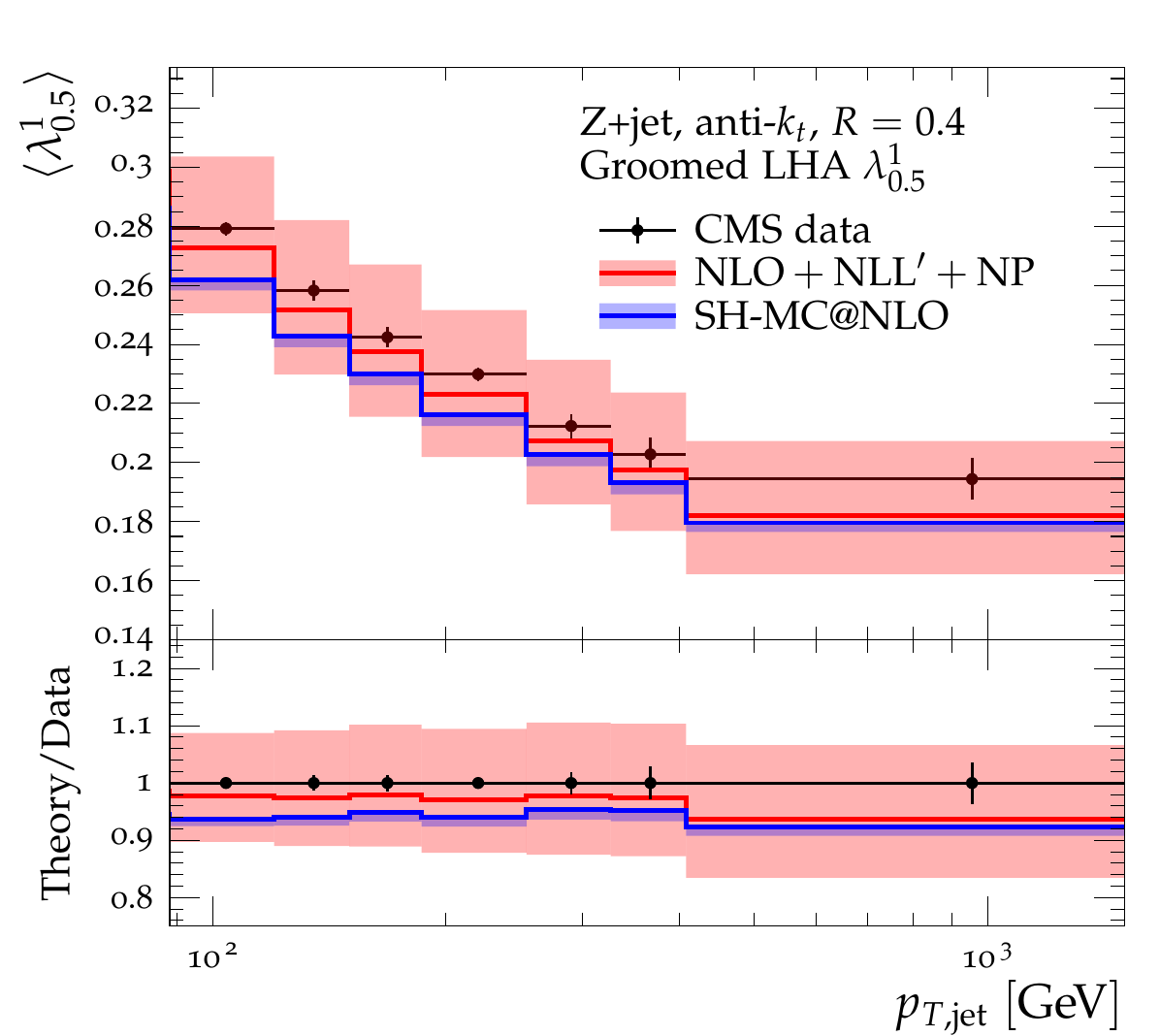}
  \caption{Results for the mean value of the Les Houches Angularity $\langle
    \lambda^1_{0.5} \rangle$ for ungroomed (top row) and groomed (bottom row)
    $R=0.4$ anti-$k_t$ jets as a function of $p_{T,\text{jet}}$. The left and
    middle panel correspond to the central and forward jet in dijet events,
    respectively, the right one to the leading jet in $Z$+jet
    production.}\label{fig:mean_LHA}
\end{figure}

Following the analysis performed by the CMS collaboration in
Ref.~\cite{CMS:2021iwu} we also consider the mean value of the
angularity distributions, as a function of
$p_{T,\text{jet}}$. Corresponding results for the case of the
$\lambda^1_{0.5}$ angularity are shown in Fig.~\ref{fig:mean_LHA}. The
layout is organised as in Fig.~\ref{fig:dist_LHA} with ungroomed jets
in the top row and \softdrop-groomed jets in the bottom row and with,
from left to right, the central and forward jet in dijet events and
the leading jet in $Z$+jet events.  Analogous results for the
jet-width and jet-thrust angularities are collected in
App.~\ref{app:other-ang}.

The comparison between the resummed and the MC predictions highlights
the feature previously discussed, namely that the estimated
uncertainties of the \NLOpNLLpNP results are significantly larger than
the ones obtained for the \SHMCatNLO simulations, with the former
being dominated by the $x_L$ variation. We observe that both
theoretical predictions consistently underestimate the experimentally
measured mean values. From the ratio plots in the bottom panels we can
read off that the theory-to-data ratios for the central values are
almost constant, exhibiting only a very mild dependence on the
transverse momentum.  This holds in particular for the \SHMCatNLO
predictions that undershoot the data by about $5-10\%$. Larger
deviations, reaching up to 18\% at the highest $p_{T,\text{jet}}$
values, are seen for the \NLOpNLLpNP results in the ungroomed
case.
In the groomed case, although our \NLOpNLLpNP predictions
systematically undershoot the data, we observe agreement, within the
theoretical uncertainties.
In Ref.~\cite{CMS:2021iwu}, similar results have been found for the LO
MC generators considered there.  Part of this effect can be explained
by the observed underestimation of large angularity values in the
theoretical predictions when comparing to data. However, in this
region the cross section is quite low, thus it contributes rather
little to the average value $\langle \lambda^1_{0.5}\rangle$. As we
have seen in section~\ref{sec:res_np}, the angularity distribution is
significantly affected by NP corrections. Accordingly, it would
clearly be interesting to include the new data on the jet angularities
in tunes of the NP models, thereby further investigating which
parameters affect their description. In turn, improved NP transfer
matrices could be derived and applied to the \NLOpNLLp predictions.

Detailed studies of jet angularities allow us to quantitatively assess how well
MC simulations, and more generally theoretical calculations, describe the particle
distribution inside jets. In turn, they offer potential to analyse the description of
QCD radiation off quarks and gluons, separately. This issue was, for instance, studied in
Refs.~\cite{Andersen:2016qtm,Gras:2017jty}. The measurement of CMS presented in
\cite{CMS:2021iwu} now allows us to test theoretical predictions on jets from
gluon-enriched and quark-enriched samples. For this purpose, the analysis identified five
interesting phase-space regions and parameter choices, detailed in
table~\ref{tab:final}, that can be classified as having an enhanced gluon- or
quark-jet contribution in the dijet or $Z$+jet process, respectively.
We refer to section~\ref{sec:results-perturbative} and
Fig.~\ref{fig:gluonfractions} in particular for a discussion on
flavour fractions in these specific channels.

\begin{table}[hbt!]
\begin{center}
	\begin{tabular}{ | c | c | c | c | c |}
    \hline
configuration & type of jet & $p_{T,\text{jet}}$ [GeV]& $g$-enriched & $q$-enriched \\
    \hline
   (1) & ungroomed $R=0.4$ &[120,150] & dijet central & $Z$+jet\\
   (2) & ungroomed $R=0.4$ &[1000,4000] & dijet central & dijet forward\\
   (3) & ungroomed $R=0.8$ &[120,150] & dijet central & $Z$+jet\\
   (4) & ungroomed $R=0.4$ (tracks only)&[120,150] & dijet central & $Z$+jet\\
   (5) & \softdrop ($\beta=0$, $z_\text{cut}=0.1$) $R=0.4$ &[120,150] & dijet central & $Z$+jet\\
\hline
\end{tabular}
\end{center}
\caption{Configurations selected in Ref.~\cite{CMS:2021iwu} to test theory predictions
  for gluon-enriched and quark-enriched samples.}\label{tab:final}
\end{table} 

\begin{figure}
  \centering
   \includegraphics[width=0.9\textwidth]{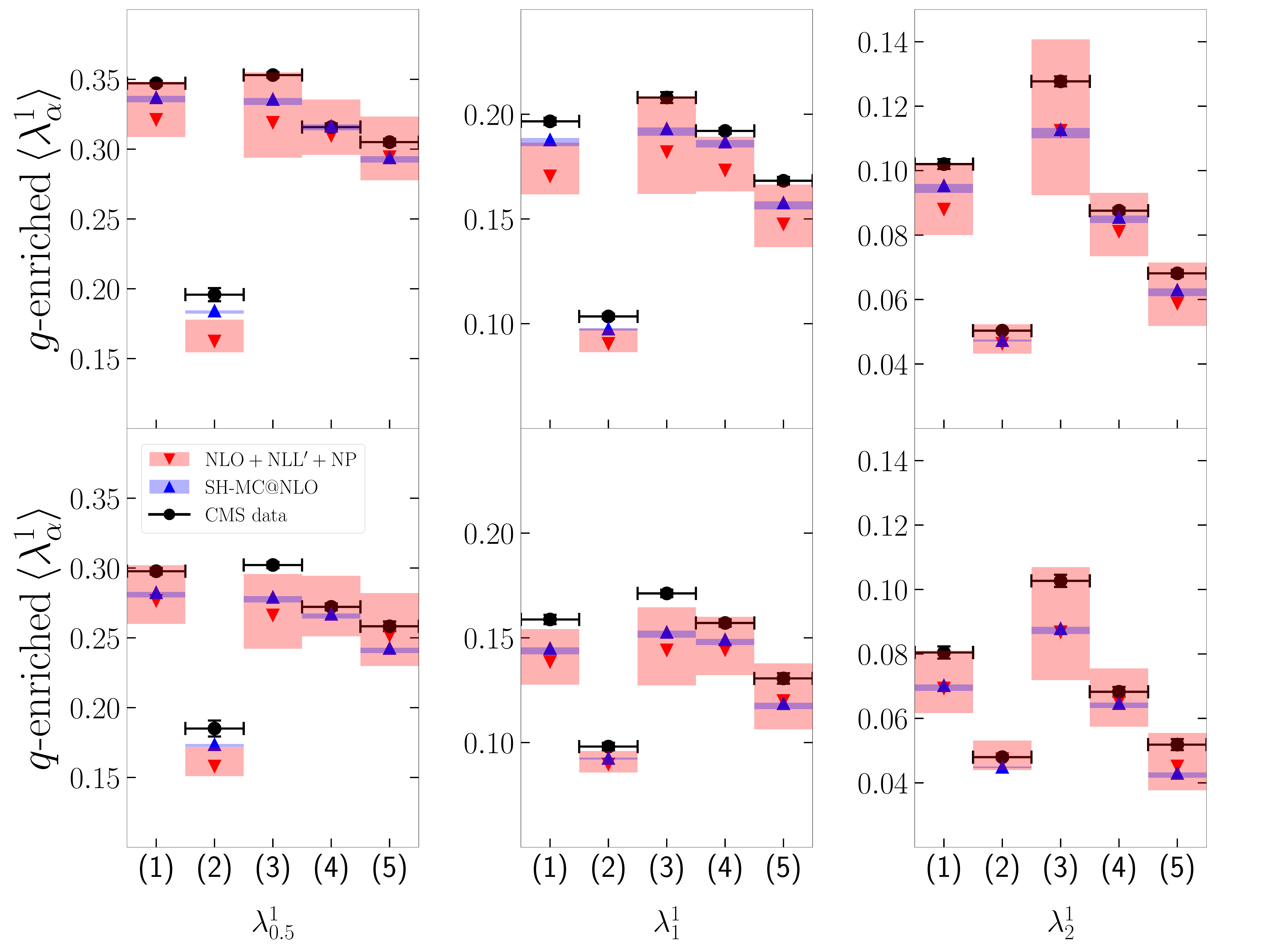}
   
   \caption{Jet-angularity mean values for the phase-space regions enriched
     with gluon- and quark jets in dijet and $Z$+jet production, respectively. See
     Tab.~\ref{tab:final} for details on the configurations and the respective
     process considered as $g$- and $q$-enriched.}	
  \label{fig:dist_avr_enriched}
\end{figure}

For these five configurations we consider the three IRC safe
angularities $\lambda^1_\alpha$ with $\alpha\in\{0.5,1,2\}$,
\emph{i.e.}\ the LHA, jet width, and jet thrust. We present in
Fig.~\ref{fig:dist_avr_enriched} our \NLOpNLLpNP and \SHMCatNLO
predictions for the mean values $\langle\lambda^1_\alpha\rangle$ of
the five configurations, separately for the gluon- (upper row) and
quark enriched (bottom row) samples and compare them with the
corresponding experimental results. Both theoretical approaches yield
results that are in fair agreement with the data. The deviations of
the central values from the measured ones are in fact rather similar
for all three observables, and for all the configurations under consideration.
Also, we do not observe a qualitative difference in the description of
the $g$- or $q$-enriched samples. Note that, both in the data and in the
predictions, we observe smaller jet angularity mean values for the
$q$-enriched samples compared to the $g$-enriched case. This is
theoretically anticipated given that gluons carry more colour charge
and accordingly radiate more. For the high-$p_T$ configuration (2)
this effect is, as expected, rather marginal,
since the fractions of gluon jets are relatively similar in the central
and forward cases.
Our theoretical calculations nicely capture this effect even
quantitatively. As seen for the LHA already, the theoretical
uncertainties on the \NLOpNLLpNP predictions appear to be much larger
than the \SHMCatNLO ones, motivating to consider the evaluation of the
NNLL, and eventually NNLO,
corrections. From these considerations we can conclude that for 
the calculations considered here, \emph{i.e.}\ \NLOpNLLpNP and
\SHMCatNLO, QCD radiation off both hard quarks and hard gluons is
well-modelled. This is interesting as it was
noted before~\cite{Andersen:2016qtm,Gras:2017jty} that general-purpose MC
event generators do not always agree in their description of QCD
radiation off gluons, while they largely do for radiation off quarks,
heavily constrained by tunes on LEP data. However, it has to
be noted that, although the considered samples are certainly $q$- and
$g$-enriched, they still contain significant contributions from the
respective other flavour channel, \emph{cf.}\
Fig.~\ref{fig:gluonfractions}.

\begin{figure}
  \centering
   \includegraphics[width=0.49\textwidth]{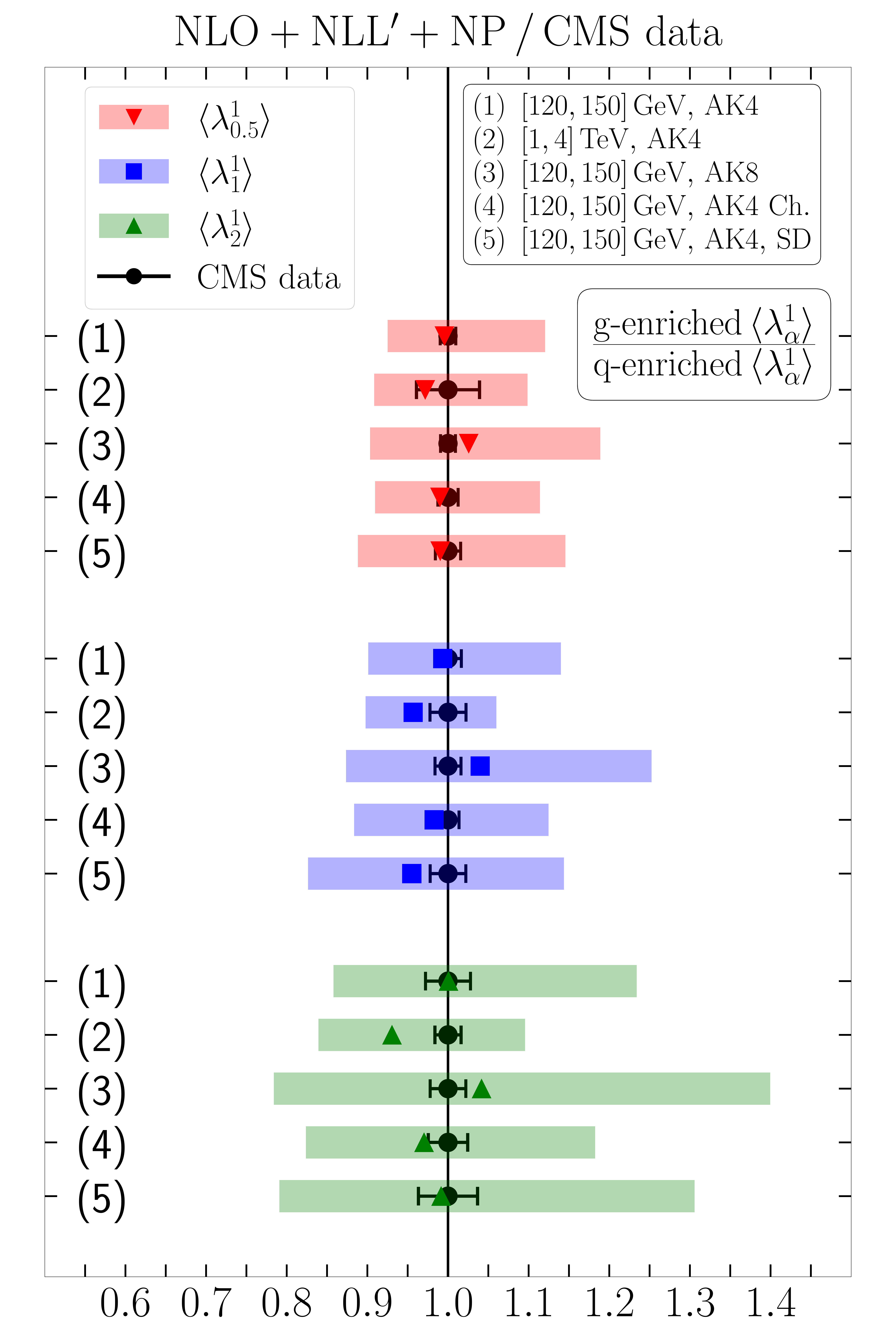}
   \includegraphics[width=0.49\textwidth]{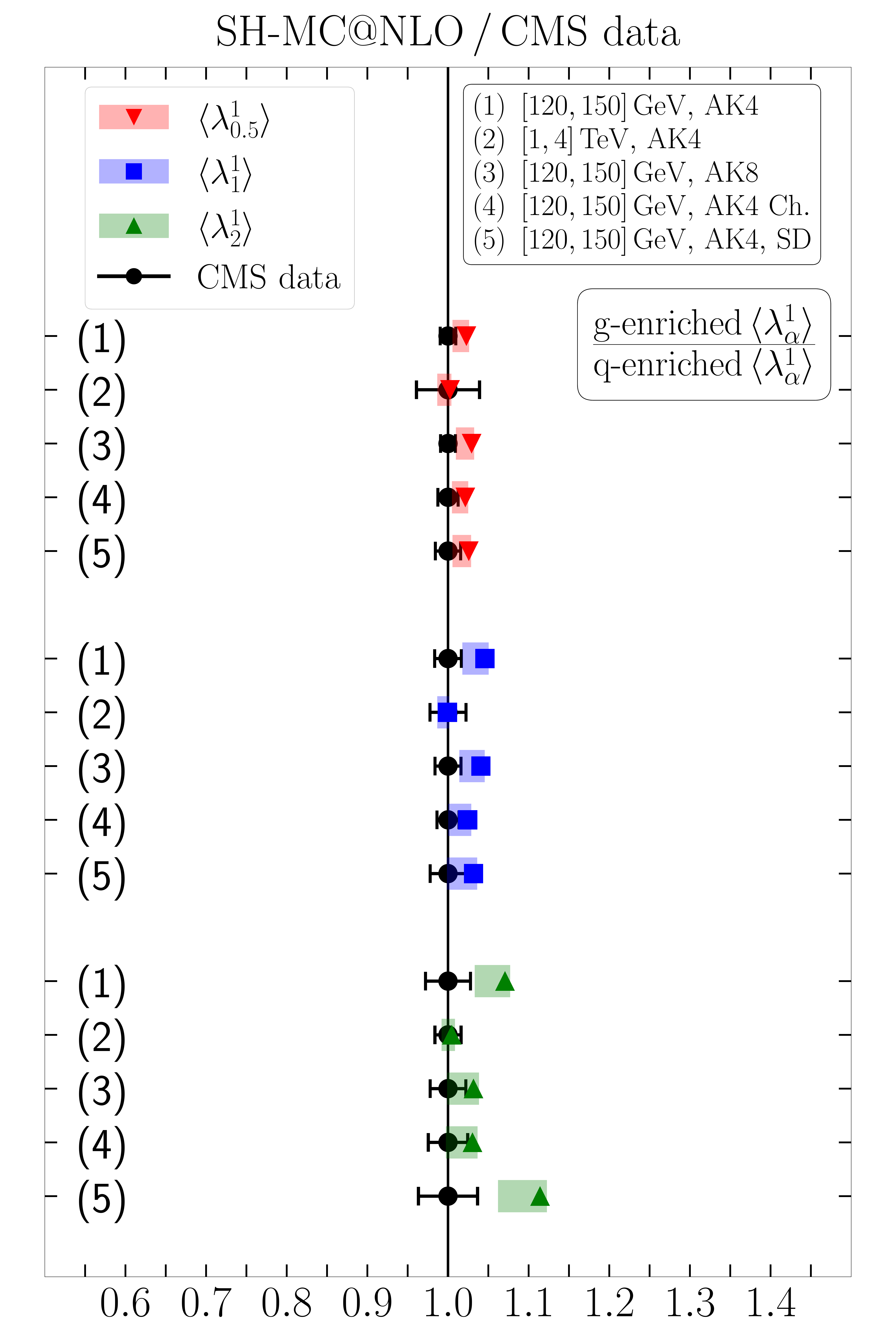}
   
   \caption{Ratio of jet-angularity mean values for the gluon- and quark-enriched
     samples in the five phase-space regions detailed in Tab.~\ref{tab:final}. The
     results obtained from the \NLOpNLLpNP calculation (left) and the \SHMCatNLO
     simulation (right) here get divided by the respective value reported by
     CMS~\cite{CMS:2021iwu}.}
  \label{fig:dist_avr_GvsQ}
\end{figure}

To further study the quality in the modelling of gluon and quark jets, we next consider
ratios of the angularity mean values in the $g$- and $q$-enriched samples for the five 
phase-space selections. In Fig.~\ref{fig:dist_avr_GvsQ} we present our corresponding theoretical
predictions, normalised to the result reported by CMS. While the left-hand plot contains the
results for the \NLOpNLLpNP calculation, the right-hand plot shows corresponding predictions
obtained from \SHMCatNLO. For both calculational methods the central values agree remarkably
well with the measured ones \emph{i.e.}\ have a ratio to data centred close to unity.
For the \SHMCatNLO simulation we observe the largest deviation, about $10\%$, for the jet-thrust
angularity for the case of \softdrop-groomed jets of moderate-$p_T$, see also Figs.~\ref{fig:dist_thrust}
and \ref{fig:mean_thrust} in App.~\ref{app:other-ang}. 

This good description reflects the fact that, despite of the notorious underestimation of the angularity mean
values seen in Fig.~\ref{fig:mean_LHA}, both our calculational approaches seem to treat quark and gluon
jets similarly well, and do not introduce an artificially-enhanced discriminative power
between both hypotheses. This is to some degree in contrast to the generator predictions
studied by CMS in~\cite{CMS:2021iwu}. There, in particular for the low-$p_{T,\text{jet}}$ regions,
predictions based on the Born matrix elements for $Z$+jet and dijet production overestimated
the difference between gluon and quark jets by up to $20-30\%$. This might to some extent
originate from the different gluon- and quark-jet decomposition they predict for the various phase-space
regions (see Fig.~\ref{fig:gluonfractions} and the discussion of flavour fractions at Born and
NLO QCD accuracy). We finally point out that, as noted before, the uncertainty estimate for the \NLOpNLLpNP predictions
is quite large, in particular for jet thrust, $\lambda^1_2$.

\section{Conclusions}\label{sec:conclusions}

Following a recent measurement accomplished by the CMS collaboration~\cite{CMS:2021iwu},
we have performed a detailed phenomenological analysis of IRC safe jet angularities,
considering both ungroomed and \softdrop jets.  
Our theoretical predictions account for all-order effects at next-to-leading logarithmic accuracy,
including non-global logarithms, and matching to next-to-leading order QCD calculations. Thanks to
a flavour-sensitive matching procedure, based on the \BSZ flavour-$k_t$ algorithm, we are able to
reach \NLOpNLLp accuracy.  Furthermore, our predictions are supplemented by non-perturbative corrections,
that account for hadronisation effects and the Underlying Event.

With respect to our previous work, see \emph{e.g.}~\cite{Caletti:2021oor}, our theoretical predictions
feature two improvements. First of all, on the perturbative side, we now include resummed and matched calculations
for dijet events, in addition to the ones for $Z$+jet production recently presented in~\cite{Caletti:2021oor}
and confronted with experimental data in~\cite{CMS:2021iwu} already. 
The resummation for dijet final states features more complicated colour structures, for both the global
and the non-global part, which is dealt with by our largely-automated implementation of colour matrices
in the \sherpa resummation framework~\cite{Gerwick:2014gya,Baberuxki:2019ifp}.

Secondly, we employ a more sophisticated approach for dealing with non-perturbative corrections
that we extract from fully exclusive hadron-level simulations at \MCatNLO accuracy with the general-purpose
event generator \sherpa.  Rather than relying on a simple bin-by-bin
rescaling, we have developed and implemented a
rather general transfer-matrix approach that can account for migrations across different kinematical
variable and observable bins. In the present case, we account for
migrations both in the jet transverse momentum
and in the angularity observable when going from parton- to hadron level. We thereby largely reduce the
sensitivity to differences in the perturbative predictions from the analytic resummation and the
employed MC simulations. This cures deficiencies of our previous approach, that showed pathological features
for observable values corresponding to scales below the parton-shower cutoff. 

We compare our final \NLOpNLLpNP results to a plethora of distributions measured by CMS, both in
dijet and $Z$+jet production. We thereby consider differential cross sections for the LHA, jet
width and thrust, as well as their respective mean values measured as
a function of the jet
transverse momentum. For comparison, we also include results from \sherpa at \MCatNLO accuracy.
Overall, we find a fair agreement between the data and the theoretical
predictions.
Our results however tend to systematically underestimate the mean
values measured by CMS. With respect to the comparisons presented in
Ref.~\cite{CMS:2021iwu} for the $Z$+jet selection, we clearly achieve an
improved description of the data, thanks to the new
transfer-matrix method used to include non-perturbative corrections.
Finally, still following the CMS analysis, we compare our theoretical predictions to the
measurements performed on quark- and gluon-enriched samples. We find that both the resummed
calculation and the \SHMCatNLO prediction are able to describe the data well.
This is interesting because jet angularities have been proposed as IRC-safe quark/gluon
taggers~\cite{Amoroso:2020lgh,Caletti:2021ysv} that can be applied not only in the context
of searches for new physics but also to Standard Model measurements that aim for extractions of
fundamental parameters, such as the strong coupling, or parton
distribution functions.
The better description of the data that we find originates to some
degree from a more accurate modelling of the flavour fractions,
\emph{i.e.}\ the relative contributions of partonic channels, that
receive sizeable NLO QCD corrections, included in both our theoretical
predictions.

Throughout this paper, we have noted that our resummed calculations
suffer from rather large theoretical uncertainties, dominated by
variations of the resummation scale. An obvious way to systematically improve on this is to promote
the resummed calculation to NNLL accuracy.
Many of the contributions that are relevant for NNLL resummation of jet shapes have already been
computed~\cite{Ellis:2010rwa}, predominately using SCET techniques, and have reached a considerable
degree of automation, see \emph{e.g.}~\cite{Bell:2020yzz,Bell:2018oqa,Bell:2021dpb}. Furthermore,
first calculations for groomed observables exist even at N$^3$LL~\cite{Kardos:2020gty} and
with an improved description of the transition between groomed and ungroomed regimes~\cite{Benkendorfer:2021unv}. 
The resummation framework we are employing in our plugin also has already been extended,
for global observables, to NNLL~\cite{Banfi:2014sua}.
The bottleneck in this enterprise is the inclusion of subleading non-global logarithms. The first resummation of these effects has been
achieved very recently~\cite{Banfi:2021owj,Banfi:2021xzn} and we look forward to investigating whether this method
can be easily interfaced with our framework, in order to perform higher-precision phenomenology of
jet angularities. 

Another obvious improvement of our calculation would be to upgrade the
fixed-order component of the angularity
distributions to NNLO accuracy.
This requires computing the hadronic production of three QCD partons,
respectively $Z$+2 partons, at two-loop accuracy. In this context, we
note that the first NNLO evaluation of three-jet observables has
recently been presented~\cite{Czakon:2021mjy}.
Finally, the two-loop corrections to the dijet and $Z$+jet processes
(see~\cite{Currie:2016bfm,Gehrmann-DeRidder:2015wbt} and references therein) are needed to achieve
NNLL$'$ accuracy.

In the context of assessing the impact of NP corrections, it would be interesting
to compare the transfer-matrix approach developed here to results obtained using first-principle
field-theoretical arguments (see for instance Refs.~\cite{Hoang:2019ceu,Pathak:2020iue} for
recent work on hadronisation corrections for \softdrop jets).
In a more generic context, it would be interesting to see the impact
that the measurement of jet angularities has on tuning NP parameters
of general-purpose Monte Carlo generators.
One could then also provide independent tunes for different
parton-shower cutoffs.
This could be, in turn, added to our transfer-matrix approach of NP
corrections to parton-level analytic calculations.
We close by noting that it would clearly be interesting to apply our calculational methods to
jets arising from different production modes, \emph{e.g.}\ involving top-quarks~\cite{CMS:2018ypj,ATLAS:2019kwg},
or at different centre-of-mass energies~\cite{ALICE:2021njq}. We leave this for future work.

\clearpage
\section*{Acknowledgements}

We would like to thank Andreas Hinzmann for fruitful discussions and comments on the manuscript.
We are grateful to Vincent Theeuwes for collaboration at an early stage of the project. 

This work is supported by funding from the European Union's Horizon 2020 research and innovation programme as part of the Marie Sk\l{}odowska-Curie Innovative Training Network MCnetITN3 (grant agreement no.~722104).
SS and DR acknowledge support from BMBF (contracts 05H18MGCA1 and 05H21MGCAB). SS acknowledges funding
by the Deutsche Forschungsgemeinschaft (DFG, German Research Foundation) - project number 456104544.
The work of SC, OF and SM is supported by Universit\`a di Genova under the curiosity-driven grant ``Using jets to challenge the Standard Model of particle physics'' and by the Italian Ministry of Research (MUR) under grant PRIN 20172LNEEZ.
We also acknowledge support from the Royal Society through the project 580986 ``Resum(e) The Path To Discovery".
SC would like to thank the Institute for Particle Physics Phenomenology (Durham University) for hospitality during the course of this work. 


\FloatBarrier

\appendix

\section{Results for jet width and jet thrust} \label{app:other-ang}
\begin{figure}
  \centering
  \includegraphics[width=0.32\textwidth]{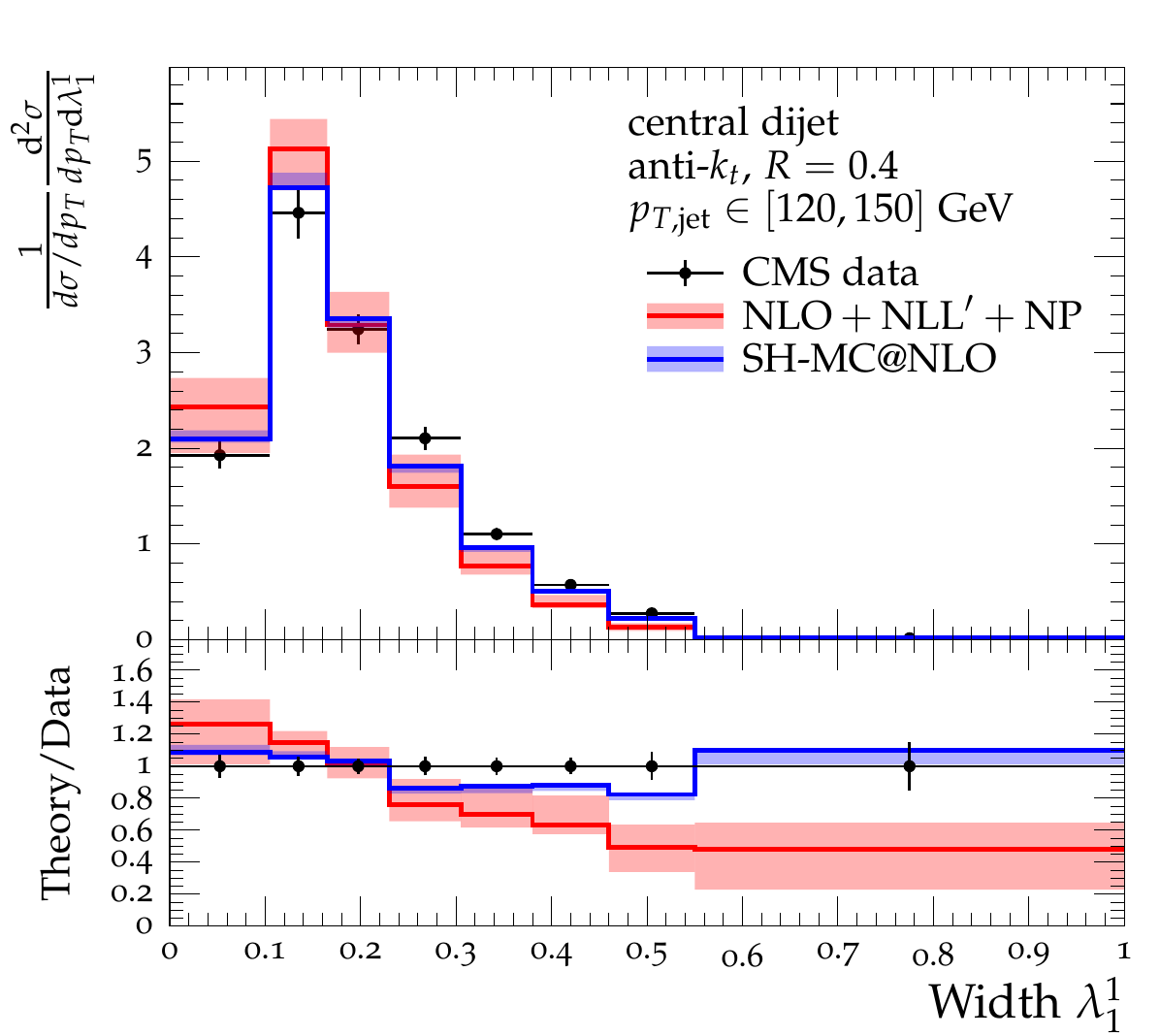}
  \includegraphics[width=0.32\textwidth]{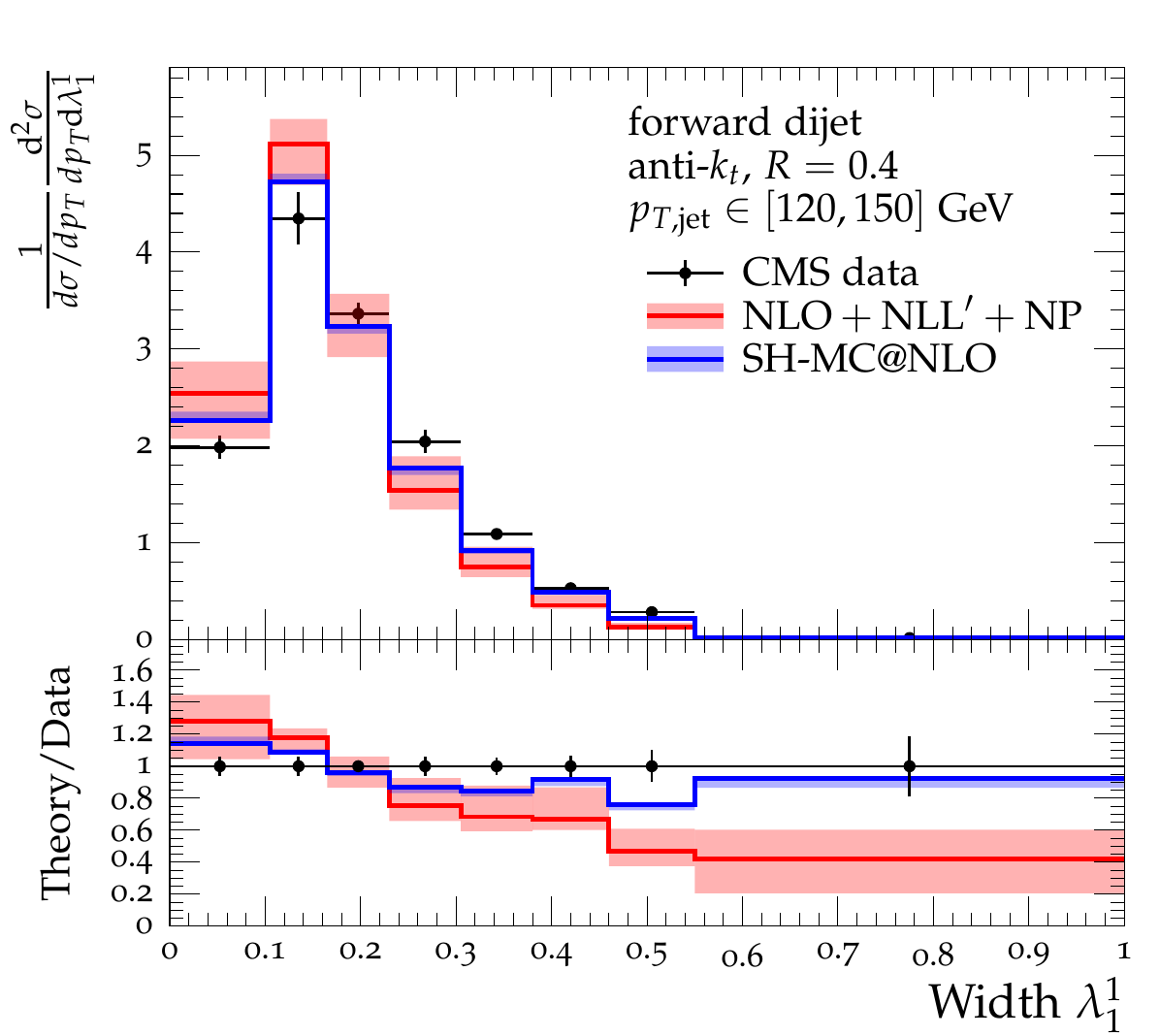}
  \includegraphics[width=0.32\textwidth]{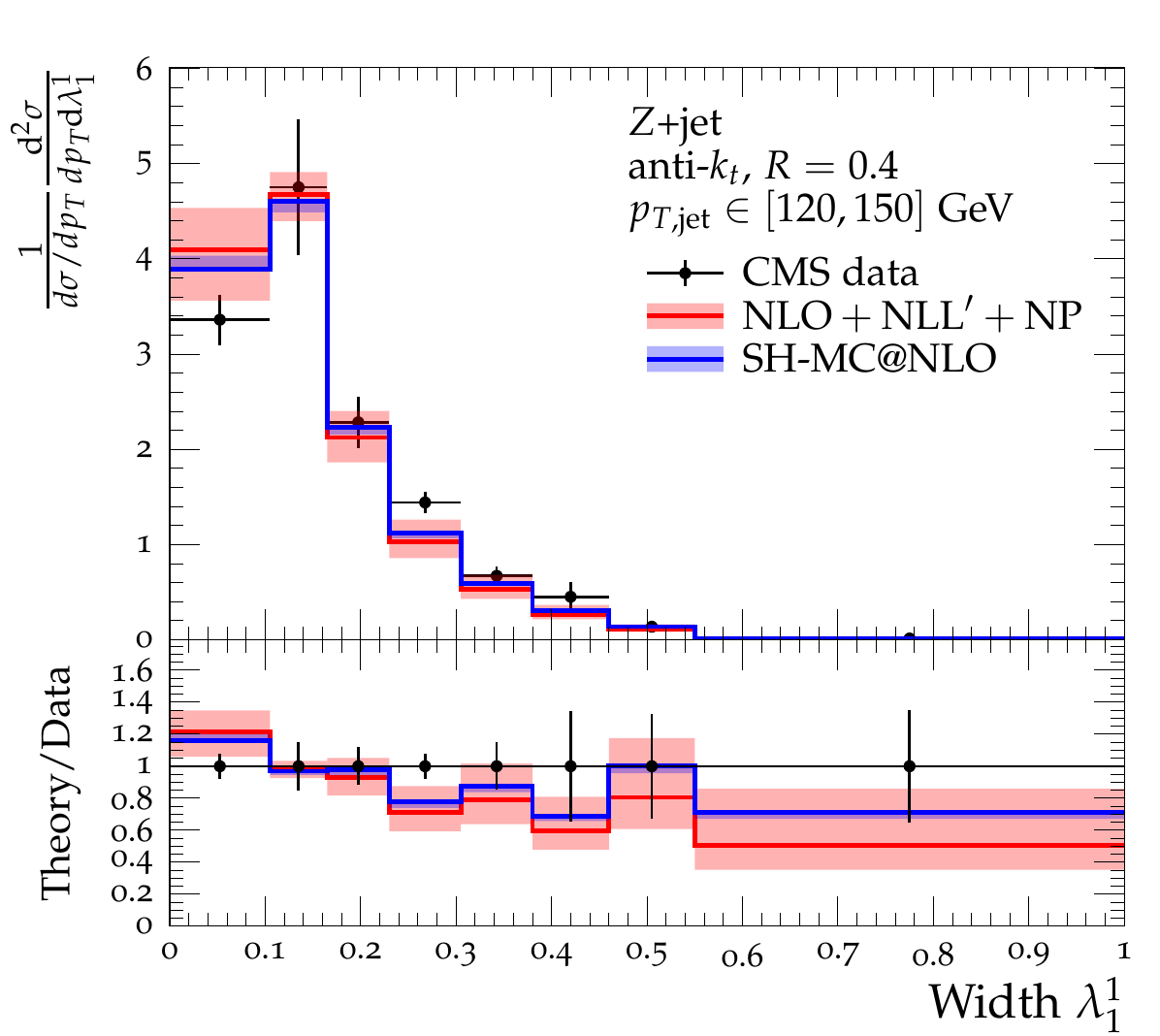}
  \includegraphics[width=0.32\textwidth]{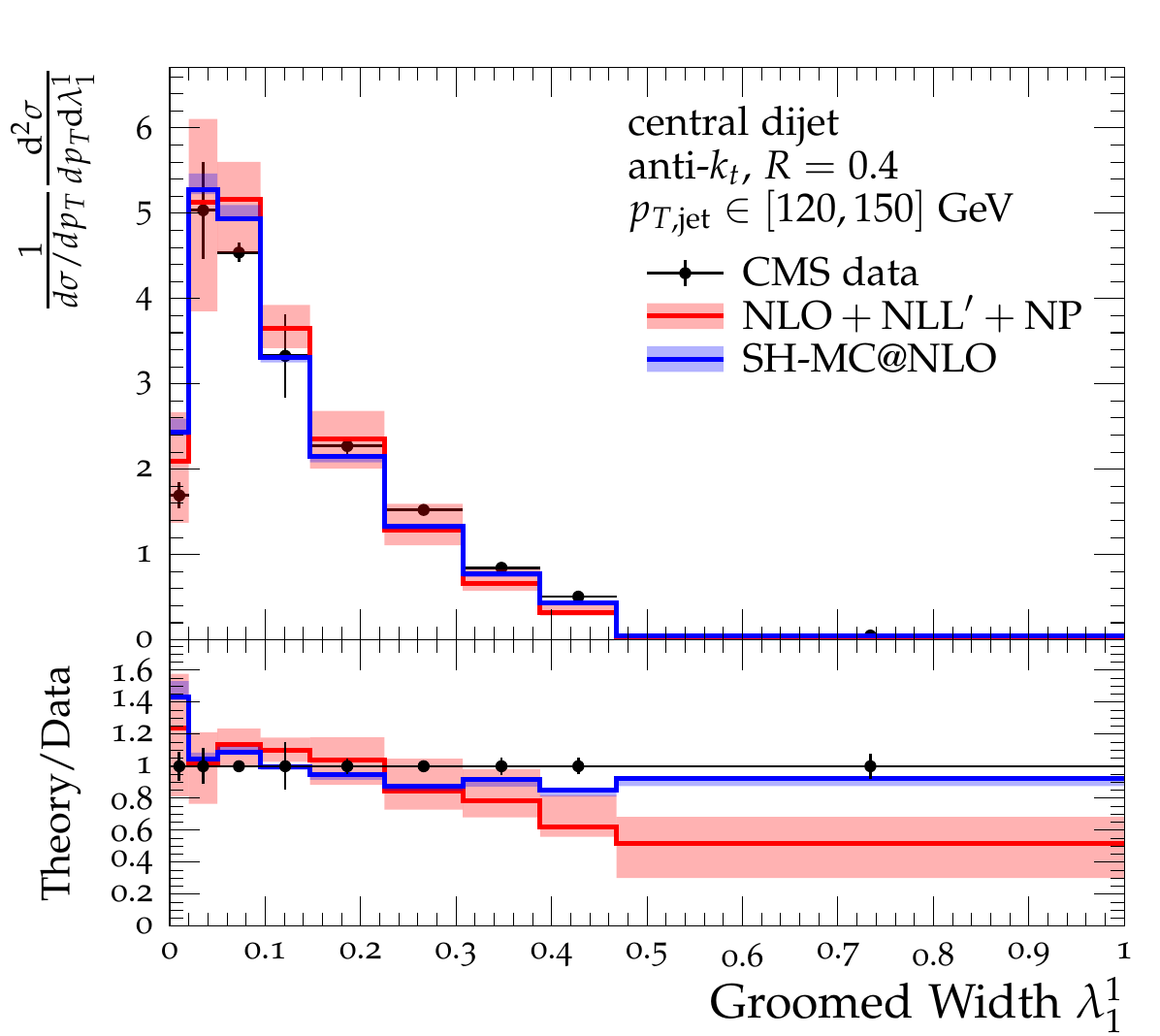}
  \includegraphics[width=0.32\textwidth]{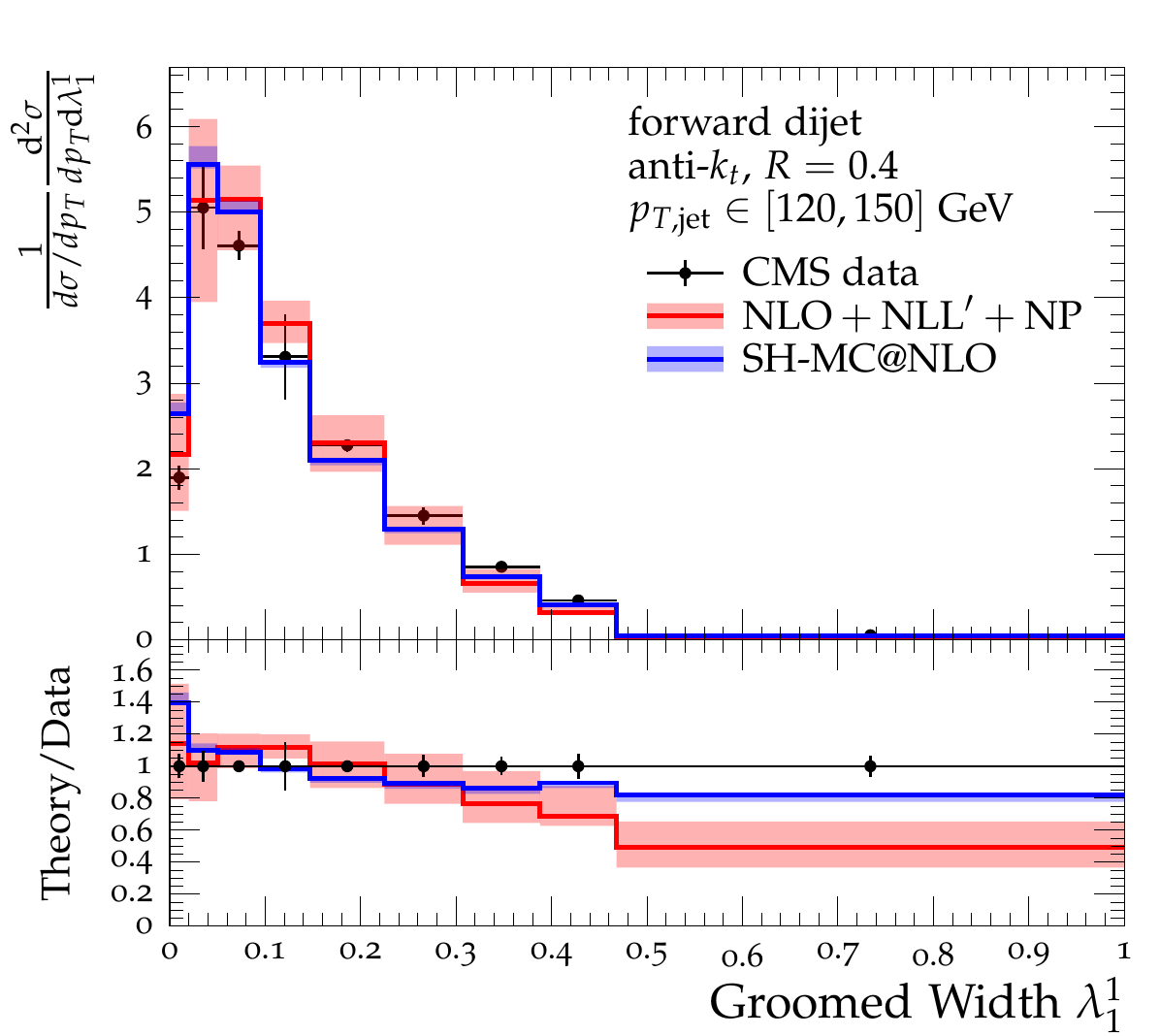}
  \includegraphics[width=0.32\textwidth]{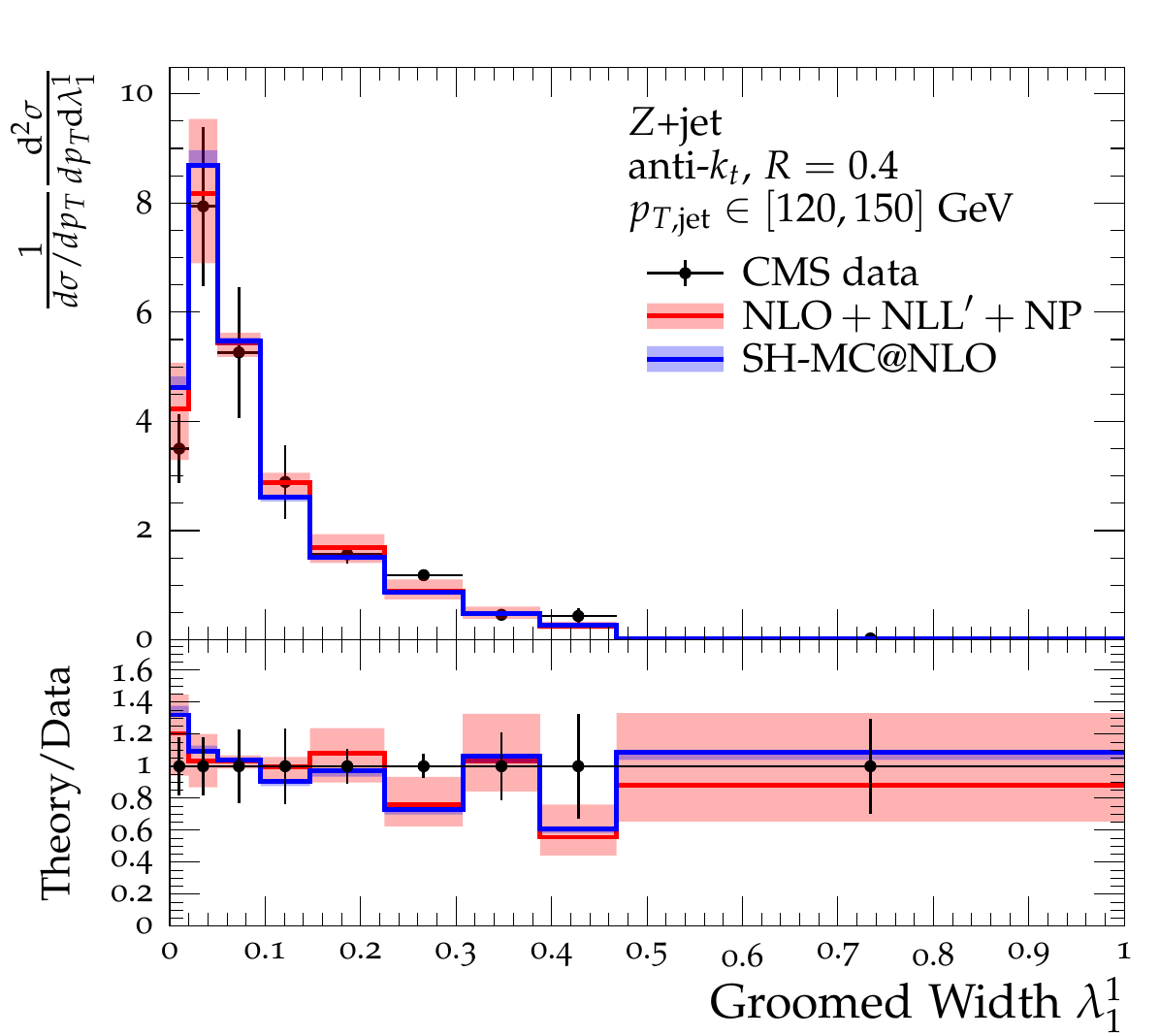}
  \caption{Results for the jet width $\lambda^1_{1}$ for ungroomed (top row)
    and groomed (bottom row) $R=0.4$ anti-$k_t$ jets with $p_{T,\text{jet}}\in[120,150]~\text{GeV}$.
    The left and middle panel correspond to the central and forward jet in dijet
    events, respectively, the right one to the leading jet in $Z$+jet production.}\label{fig:dist_width}
\end{figure}

\begin{figure}
  \centering
  \includegraphics[width=0.32\textwidth]{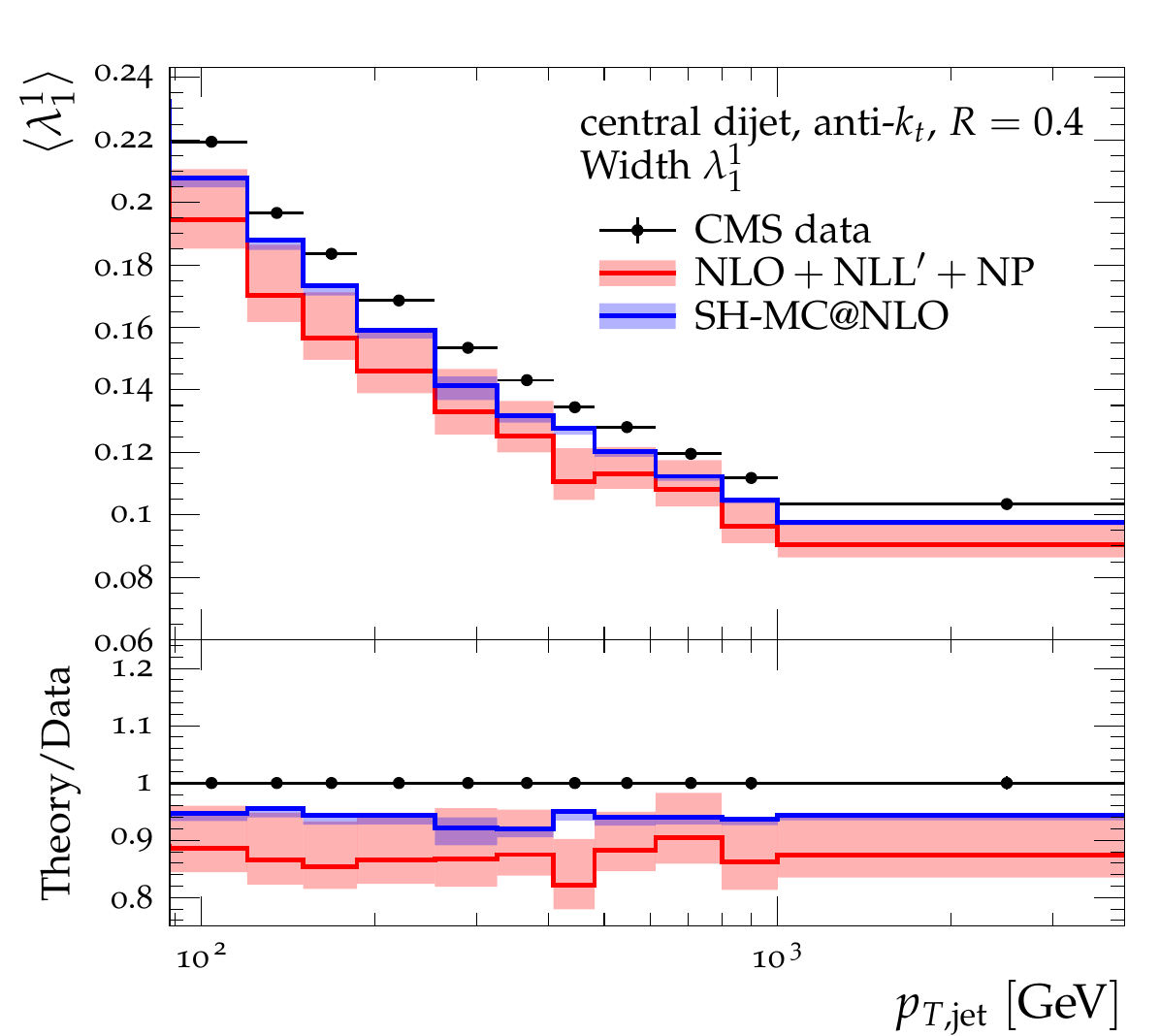}
  \includegraphics[width=0.32\textwidth]{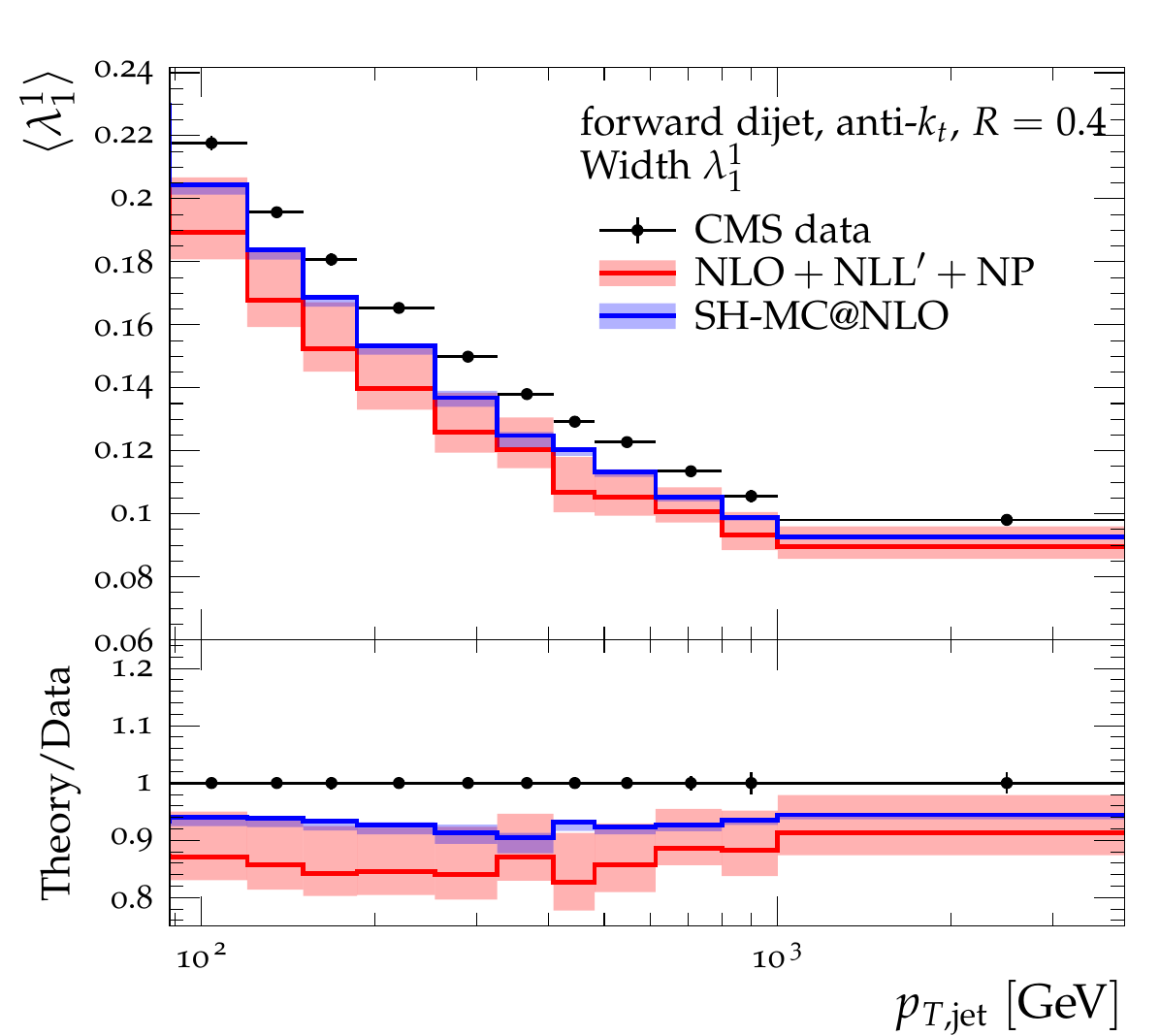}
  \includegraphics[width=0.32\textwidth]{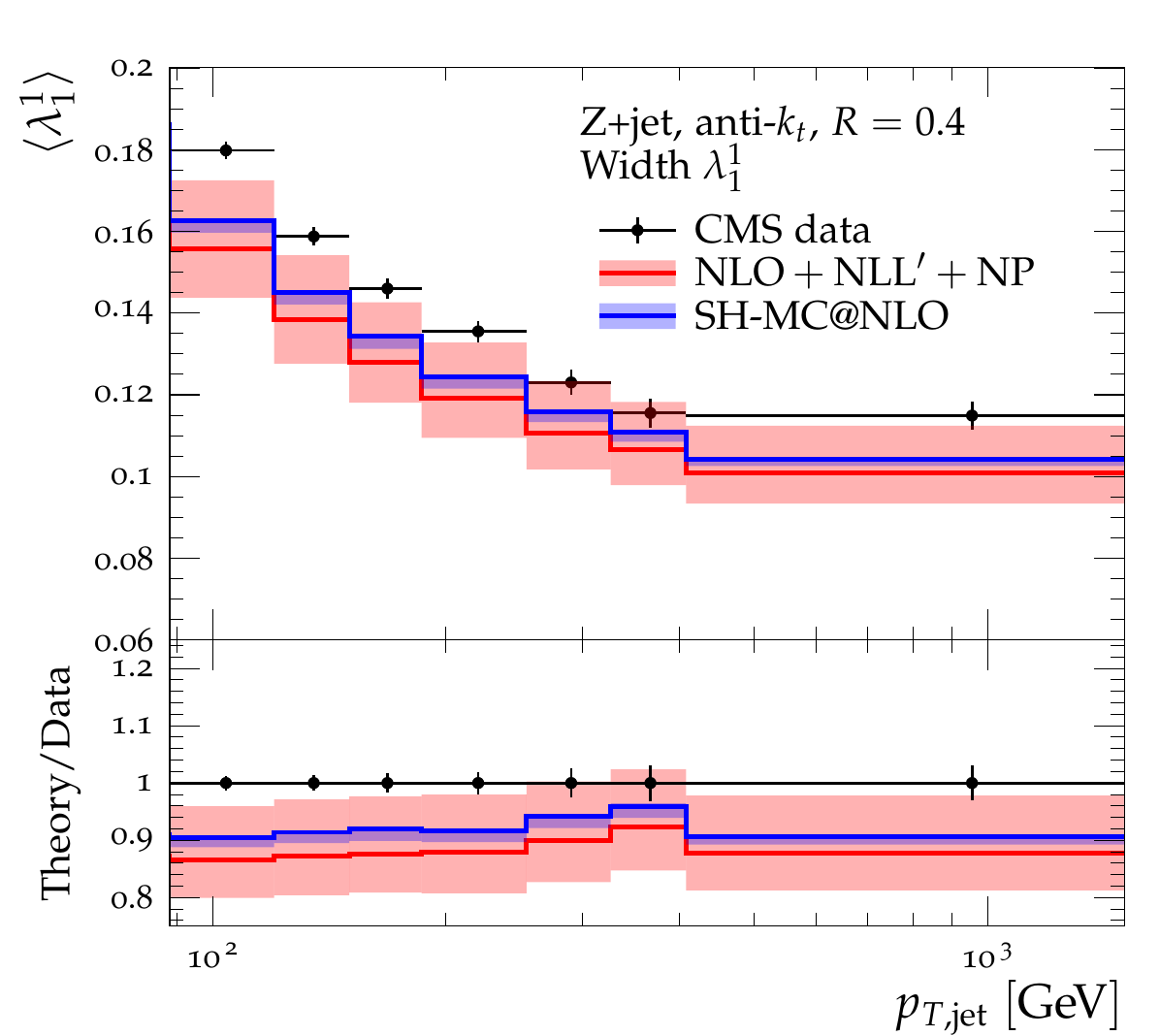}
  \includegraphics[width=0.32\textwidth]{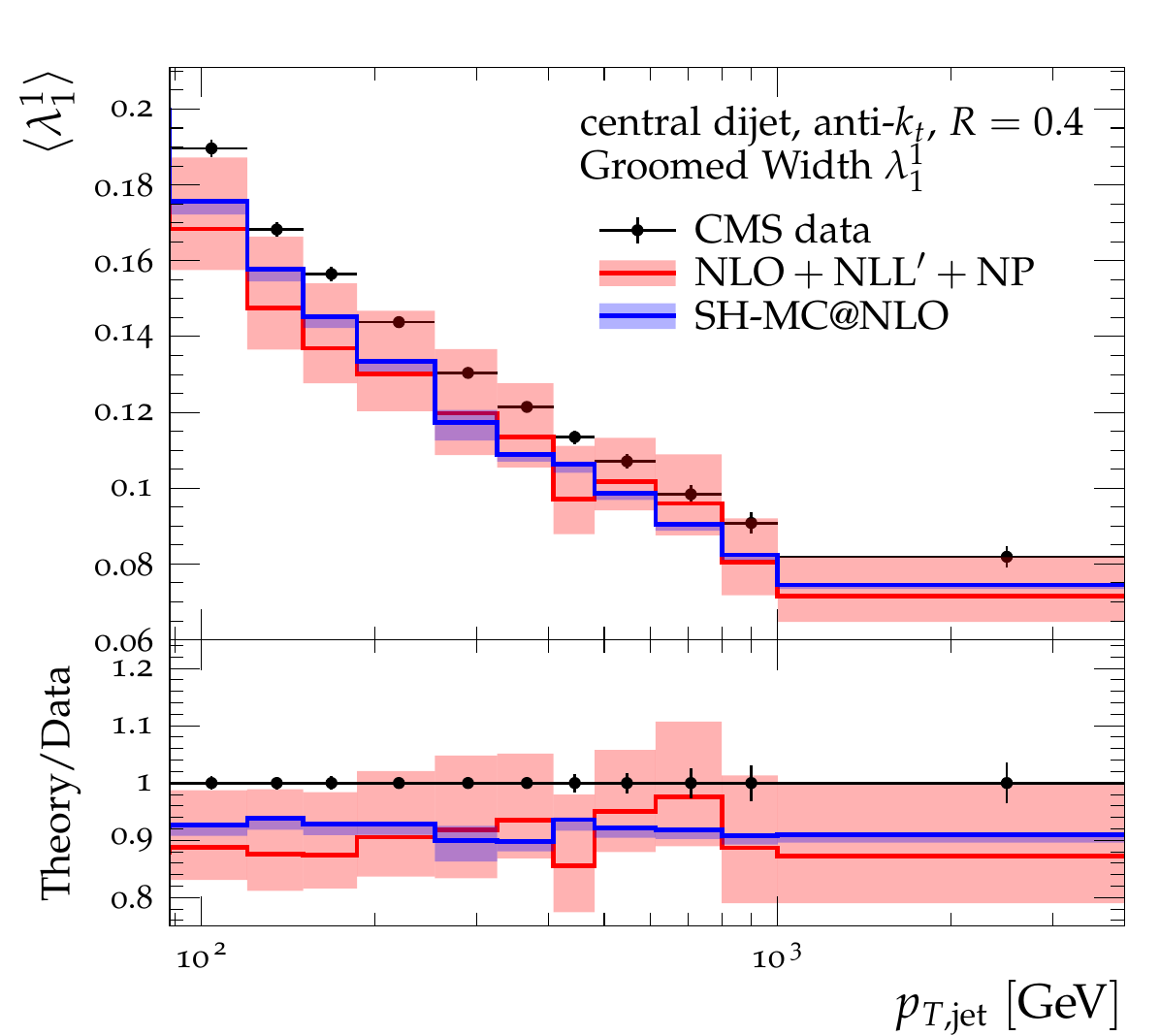}
  \includegraphics[width=0.32\textwidth]{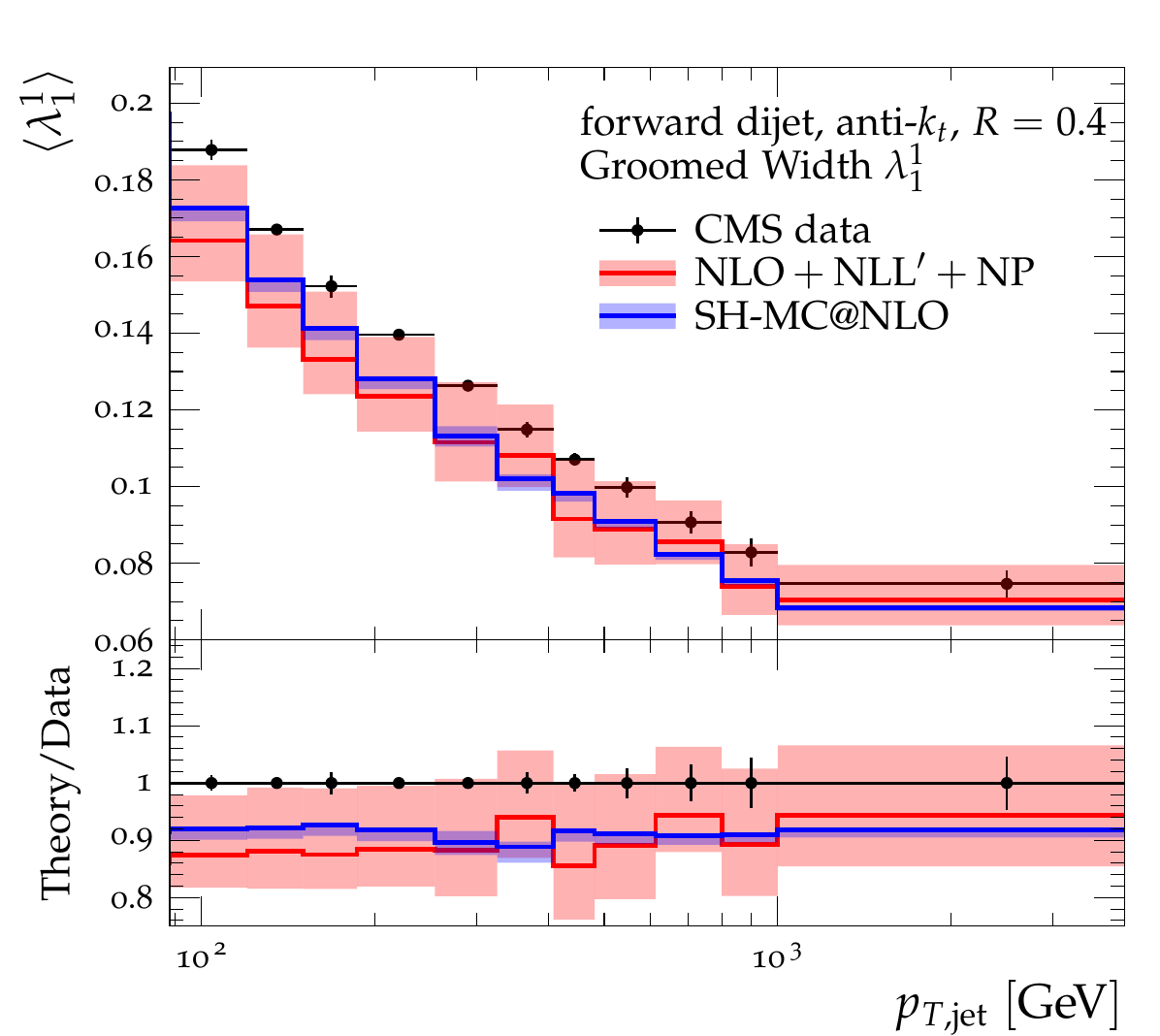}
  \includegraphics[width=0.32\textwidth]{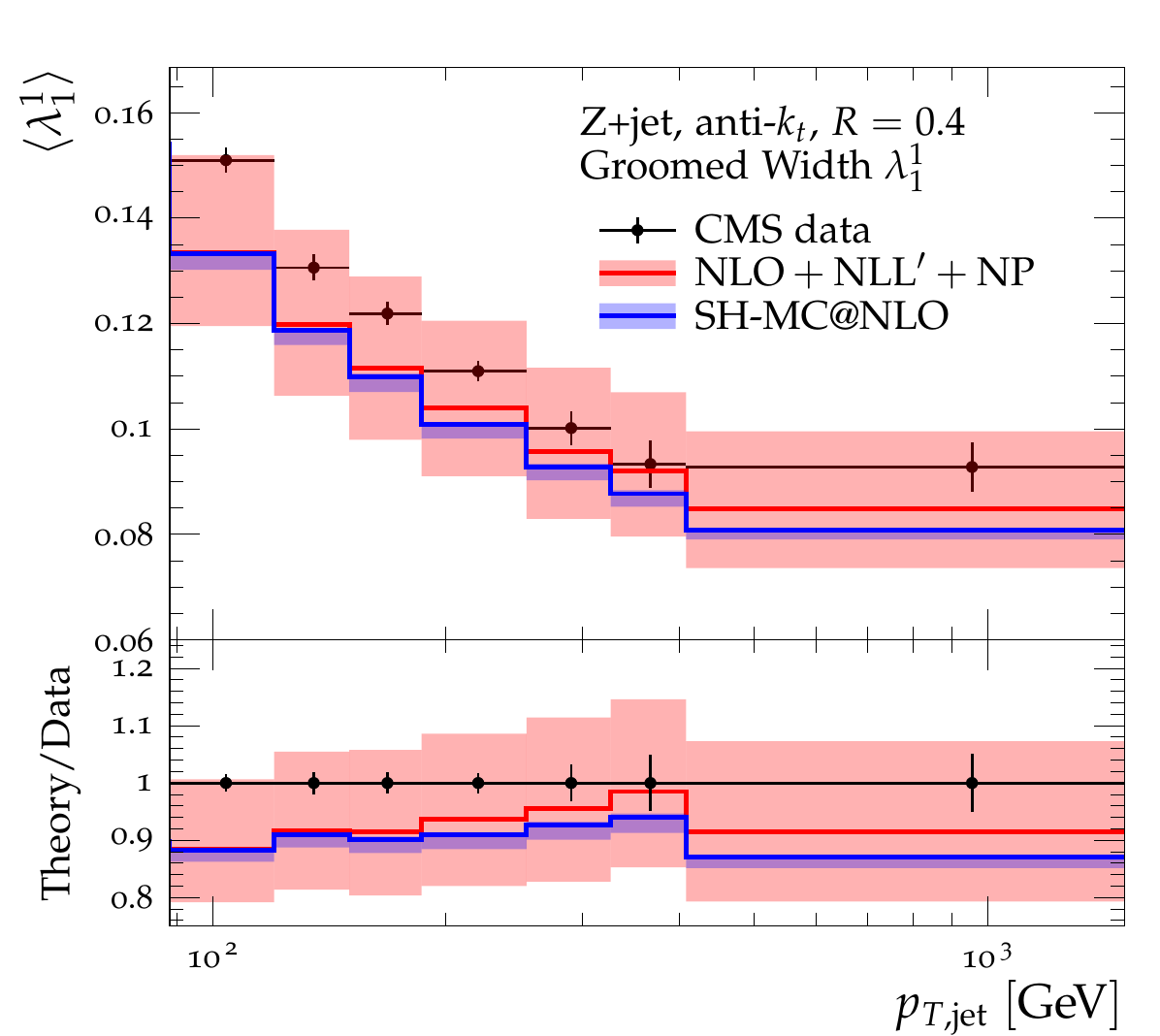}
  \caption{Mean values of jet width, \emph{i.e.}\ $\langle\lambda^1_{1}\rangle$, for ungroomed (top row)
    and groomed (bottom row) $R=0.4$ anti-$k_t$ jets in dependence of $p_{T,\text{jet}}$.
    The left and middle panel correspond to the central and forward jet in dijet
    events, respectively, the right one to the leading jet in $Z$+jet production.}\label{fig:mean_width}
\end{figure}

\begin{figure}
  \centering
  \includegraphics[width=0.32\textwidth]{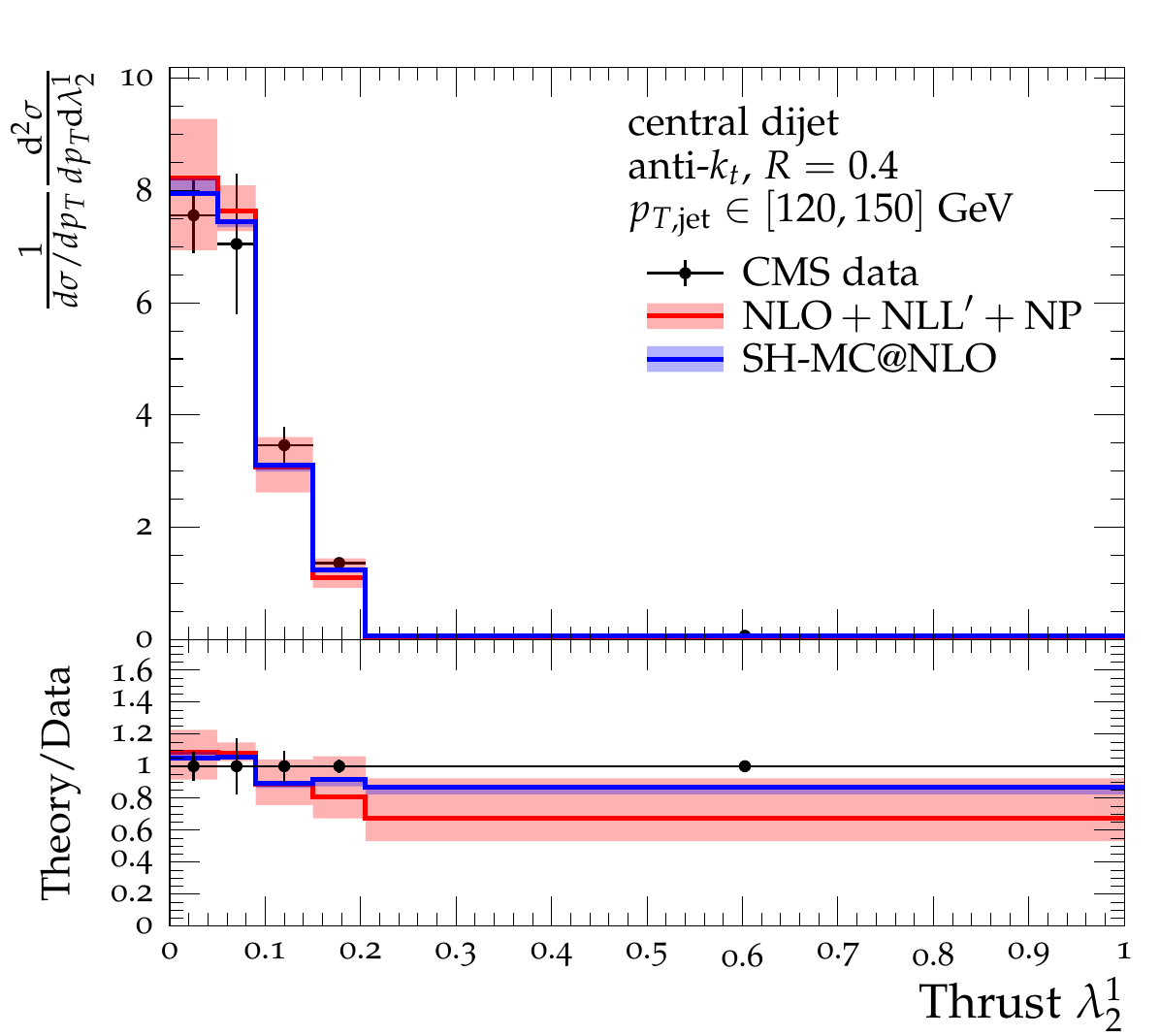}
  \includegraphics[width=0.32\textwidth]{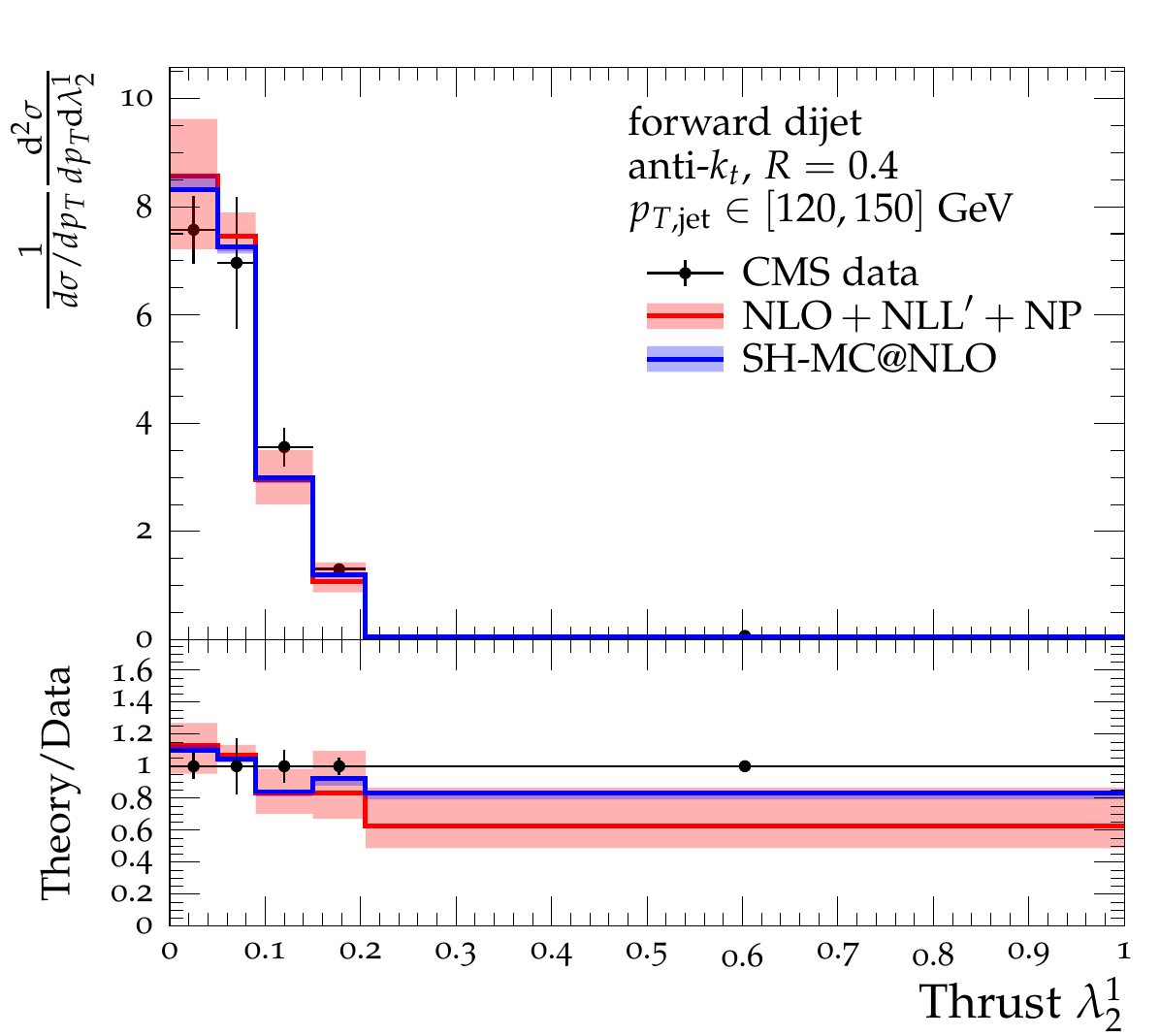}
  \includegraphics[width=0.32\textwidth]{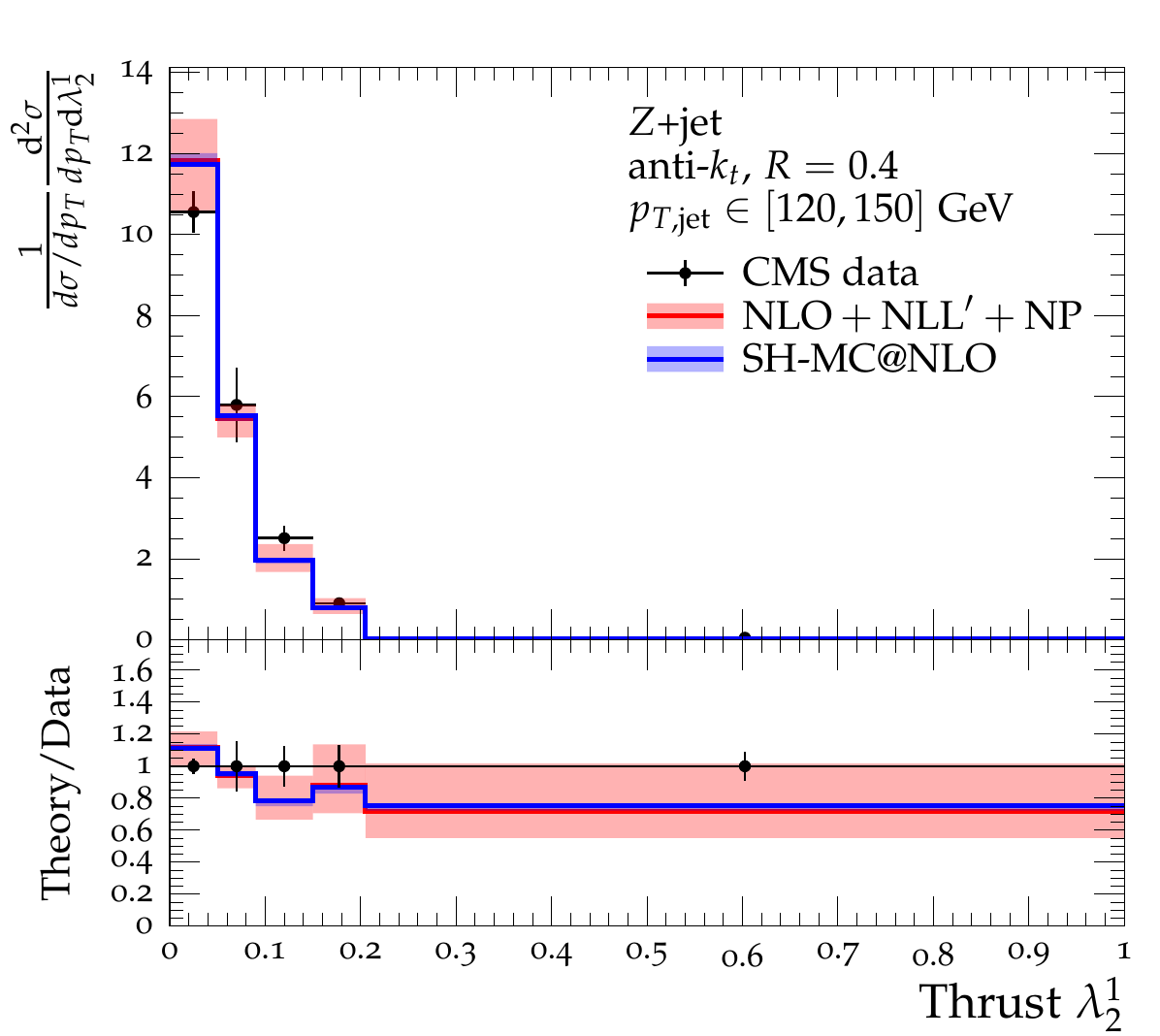}
  \includegraphics[width=0.32\textwidth]{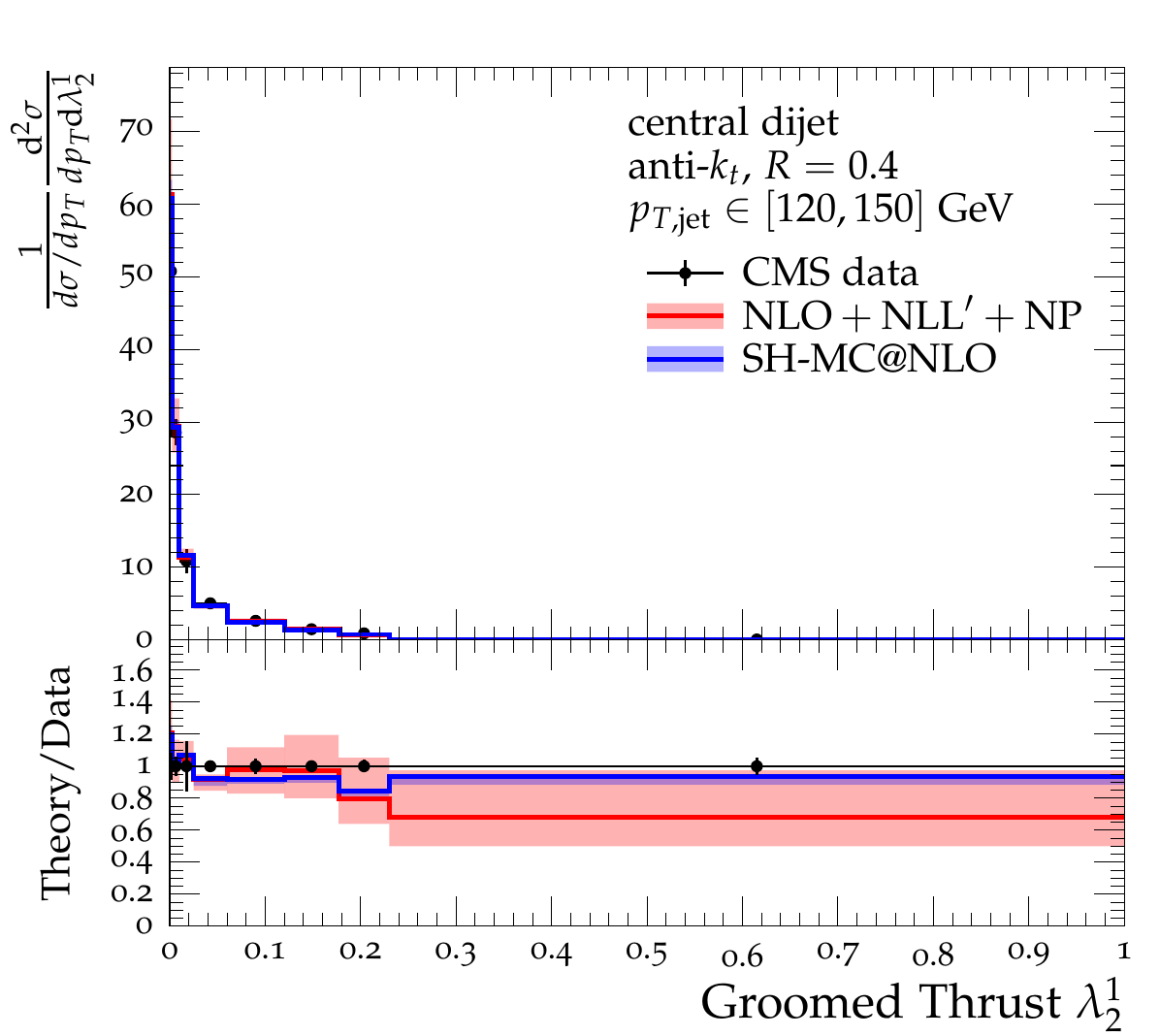}
  \includegraphics[width=0.32\textwidth]{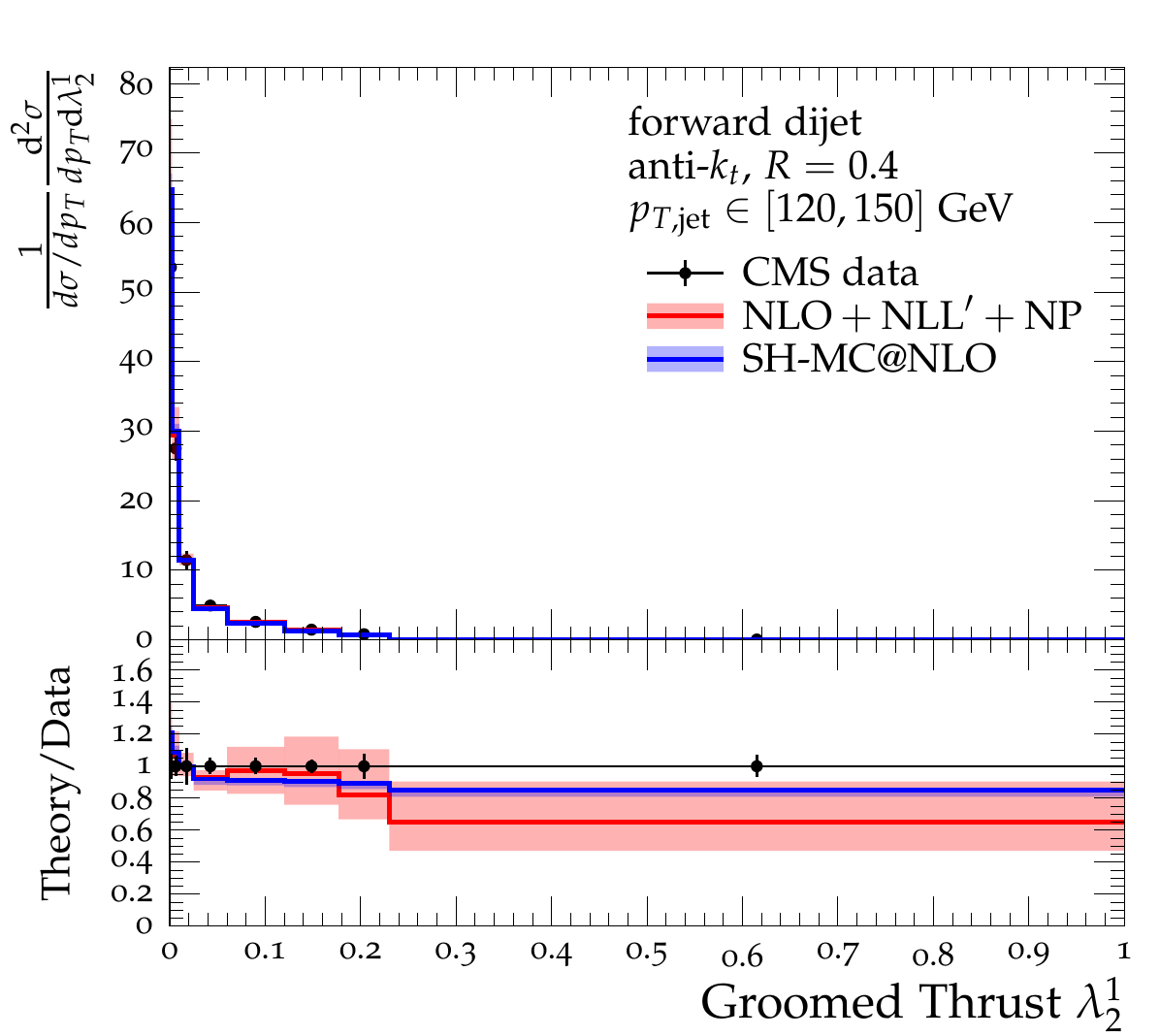}
  \includegraphics[width=0.32\textwidth]{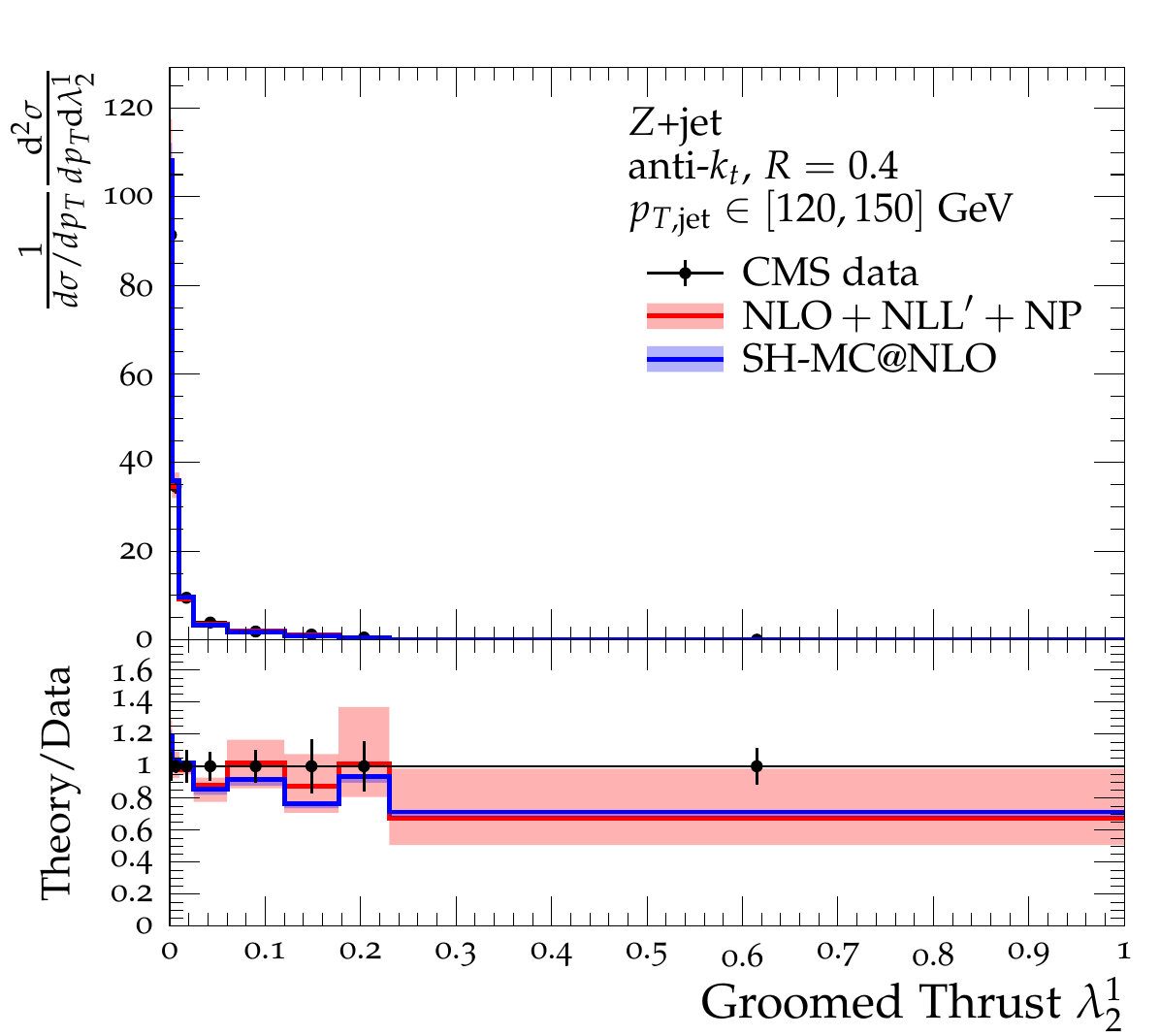}
  \caption{Same as Fig.~\ref{fig:dist_width} but for jet thrust $\lambda^1_2$.}\label{fig:dist_thrust}
\end{figure}

\begin{figure}
  \centering
  \includegraphics[width=0.32\textwidth]{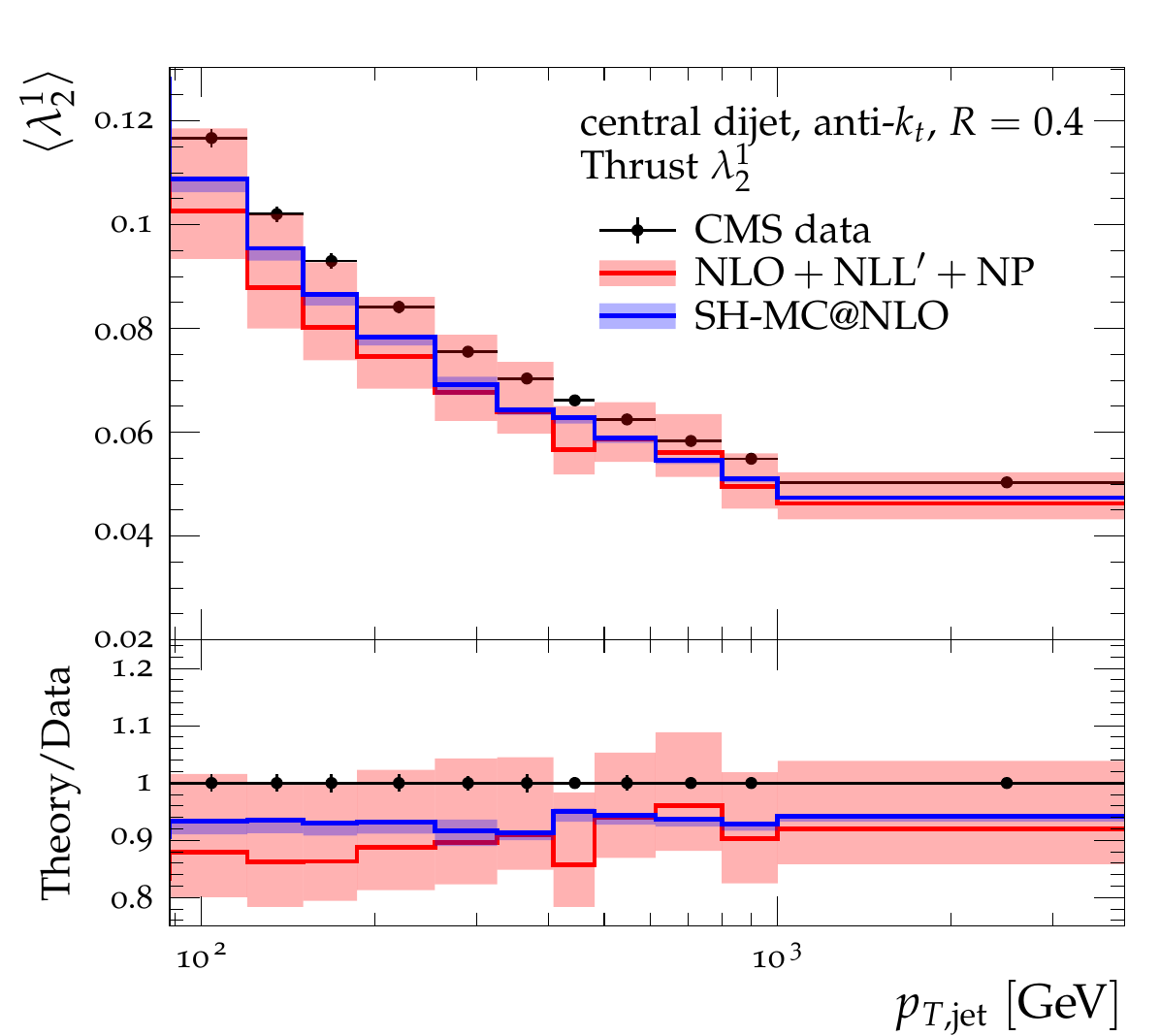}
  \includegraphics[width=0.32\textwidth]{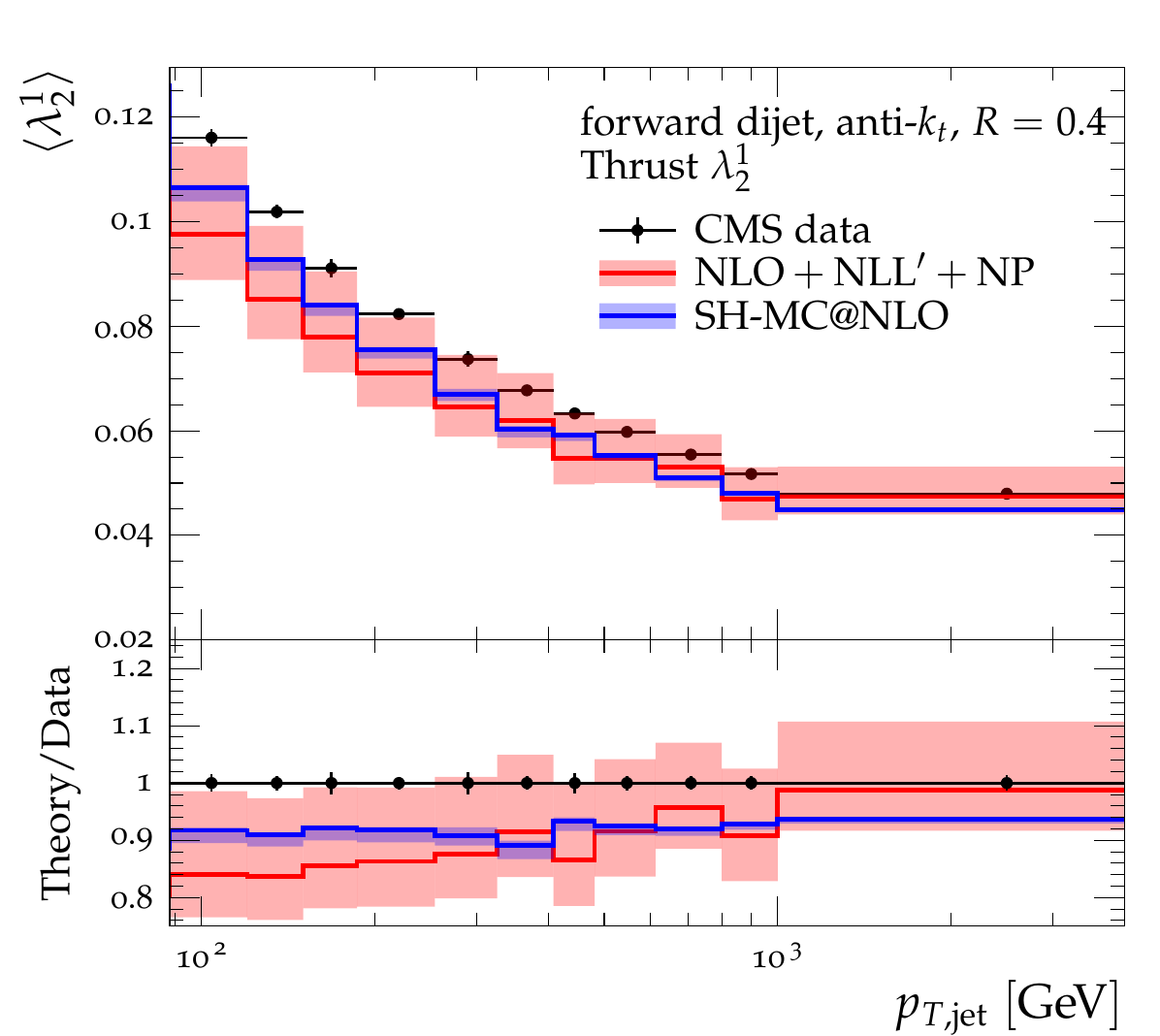}
  \includegraphics[width=0.32\textwidth]{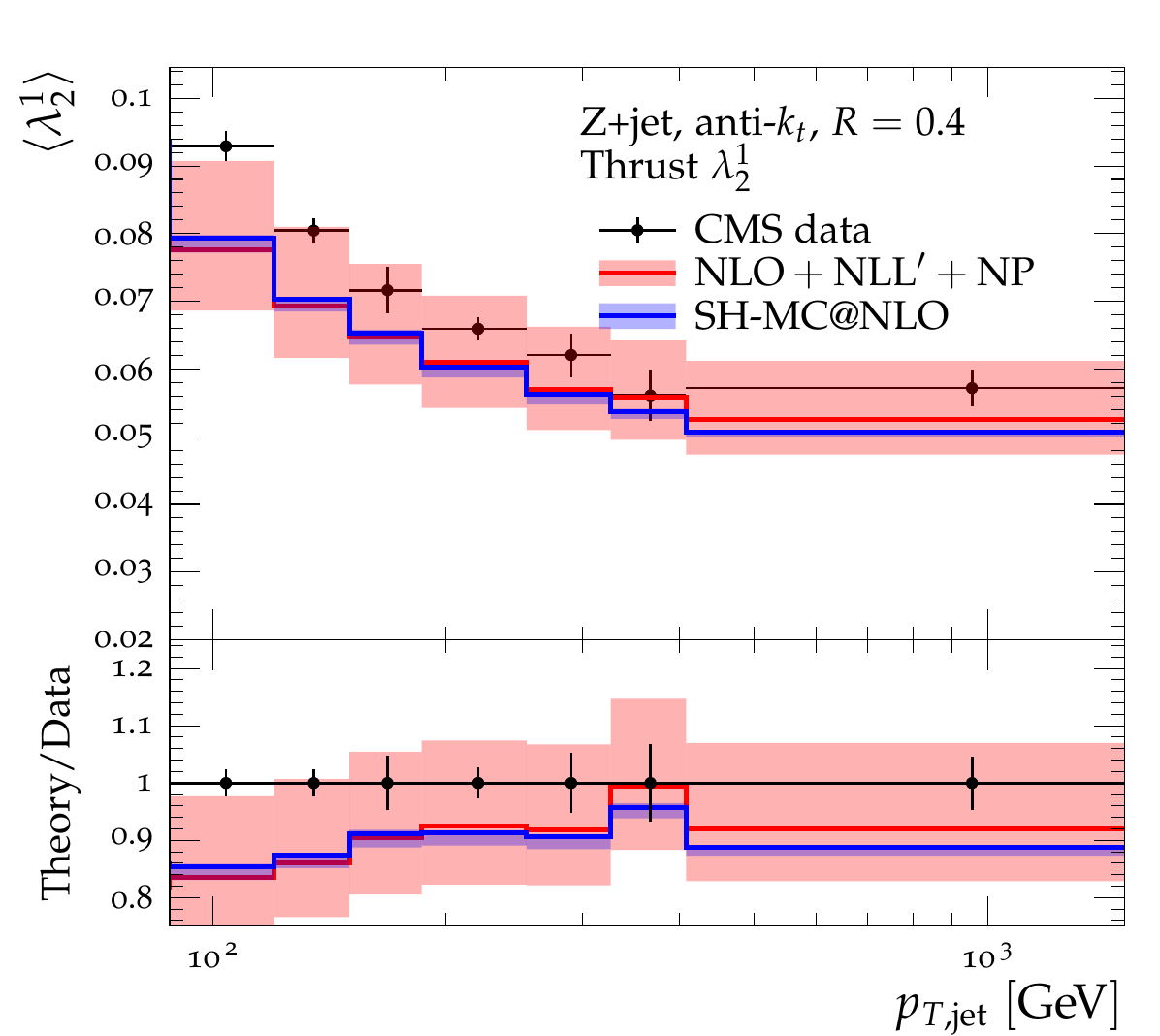}
  \includegraphics[width=0.32\textwidth]{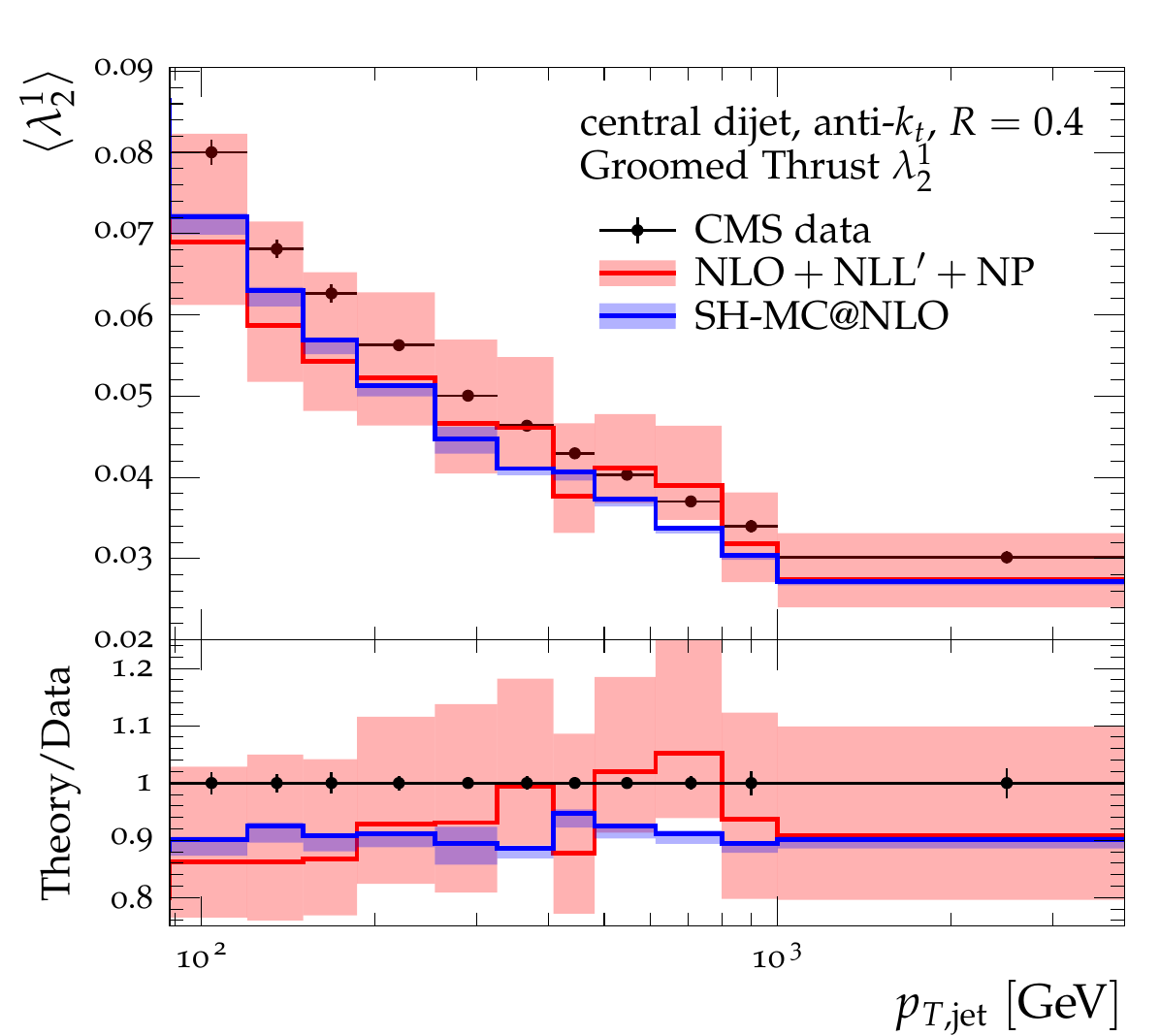}
  \includegraphics[width=0.32\textwidth]{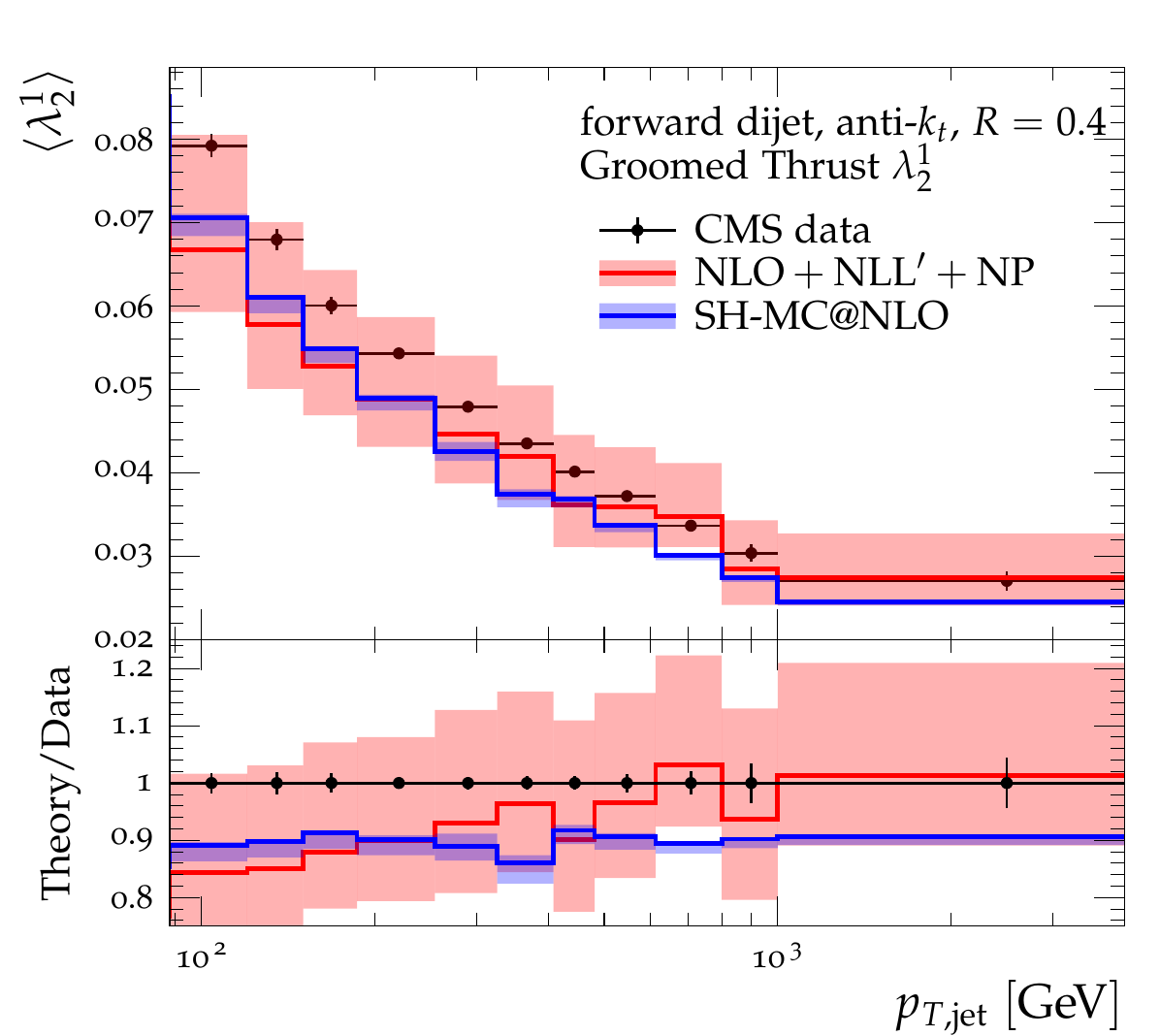}
  \includegraphics[width=0.32\textwidth]{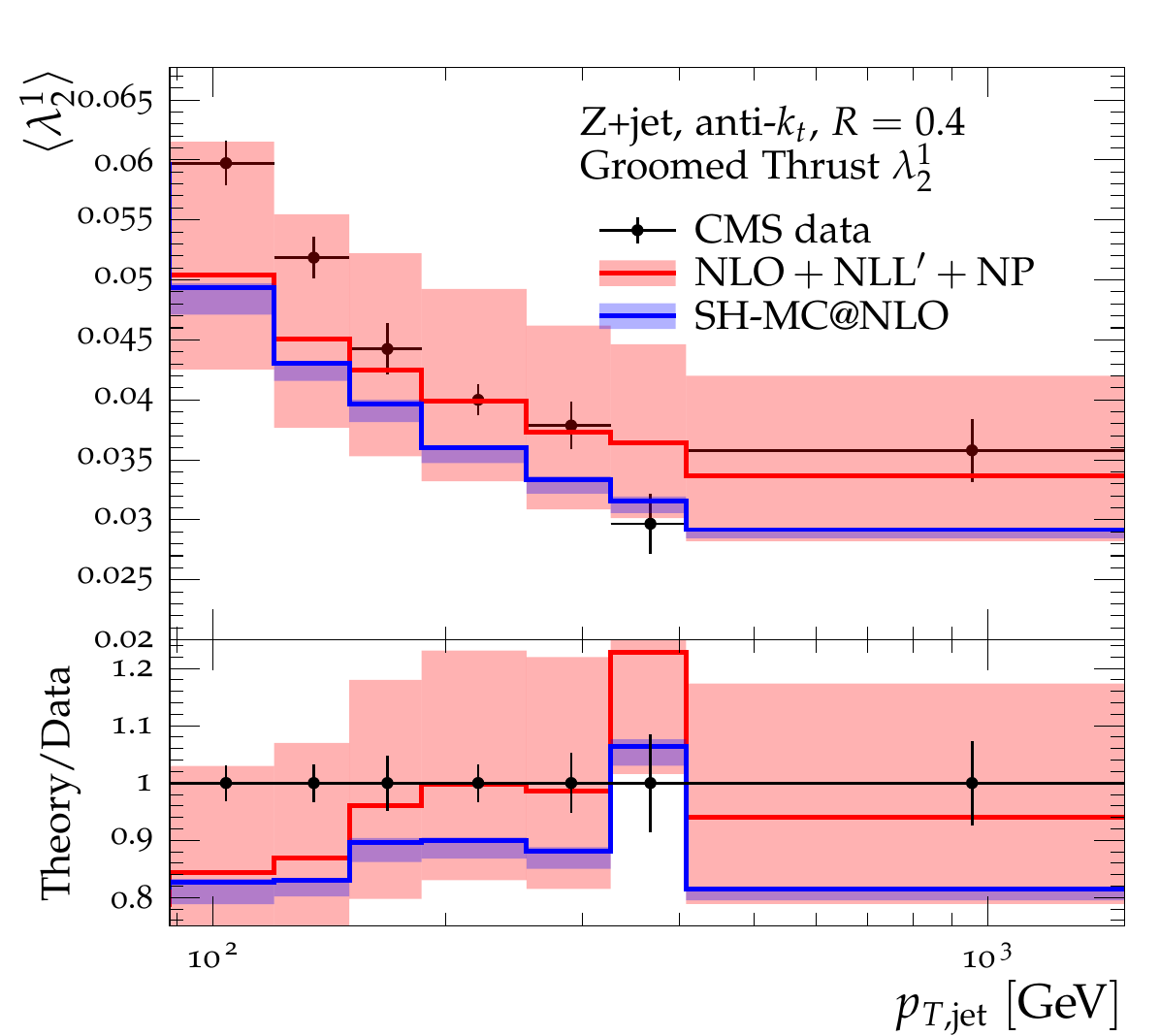}
  \caption{Same as Fig.~\ref{fig:mean_width} but for the mean values of jet thrust, \emph{i.e.}\ $\langle\lambda^1_2\rangle$.}\label{fig:mean_thrust}
\end{figure}

In this appendix, we collect additional results obtained with our \NLOpNLLpNP calculation and \SHMCatNLO
for the jet-width ($\lambda_1^1$) and jet-thrust ($\lambda_2^1$) angularities, for anti-$k_t$, $R=0.4$ jets.
In Fig.~\ref{fig:dist_width} we show the jet-width distributions for $p_{T,\text{jet}}\in[120,150]~\text{GeV}$,
for the central and forward jet in dijet events, as well as the leading jet in $Z$+jet production, for
ungroomed and \softdrop jets. In Fig.~\ref{fig:mean_width} we compile results for the mean value of the
jet-width distributions as a function of the jet transverse momentum. Figs.~\ref{fig:dist_thrust}
and~\ref{fig:mean_thrust} show analogous results for the jet-thrust observable. 

\FloatBarrier

\bibliographystyle{amsunsrt_modp}
\bibliography{references}

\end{document}